\documentclass[letterpaper,11pt]{article}
\pdfoutput=1 % to force arxiv to pdflatex

\usepackage{jheppub}
\usepackage{graphicx}
\usepackage{hyperref}
\usepackage{caption,subcaption} % Subfigures
\usepackage[section]{placeins}

\newcommand{\comment}[1]{}

\title{Classes of Stable Initial Data for Massless and Massive Scalars in 
Anti-de Sitter Spacetime}

\author[a]{Nils Deppe}
\author[b]{and Andrew R.~Frey}

\affiliation[a]{Department of Physics, Cornell University\\
Ithaca, New York 14853, USA}

\affiliation[b]{Department of Physics and Winnipeg Institute for
Theoretical Physics, University of Winnipeg\\
Winnipeg, Manitoba R3B 2E9, Canada}

\emailAdd{nd357@cornell.edu}
\emailAdd{a.frey@uwinnipeg.ca}

\abstract{Since horizon formation in global anti-de Sitter spacetime is dual 
to thermalization of a conformal field theory on a compact space, 
whether generic initial data is stable or unstable against gravitational
collapse is of great interest.  We argue that all the known stable initial
data for massless scalars are dominated by single scalar eigenmodes,
specifically providing strong numerical evidence consistent with the
interpretation that initial data with equal energies in two modes collapse
on time scales of order the inverse square of the amplitude.  We further
scan the parameter space for massive scalar field initial data
and present evidence for a novel class of stable or
quasi-stable solutions for massive scalars with energy spread through 
several eigenmodes.}


\begin{document}
\maketitle

\section{Introduction}\label{s:intro}

Through the AdS/CFT correspondence, gravitational physics on (global) 
$d+1$-dimensional anti-de Sitter spacetime (AdS$_{d+1}$) is 
dual to a conformal field theory (CFT) on $\mathbb{R}\times S^{d-1}$
(the boundary of AdS), with
classical gravity valid in the strong coupling regime of the CFT.  
Thermal states of the CFT are dual to black holes in the bulk AdS 
spacetime,\footnote{classically --- semi-classically, black holes smaller
than the AdS curvature radius are thermodynamically unstable to Hawking
radiation, so the correct dual to a low-temperature thermal CFT state can be
either a gravitational radiation gas or a small black hole in equilibrium
with such a gas.} so thermalization of an initial distribution of energy in
the CFT is dual to horizon formation, ie, gravitational collapse, in the
bulk AdS.  It seems natural to expect that even small amounts of added 
energy confined to a compact space generically eventually thermalize, so
we are led to a perhaps surprising conclusion that generic initial data
for matter in AdS should lead to black hole formation.  On the gravitational
side of the correspondence, the reason is that massless fields can reach
spatial infinity and reflect off the boundary in finite time, so energy
cannot disperse as it would in asymptotically flat spacetime (similarly,
massive fields are confined by the effective gravitational potential of AdS).

From this point of view, there are two natural questions.  First,
are there classes of initial data that are stable against gravitational
collapse to a black hole in the bulk AdS (especially with measure greater
than zero), and what is the dual CFT interpretation of this initial data?
Second, in the case of gravitational collapse, how long does given initial
data take to form a horizon --- or, in CFT terms, how long does given initial
data take to thermalize?  Strictly speaking, both bulk horizon formation 
and CFT thermalization take infinite boundary time, so we can formulate
the question more precisely by asking how long it takes for energy to spread
through a large number of frequency modes, which typically happens just 
before an approximate horizon forms in the bulk.\footnote{Again, it is not
possible for strict equipartition of a finite energy through an infinite
number of modes, so we have an approximate notion in mind.}
We therefore take bulk horizon formation, as we will define approximately 
in section \ref{s:methods} below, as an indication of boundary CFT 
thermalization as well.  This second question is actually easier to answer
in the unstable case: at low perturbation amplitudes, self-gravitation can only
become important on time scales of order $\epsilon^{-2}$, where $\epsilon$
is the amplitude.  In fact, the first question is often phrased in terms 
of stability over times of order $\epsilon^{-2}$.

In 2011, \cite{Bizon2011} pioneered numerical work on these
questions (see also \cite{Garfinkle:2011hm,Jalmuzna2011,Garfinkle:2011tc}).  
Specifically, \cite{Bizon2011} presented intriguing numerical
evidence that perturbations of AdS are generically unstable to black hole 
formation, at the expected time scale for low amplitudes.
The original studies of massless scalar field matter have since been 
followed by studies of complex scalar fields \cite{Buchel2012}
and modified theories of gravity \cite{Deppe2015}.  Except when there is a 
mass gap in the black hole spectrum (such as AdS$_3$ 
\cite{Bizon2013,Jalmuzna:2013rwa} or AdS$_5$ Einstein-Gauss-Bonnet gravity
\cite{Deppe2015}), these numerical studies have pointed to instability 
to horizon formation at
arbitrarily small amplitudes for generic initial data, which occurs along
with a cascade of energy into high frequency modes.  On the other
hand, certain initial data appear stable at low amplitudes, and additional
types of initial data are the subject of some disagreement
(see \cite{Balasubramanian2014,Bizon2014a,Buchel2015}).

The last several years have seen a simultaneous effort to understand AdS
gravitational collapse in perturbation theory, particularly with the
development of an expansion in terms of the scalar eigenmodes on a fixed
AdS background \cite{Balasubramanian2014,Craps2014,Craps2015,Buchel:2014xwa}.
Among other symmetries, the perturbation theory obeys a scaling law
$A(t)\to\epsilon A(t/\epsilon^2)$ (where $A$ is the coefficient of an eigenmode 
in the expansion), which is also observed in numerical
calculations at low amplitude \cite{Bizon2011} and, as suggested above,
leads to a universal prediction that horizon formation takes a time of
order $\epsilon^{-2}$ in the perturbative regime regardless of field mass or 
higher curvature terms in the gravitational action \cite{Buchel:2014dba}.  
The same scaling symmetry is also apparent in a perturbative calculation
for a thin-shell scalar field profile \cite{Dimitrakopoulos:2014ada}.
However, the implications of the perturbation theory for stability of 
generic initial data at arbitrarily small amplitude are as yet unclear.
Conservation laws of the perturbation theory imply the possibility of 
inverse cascades of energy returning to low-frequency modes, which have
been verified in numerical simulations (see \cite{Balasubramanian2014},
for example, and our results) and which may be associated with stability
at low amplitudes.  Further,
\cite{Dimitrakopoulos:2014ada,Dimitrakopoulos:2015pwa} have argued that
(quasi-)periodic solutions of the perturbation theory should persist as
stable solutions of the full theory at arbitrarily small amplitudes.
On the other hand, 
\cite{Bizon2015} have argued for a singularity in the time derivative
of the phases $B_n$ in generic
solutions of the perturbation theory,\footnote{\cite{Craps:2015iia} appeared while
this work was in the final stages of preparation and argues the
singularity is not present in a different gauge, specifically where
the time coordinate is boundary time.} which may be related to horizon formation
in the full nonlinear theory.
As a result, stability, instability, or both may be generic in the space
of initial data (in the sense of being open sets).

Given the existing tension, it seems useful to take stock of the
characteristics of initial data that are agreed to be stable at low amplitudes.
From the initial studies, both numerical and perturbative analysis agree
that (nonlinearly modified) scalar eigenmodes are stable
\cite{Bizon2011,Buchel2013}; these are known as oscillons or, in the complex
scalar case, boson stars.\footnote{In the case of metric perturbations,
geons as described in \cite{Dias:2011ss} are the equivalent stable eigenmodes.}
In fact, perturbations around the oscillon solutions are expected to be
stable as well \cite{Dias:2012tq}, in agreement with numerical studies.
We note here that all subsequently developed solutions for which there is
numerical evidence of stability on 
long time scales are perturbed oscillons in the sense that their energy
spectrum is dominated by the contributions of a single perturbative 
eigenmode;
more precisely, one mode has at least twice the energy of any 
other individual mode.
For example, initial data of Gaussian shape and width
near the AdS scale $\ell$ is apparently stable 
\cite{Buchel2013,Maliborski:2013ula,Maliborski2013b} because it is 
actually dominated by a single scalar eigenmode; somewhat wider initial 
data requires a stronger admixture of other modes and is not quasi-stable.
Likewise, the periodic solutions of \cite{Maliborski2013,Fodor2015}
are dominated by a single mode as shown by their spectra, and, though 
\cite{Balasubramanian2014,Green:2015dsa} present 
a more general ansatz for quasi-periodic solutions to the perturbation theory, 
the physical solutions presented are all dominated by one mode.
The highest temperature quasi-periodic solution shown in
figure 1 of \cite{Green:2015dsa} still has the $j=0$ mode having
approximately twice as much energy as the $j=1$ mode.  
It is also worth noting that, in most cases, a single mode dominates the
spectrum even more strongly in the sense that it has more than 
60\% of the total energy.  Those initial data for which apparently stable
numerical solutions of the full theory have been presented in the literature 
are all of the latter type.  While high temperature quasi-periodic
solutions of the perturbation theory do exist with considerably less energy
in the dominant mode, the behavior of these has not to our knowledge been
evaluated in the full theory.
As the reader may suspect, there are also controversial cases: 
\cite{Balasubramanian2014,Buchel2015} hinted at stability of initial data
superposing two eigenmodes based on a perturbative calculation, which 
has been criticized by \cite{Bizon2014a}.

In this paper, we take a numerical approach to the stability of
scalar fields in AdS at small amplitude, specifically looking at two
questions.  First, in section \ref{s:AdS4}, we consider massless scalars
in AdS$_4$ and provide an independent numerical
analysis of the two-mode initial data of \cite{Balasubramanian2014}
as well as the superposition of three Gaussians used in \cite{Okawa2015}.
We find that, when the field is stable against gravitational collapse 
over times that grow faster than $\epsilon^{-2}$ with decreasing amplitude,
the energy spectrum is dominated by a single scalar eigenmode as described
above.  (This condition is similar to the condition of \cite{Green:2015dsa}
that stable solutions are ``close'' to a quasi-periodic solution of the
perturbation theory.)
Our calculations also point out difficulties in determining the
reliability of the perturbative expansion even in the low amplitude regime
when analytical arguments are not available.

In section \ref{s:massive},
we then turn our attention to the case of scalars with mass $\mu\neq 0$
(corresponding to backgrounds for irrelevant operators in the CFT), which
have so far only been studied numerically in \cite{Okawa2015}.
(We do not consider tachyonic scalars which are allowed for mass squared
above the Breitenlohner-Freedman bound \cite{Aharony1999}.)
Given that massive fields have different stability properties in 
asymptotically flat spacetime, this is an important test case for stability
in AdS.

As a review, consider the gravitational
collapse of massive scalars in asymptotically flat spacetime, where
an initial configuration can either form a black hole or disperse to 
infinity (ignoring self-interactions that, for example, lead to star birth
in astrophysics).  Given initial data of characteristic width $\lambda$
and $\lambda\mu\ll 1,$
\cite{Brady1997} found that collapse proceeds as for a massless scalar,
with power-law scaling behavior for the horizon radius near the
critical point between black hole formation and dispersion
\cite{Choptuik:1992jv} (known as a type II transition, see 
\cite{Choptuik:1996yg}).
However, for $\lambda\mu\gg 1$, black holes form only above a finite mass
gap (a type I transition).  In essence, the difference in these two
types of transition is whether the scalar potential or gradient energy
dominates the energy of the initial data.
For a massive scalar field in asymptotically AdS spacetime with curvature
radius $\ell$, the different ratios to be considered are
$\lambda\mu$, $\ell\mu$ and $\lambda/\ell$.  In principle, the stability
properties of the system can change whenever these ratios pass through
unity.

We present an overview of
the behavior of horizon formation time as a function of these ratios
in section \ref{s:massive}, followed by a more detailed analysis of 
the behavior at low amplitudes.  For widths $\lambda$
between the Compton wavelength and AdS scale, we find two types of 
particularly interesting behavior, depending on whether the scalar is 
light or heavy compared to the AdS scale.  For light fields, we find a
discontinuity in the horizon radius at formation as a function of the
amplitude of initial data, possibly corresponding to a change in
efficiency of thermalization in the boundary CFT.  We analyze this behavior
in more detail in section \ref{s:shrink}.
Of particular interest, for heavy scalars, we find a 
class of stable (over time scales of order $\epsilon^{-2}$) 
initial data; we present evidence in section 
\ref{s:quasistable} that these solutions are qualitatively
distinct from the single-eigenmode-dominated solutions discussed above
in that the energy is distributed roughly evenly through several modes.

We conclude with a discussion of our results and begin here with an
overview of our methods.

\section{Methods}\label{s:methods}
The evolution of a massive scalar field in AdS
is governed by the Einstein equations and the wave equation,
\begin{align}
  G_{ab}+\Lambda g_{ab}=&8\pi\left(\phi_{;a}\phi_{;b}-
  \frac{1}{2}g_{ab}(\phi_{;c}\phi^{;c}+\mu^2\phi^2)\right),\quad\phi_{;a}{}^a-\mu^2\phi=0,
\end{align}
where the mass of the scalar field $\phi$ is $\mu$. \comment{We use commas
for partial derivatives and semi--colons to denote covariant
differentiation.} Following \cite{Bizon2011}, 
we choose a spherically symmetric metric ansatz
in Schwarzschild-like coordinates
\begin{align}
  ds^2=\frac{\ell^2}{\cos^2(x/\ell)}\left(Ae^{-2\delta}dt^2+
  A^{-1}dx^2+\sin^2(x/\ell)d\Omega^{d-1}\right),
\end{align}
where $d$ is the number of spatial dimensions.  The areal radius
is $R(x)=\ell\tan(x/\ell)$, and we henceforth work in units of 
the AdS scale $\ell$ (ie, $\ell=1$).

The equations of motion are calculated using a Hamiltonian analysis,
as in \cite{Kunstatter2013}. The evolution
equations governing the nonlinear system are
\begin{align}
  \label{eq:scalarField}
  \phi_{,t}=Ae^{-\delta}\Pi,\quad\Phi_{,t}=\left(Ae^{-\delta}\Pi\right)_{,x},
    \quad\Pi_{,t}=\frac{(Ae^{-\delta}\tan^{d-1}(x)\Phi)_{,x}}{\tan^{d-1}(x)}-
  \frac{e^{-\delta}\mu^2\phi}{\cos^2(x)}\ ,
\end{align}
where $\Pi=A^{-1}e^\delta\phi_{,t}$ is the canonical momentum and 
$\Phi=\phi_{,x}$ is an auxiliary variable.  
The metric functions are determined by
\begin{align}
  \label{eq:deltaDeriv}
  \delta_{,x}=&-\sin(x)\cos(x)(\Pi^2+\Phi^2)\\
  \label{eq:massDeriv}
  M_{,x}=&\left(\tan(x)\right)^{d-1}\left[
    A\frac{\left(\Pi^2+\Phi^2\right)}{2}+
    \frac{\mu^2\phi^2}{2\cos^2(x)}\right],\quad
  A=1-2\frac{\sin^2(x)}{(d-1)}\frac{M}{\tan^{d}(x)},
\end{align}
where $M$ is the mass function and $M(t,x=\pi/2)_{,t}=0$.
We choose $\delta(t,x=0)=0$.  

The linearized system is given by
\begin{align}
  \label{eq:linearL}
  \phi_{,tt}=\frac{d-1}{\sin(x)\cos(x)}\phi_{,x}+\phi_{,xx}-
  \frac{\mu^2}{\cos^2(x)}\phi:=-L\phi
\end{align}
(see \cite{Ishibashi2004} for the mathematical properties of this system).
The eigenfunctions of the operator $L$ are given by
Jacobi polynomials (see \cite{Aharony1999} for a review)
\begin{align}
  e_j(x)&=\kappa_j\cos^{\lambda_\pm}(x)
  P^{(d/2-1,\pm\sqrt{d^2+4\mu^2}/2)}_j(\cos(2x)) .
\end{align}
The normalization and eigenvalues are given by
\begin{equation}
  \kappa_j=\sqrt{\frac{2(2j+\lambda_\pm)j!\Gamma(j+\lambda_\pm)}{
      \Gamma(j+d/2)\Gamma(j\pm\sqrt{d^2+4\mu^2}/2+1)}},\quad
  \omega_j=\lambda_\pm+2j,\quad
  \lambda_\pm=\frac{d}{2}\pm\frac{1}{2}\sqrt{d^2+4\mu^2}.
\end{equation}
In this work we choose $\lambda_+$, corresponding to the normalizable
modes according to the inner product
\comment{on the Hilbert space $L^2([0,\pi/2],\tan^{d-1}(x)dx)$ by}
$(f,g):=\int_{0}^{\pi/2}f(x)g(x)\tan^{d-1}(x)dx$.

As in \cite{Bizon2011}, we define an ``energy per mode'' by projecting
the field onto eigenmodes.  Specifically, we define 
$\Pi_j:=\left(\sqrt{A}\Pi,e_j\right)$,
$\phi_j:=\left(\phi,e_j\right)$, and\\
$\ddot{\phi}_j:=\left(\tan^{-(d-1)}(x)\partial_x\left(\tan^{d-1}(x)
A\Phi\right)-\mu^2\phi\cos^{-2}(x),e_j\right)$, so 
the energy spectrum is
\begin{align}
  \label{eq:modeEnergy}
  E_j:=\frac{1}{2}\left(\Pi_j{}^2-\phi_j\ddot{\phi}_j\right).
\end{align}
The total ADM energy is $M_{\text{ADM}}=\sum_{j=0}^{\infty}E_j$.
We use these projections to study how much
energy is captured in modes up to some $j_{max}$ and how
energy is transferred between modes throughout the
evolution.

As far as we are aware, only initial data (ID) that is either one or
more Gaussians in $\Pi$ \cite{Bizon2011,Jalmuzna2011,
Buchel2012,Buchel2013,Bizon2013,Maliborski2013b,Deppe2015,Okawa2015},
a superposition of two eigenmodes \cite{Balasubramanian2014,Bizon2014a}, 
(fake) boson stars \cite{Buchel2013}, or a
specially constructed time--periodic solution \cite{Maliborski2013} has
been studied. We consider four types of initial data.
We primarily study Gaussian data in $\Pi$,
\begin{align}\label{eq:PiGaussianID}
\Pi(t=0,x)=\epsilon\exp\left(-\frac{\tan^2(x)}{\sigma^2}\right),
\quad\phi(t=0,x)=0,
\end{align}
similar to that originally studied in \cite{Bizon2011}.
For a Gaussian pulse of width $\sigma$, like eq.~(\ref{eq:PiGaussianID}), 
we define the wavelength to be $\lambda=2\sigma$.  As discussed in the
introduction, one goal of this work is to explore the interplay among
the length scales $\lambda$, $\ell$, and $1/\mu$.
To this end, we have
considered several different values of $\sigma$ and $\mu$.

To explore the universality of our results, we also consider initial data
in the form of an ingoing pulse
\begin{align}
  \label{eq:phiGaussianID}
  \phi(t=0,x)=\epsilon\tan^2(x)\exp\left(-\frac{\tan^2(x)}{\sigma^2}\right),
\quad\Pi(t=0,x)=\phi(t=0,x)_{,x}.
\end{align}
This data is more difficult to evolve numerically, specifically
as collapse ensues, so we do not perform simulations for all the
values of $\sigma$ and $\mu$ used for the Gaussian initial
data in $\Pi$. Nonetheless, as we present in section \ref{s:profiles}
below, our results appear robust against changes in initial data.

For comparison to existing literature, we also consider two other classes
of initial data.  First, \cite{Balasubramanian2014} studied 
two-eigenmode initial data in AdS$_4$, specifically,
\begin{align}
  \label{eq:twoModeID}
  \phi(t=0,x)=\epsilon(e_0(x)+\kappa e_1(x))/3,
\end{align}
where $\kappa$ is chosen freely.
Finally, \cite{Okawa2015} considered a superposition of three
Gaussian wavepackets in $\Pi$
\begin{align}
  \label{eq:manyGaussID}
  \Pi(t=0,x)=\frac{2\epsilon}{\pi}\sum_{i=1}^{3}a_i\exp
  \left(-\frac{4(\tan(x)-R_i)^2}{\pi^2\sigma_i{}^2}\right),\quad\phi(t=0,x)=0,
\end{align}
for AdS$_4$.  We comment on both these classes of initial conditions 
as they appear in the previous literature in addition to studying them
ourselves.

Our numerical methods are similar to those used in \cite{Deppe2015} and
are described in greater detail in \cite{Maliborski2013b}.
We use a fixed grid of $2^n$ grid points (with the option to restart at 
higher resolution), mainly with a 4th order Runge-Kutta (RK4) 
spatial integrator.  For some initial data we find it necessary
to use a fifth-order Dormand-Prince (DoPr) method for the spatial integration.
The DoPr method is typically only needed when the solution
approaches collapse and the finer spatial scales need to be resolved 
more accurately.  We terminate the simulation and determine
that a horizon forms at $A(t_H,x_H)\leq 2^{k-n}$ with $k=7$ for $\mu\leq20$,
while for larger masses $k=8$.  We have found that
requiring smaller values of $k$ in some cases requires the formation of
very narrow troughs in the horizon function $A(t,x)$ which are not 
resolved by the spatial grid, so we have chosen these values to reliably
report horizon formation times.
A discussion of the convergence properties of our code is in
appendix \ref{s:convergence}.

\section{Stability of massless scalars in AdS$_4$}\label{s:AdS4}

In this section, we present results of simulations in AdS$_4$ as a 
comparison to the previous literature.  We are particularly concerned with
two sets of initial data which have been claimed to lead to stable
solutions at low amplitude: two-mode initial data and three-Gaussian 
initial data.

\subsection{Two-mode initial data}\label{s:twomode}

\begin{figure}[!t]
  \centering
  \begin{subfigure}[t]{0.47\textwidth}
    \includegraphics[width=\textwidth]{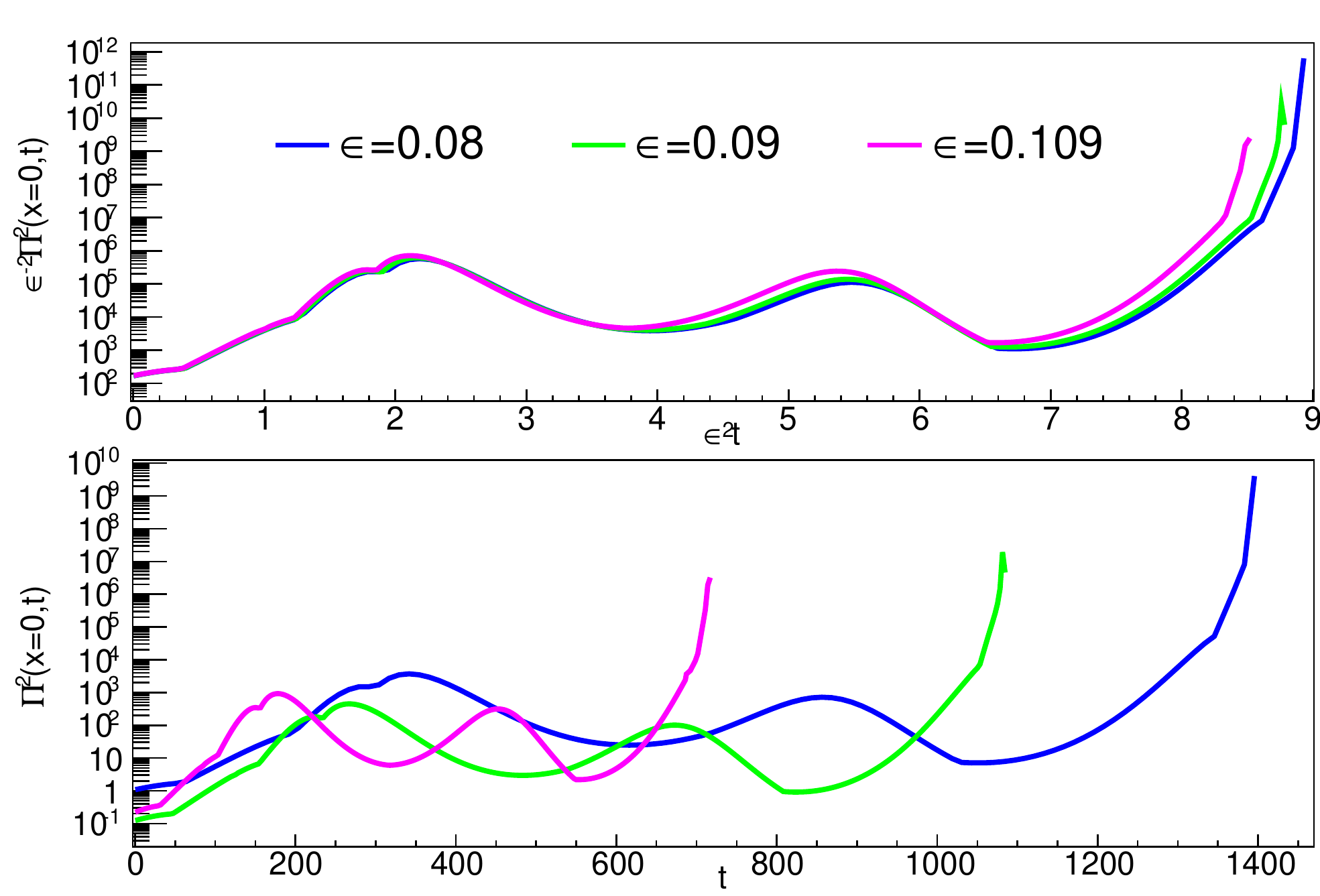}
    \caption{Top panel: rescaled $\epsilon^{-2}\Pi^2(\epsilon^2t,x=0)$
      vs rescaled time $\epsilon^2t$.  Bottom panel: $\Pi^2(t,x=0)$.}
    \label{fig:twoModeAds4Pi}
  \end{subfigure}\hfill
  \begin{subfigure}[t]{0.47\textwidth}
    \includegraphics[width=\textwidth]{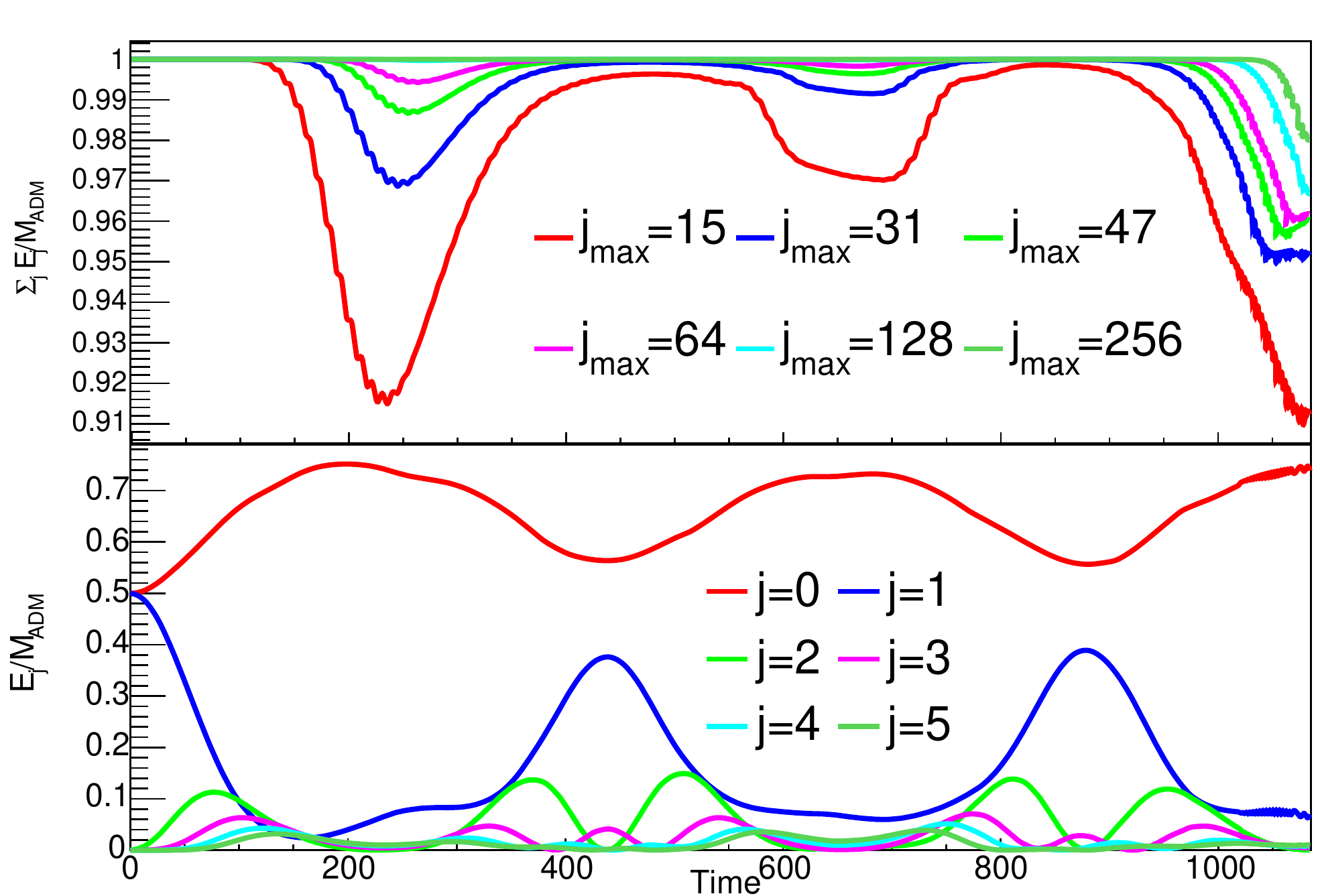}
    \caption{Top panel: sums of
      $\hat{E}_j$ up to $j_{max}$.
      Bottom panel: evolution of $\hat{E}_j$ in
      the lowest six modes. Initial data has $\epsilon=0.09$.}
    \label{fig:twoModeAds4Energy}
  \end{subfigure}
  \caption{Results from the evolution of two-mode initial
    data (\ref{eq:twoModeID}) with $\kappa=3/5$.}
  \label{fig:twoModeDataAds4}
\end{figure}

First, we consider massless scalars with
two-mode initial data as in eq.~(\ref{eq:twoModeID}) and
$\kappa=3/5$, following \cite{Balasubramanian2014,Buchel2015,Green:2015dsa} and
\cite{Bizon2014a}, which is a point of disagreement between the two sets of
publications.\comment{\footnote{A reply to
\cite{Bizon2014a} was posted during preparation of this work\cite{Buchel2015}
as was further discussion of two-mode initial data by many of the same .}}
Specifically, \cite{Balasubramanian2014} suggests stability of 
sufficiently small amplitude two-mode data for the length of their simulations,
including at $\epsilon=0.09$, while \cite{Bizon2014a}
found horizon formation at a long but finite time.
We undergo an independent study of this initial
data for amplitudes $\epsilon=0.109,0.09$ and $0.08$ and find that
all three simulations end in collapse.  
Namely, comparing the bottom panel of figure~\ref{fig:twoModeAds4Pi} to
figures 3 and 4 of \cite{Balasubramanian2014} and 
figure 1 of \cite{Bizon2014a}, shows agreement with Bizo\'n
and Rostworowski \cite{Bizon2014a}.  Below we provide evidence
that the discrepancy is due to insufficient resolution in 
\cite{Balasubramanian2014,Buchel2015}.

Our simulations also show the same scaling properties first noticed 
in \cite{Bizon2011}, namely that $\epsilon^{-2}\Pi^2(\epsilon^2t,x=0)$
is an approximate universal function for a given initial field profile
as long is the simulation is still in the perturbative regime.
This scaling symmetry is manifest in the improved perturbation theory
of \cite{Balasubramanian2014,Craps2014,Craps2015} (also known as ``two-time
formalism'' or TTF), as discussed in the introduction. 
The close agreement after rescaling, as seen in figure
\ref{fig:twoModeAds4Pi}, provides evidence that the numerical 
solutions are still perturbative until shortly before horizon formation.
Nonetheless, the perturbative TTF solution diverges from the nonlinear
numerical solution with the same initial data (including our calculation
and those of \cite{Balasubramanian2014,Buchel2015,Green:2015dsa,Bizon2014a}).
Specifically, in figure 3 of \cite{Balasubramanian2014},
$\Pi^2(t,x=0)$ in the TTF evolution including modes up to $j=47$
falls below the numerical result
by approximately an order of magnitude at $t\approx 200$, while the
numerical solution still agrees with \cite{Bizon2014a} and our results.
Even the TTF solution including modes up to $j=200$ \cite{Green:2015dsa}
diverges strongly from the numerical solution; $\Pi^2(t,x=0)$ is about 
two orders of magnitude too small for $t$ just below 1100.  To be clear, it 
is possible that at smaller $\epsilon$ the fully nonlinear evolution could survive to a larger
value of $\epsilon^2 t$ if $\Pi^2(t,x=0)$ turns over, forming a local
maximum; whether all amplitudes collapse after two local maxima in 
$\Pi^2(t,x=0)$ or exhibit a larger number is an open question.

To gain a better understanding of why the numerical and TTF solutions diverge, 
we perform a spectral decomposition of the energy for
$\epsilon=0.09$. The top panel of figure \ref{fig:twoModeAds4Energy}
shows the total energy in modes up to $j_{max}$, while the bottom panel shows 
the energy in the lowest six modes.\footnote{We actually plot the
normalized energy in the modes, $\hat{E}_j=E_j/M_{ADM}$.}
We first see a cascade of energy into
higher modes, followed by an inverse cascade of energy back into the 
lower modes, in agreement with \cite{Balasubramanian2014}.  
There is a strong cascade into higher modes just before
horizon formation.  Figure \ref{fig:twoModeAds4Energy}
shows that at this point about 4\% of the energy is in
modes greater than $j=47$ and almost 2\% in $j>256$. 
Because higher modes are more
sharply peaked around the origin, even a small amount of
energy in higher modes can cause large differences in the
magnitude of $\Pi^2(x=0,t)$ and in determining whether or not a collapse
occurs. The lesson is that perturbative solutions
may only be reliable if a large number of eigenmodes are considered,
even when the full solution is well within the perturbative regime.
The correct claim of \cite{Buchel2015} that the TTF evolution with 
$j_{max}=47$ accurately captures the dynamics of the lowest modes is 
immaterial to the question of horizon formation because this is a local
question near $x=0$ in the small amplitude limit.  It
appears that a small amount of energy transferred to higher modes is 
sufficient to cause gravitational collapse/thermalization.  

\begin{figure}[!t]
  \centering
  \begin{subfigure}[t]{0.48\textwidth}
    \includegraphics[width=\textwidth]{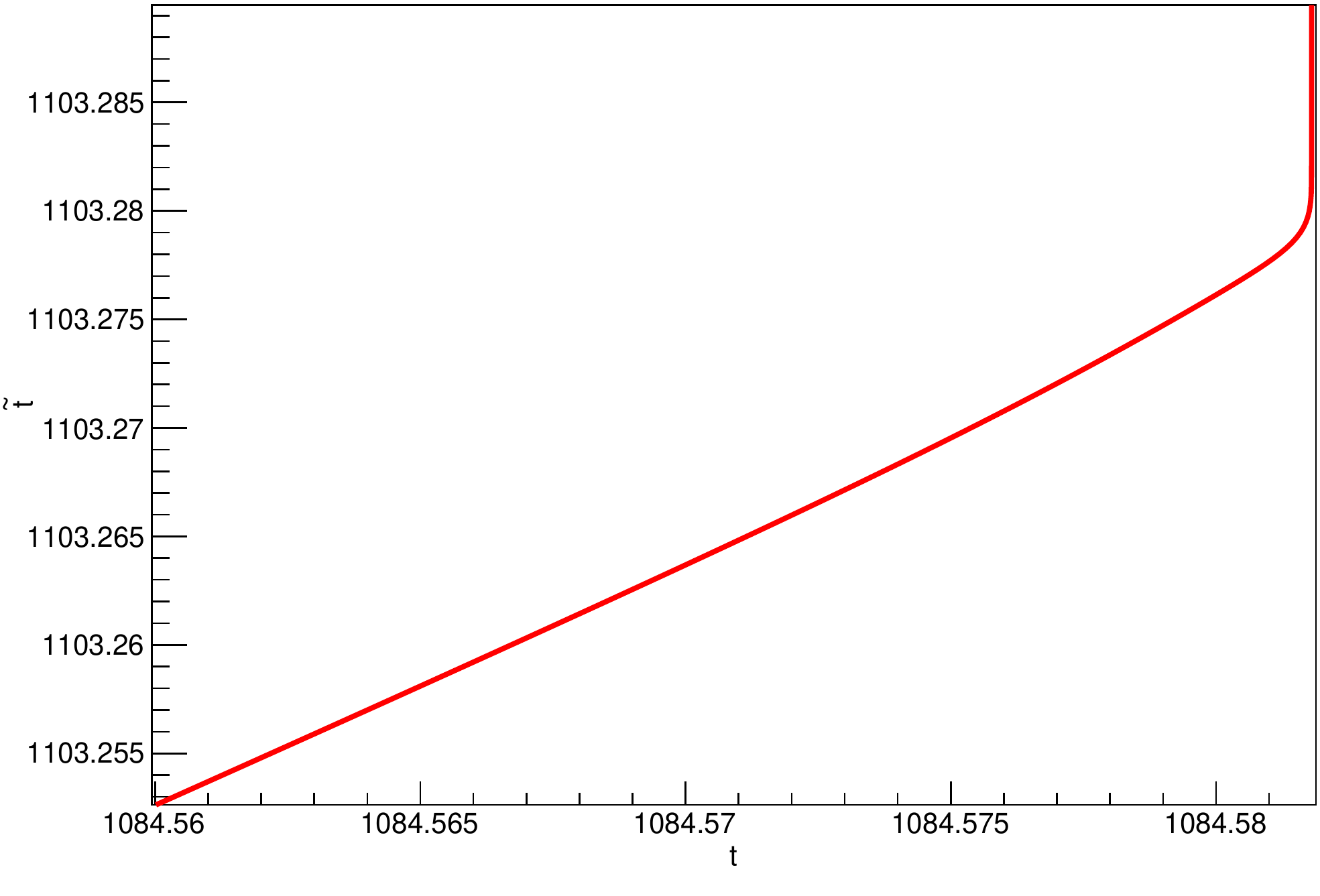}
    \caption{Boundary time $\tilde{t}$ vs
      coordinate time $t$ at resolution $n=19$.}
    \label{fig:timeChange}
  \end{subfigure}\hfill
  \begin{subfigure}[t]{0.48\textwidth}
    \includegraphics[width=\textwidth]{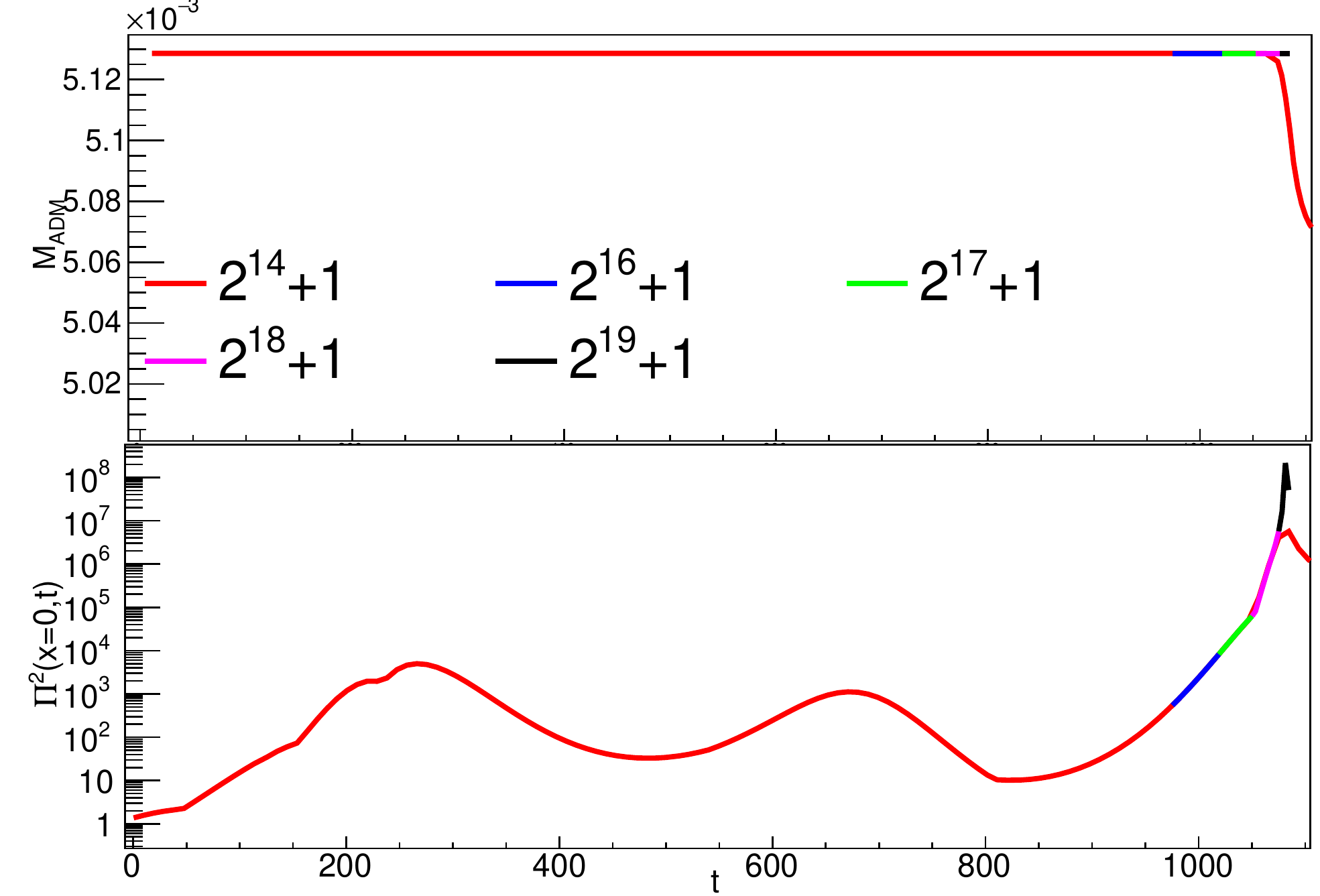}
    \caption{Top panel: ADM mass vs time
      at various resolutions. Bottom
      panel: upper envelope of $\Pi^2(t,x=0)$ for the same simulations.}
    \label{fig:massloss}
  \end{subfigure}
  \caption{Coordinate transformation and consistency of our code at
    several resolutions for the $\epsilon=0.09$ evolution.}
  \label{fig:twoModeFigure}
\end{figure}

We have, however, not yet explained the difference in the non-perturbative
numerical solutions of \cite{Balasubramanian2014} and \cite{Bizon2014a}
at long times.  In principle, one possibility is that these two
references use different time coordinates (ours matches that of
\cite{Bizon2014a}).\footnote{Note, though, that \cite{Green:2015dsa} uses
the same time coordinate we do.}  For comparison,
we perform a numerical coordinate transformation to 
the time coordinate $\tilde{t}$ of \cite{Balasubramanian2014,Buchel2015},
which is chosen such that $\delta(\tilde{t},x=\pi/2)=0$ (this is conformal
time on the boundary of AdS).
The two coordinate times are related by
\begin{align}\tilde{t}=\int_{0}^{t}\exp\left[-\delta(t,x=\pi/2)\right]dt\ ; 
\label{eq:bdryTime}\end{align}
we show the 
relationship between $t,\tilde t$ for late times in figure
\ref{fig:timeChange} for the $\epsilon=0.09$ initial data. 
Because the system is weakly gravitating until collapse,
$\Delta\tilde{t}/\Delta t\approx 1.02$ for most of the simulation; however, as 
the horizon starts to form, time dilation becomes more significant, and the
boundary time passes more quickly. 
We find that collapse is delayed by $\tilde{t}_{H}-t_{H}\approx 20$ for the
$\epsilon=0.09$ evolution for resolution and horizon formation threshold
given by $n=19$.  This is insufficient to account for the lifetime 
$\tilde{t}\approx1500$ of the simulations of 
\cite{Balasubramanian2014}.\footnote{Technically speaking, a lower 
threshold for horizon formation results in greater time dilation.  However,
even though we have been unable to locate a discussion of the threshold
used in \cite{Balasubramanian2014}, we think this is an unlikely source of
the error simply because our threshold is quite low.}  Furthermore,
time dilation only stretches the time axis of figures 
\ref{fig:twoModeDataAds4} and \ref{fig:massloss} and would not add extra
oscillations as in figures 3 and 4 of \cite{Balasubramanian2014}.

Instead, the difference between \cite{Balasubramanian2014,Buchel2015} and
\cite{Bizon2014a} appears entirely due to insufficient resolution in the
simulations of \cite{Balasubramanian2014}.  Specifically,
the consistency and convergence
tests presented in \cite{Balasubramanian2014} were stopped
much earlier than the times in the evolution that are in question,
the longest ending at $\tilde{t}\approx 600$.  As an illustration,
the simulation depicted in red in figure \ref{fig:massloss}
has grid spacing $\Delta\approx9.6\times10^{-5}$, a higher resolution
than used in any of the tests in \cite{Buchel2015}. The rapid decrease
in the ADM mass shows that late in the evolution our code also
suffers from loss of convergence if insufficient resolution is used, and
the evolution experiences a similar ``afterlife'' to those of
\cite{Balasubramanian2014,Buchel2015}.
We have determined that this is due to the largest amplitude pulse
sharpening and eventually being squeezed between grid points.
Figure \ref{fig:massloss} also shows higher resolution simulations
where the relative change in ADM mass is $\Delta M/M\approx 7\times10^{-8}$
for the entirety of the evolution.

Unfortunately, the convergence tests presented in \cite{Buchel2015} are
insufficient to evaluate whether or not
the numerical solution is converging to the continuum solution.
Typically when performing convergence tests either the number of 
grid points is increased
by a constant amount or the resolution is increased by a constant factor.
The convergence test presented in figure 1 of \cite{Buchel2015}
does not do either of these.  Rather, what the figure shows is that 
their code converges to the solution with $\Delta\approx 1.2\times10^{-4}$ 
as $\Delta$ asymptotically approaches this value.
Meanwhile, we have also
compared numerical values of $\Pi^2(t,x=0)$ with \cite{Bizon2014a}
and obtain the same results to several significant figures; an exact
match is not expected due to slight differences in the algorithms used.
The results of our study suggest that the disagreement between
\cite{Balasubramanian2014} and \cite{Bizon2014a} would be resolved if
\cite{Balasubramanian2014} used sufficient resolution to be in the convergent
regime late in the evolution.

At this point, it is worth assessing the importance of analyzing such
apparently well-trodden ground (as \cite{Buchel2015} seemingly wonders).
As we stated in the introduction above, one of the key questions about 
gravitational collapse in AdS is what classes of initial data can lead to
stable evolution at small amplitude.  While numerical studies cannot answer
this question definitively, it is still critical to be careful about 
whether numerics have actually provided evidence of stability or not.
In this case, equal-energy 
two-mode initial data, if stable, would present a qualitatively
distinct class of stable initial data --- all the apparently stable 
initial data for massless scalars is dominated by a single eigenmode.
Indeed, some of the authors of \cite{Balasubramanian2014} postulated
in \cite{Buchel:2014xwa} that the two-mode initial data is close to a 
quasi-periodic solution (of a type formulated in \cite{Balasubramanian2014}),
but even the quasi-periodic solution closest to their two-mode initial data
has 70\% of its energy in the $j=0$ eigenmode.  It is also important to
realize that, while perturbative methods are very powerful, they also
suffer from stringent resolution requirements (ie, many scalar eigenmodes
must be included) because gravitational collapse is an essentially local
question at low amplitude.  In other words, as \cite{Green:2015dsa}
points out, it is the behavior of the very high $j$ tail of the energy 
spectrum that determines the ultimate fate of the system.
In summary, while it is impossible at this time to rule out stability of
equal-energy two-mode initial data as $\epsilon\to 0$, all current 
numerical evidence is consistent with instability to horizon formation 
over times of order $\epsilon^{-2}$ precisely due to the contributions of
high $j$ modes.

\subsection{Multiple-Gaussian initial data}\label{s:multiGauss}

\begin{figure}[!t]
  \centering
  \begin{subfigure}[t]{0.48\textwidth}
    \includegraphics[width=\textwidth]{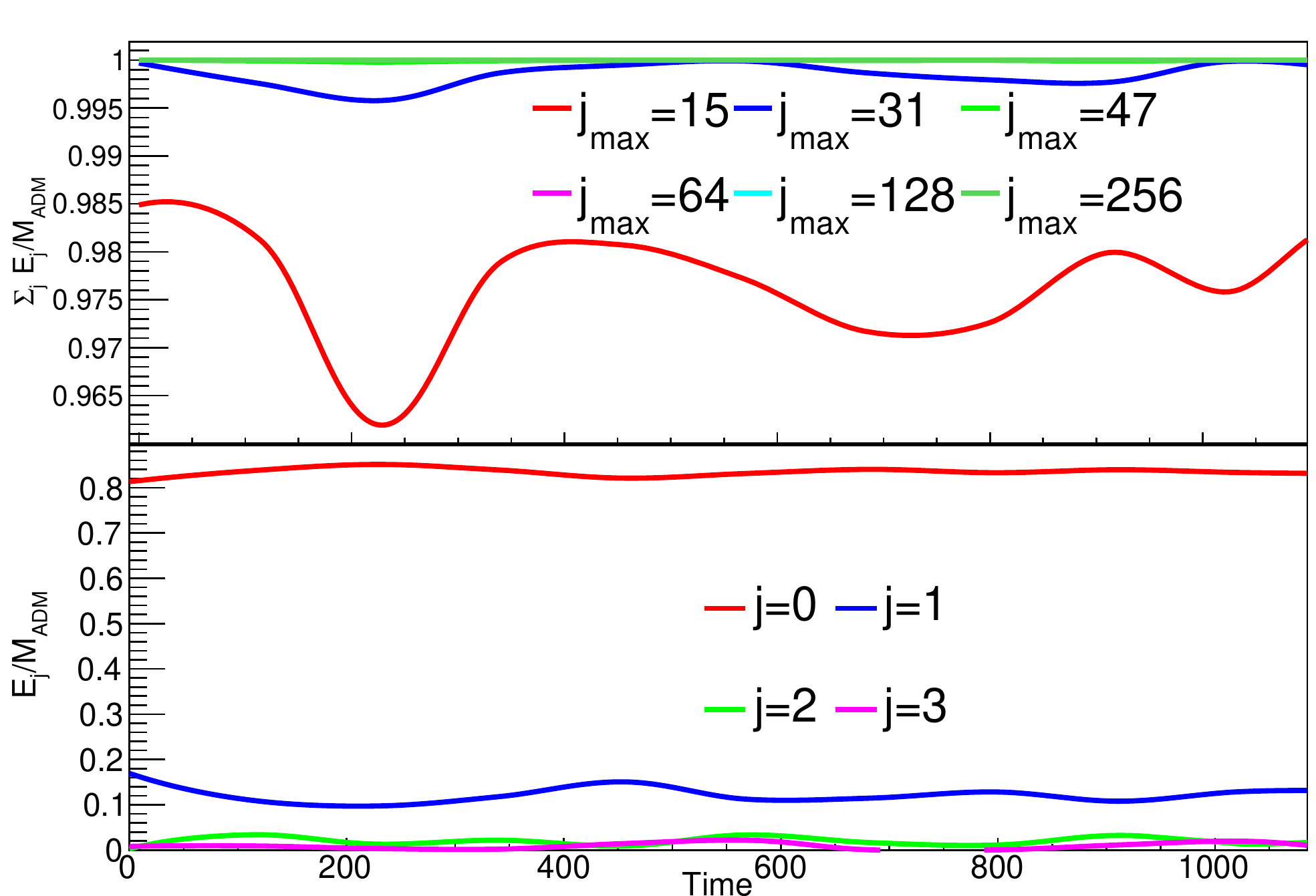}
    \caption{Top panel: sums of the $\hat E_j$ up to $j_{max}$.
      Bottom panel: evolution of the energy in
      the lowest four modes.  $\epsilon=1$. }
    \label{fig:CardosoEnergy}
  \end{subfigure}\hfill
  \begin{subfigure}[t]{0.48\textwidth}
    \includegraphics[width=\textwidth]{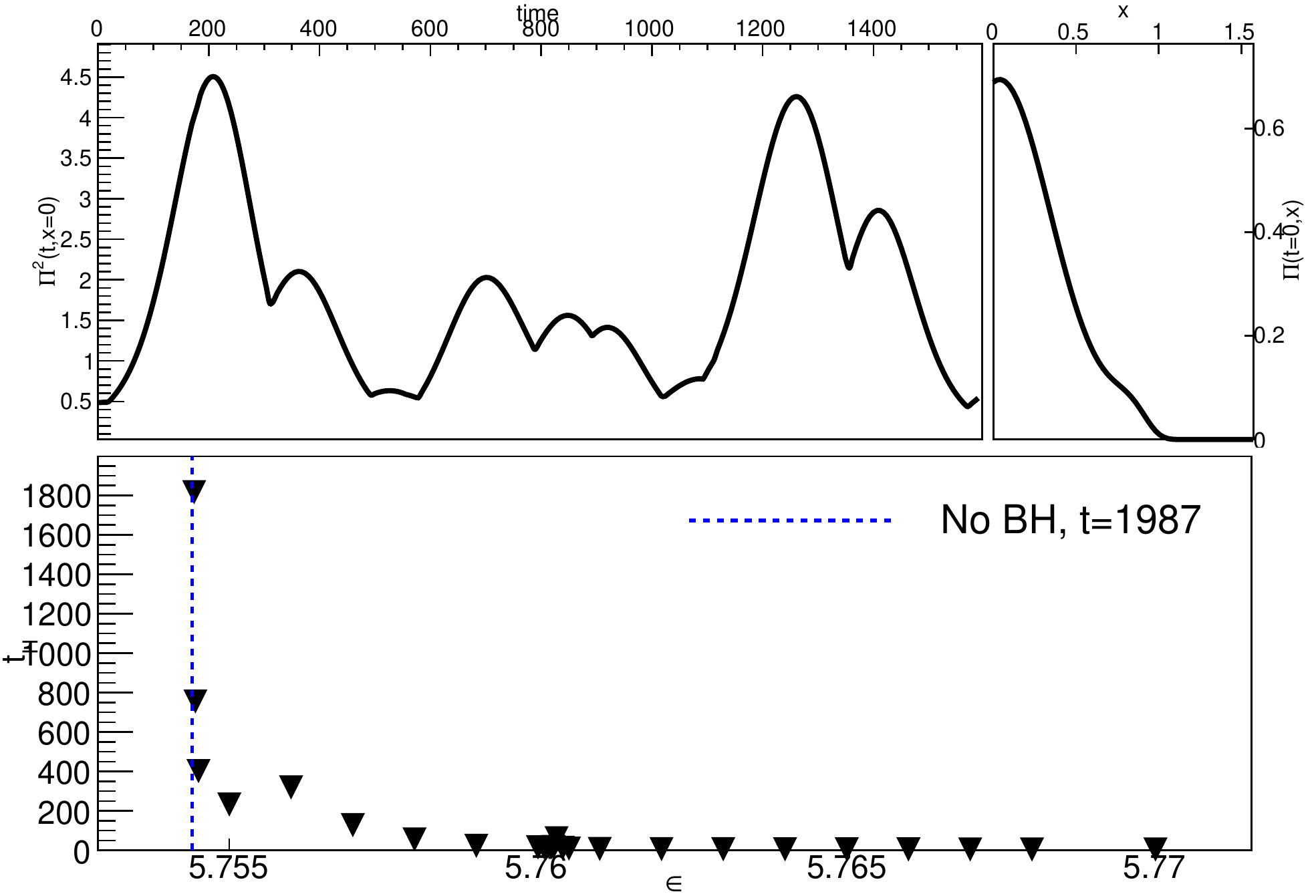}
    \caption{Top panel: $\Pi^2(t,x=0)$ (left) and
      the initial profile (right) for $\epsilon=1$.
      Bottom panel: $t_H$ vs $\epsilon$.}
    \label{fig:CardosoPi}
  \end{subfigure}
  \caption{Results for three-Gaussian
    initial data (\ref{eq:manyGaussID}). Below an amplitude of 
$\epsilon_*=5.756...$, we find no horizon formation out to a simulation
time of $t=1987$. }
  \label{fig:3GaussCardoso}
\end{figure}

We also study the multiple Gaussian initial data of eq.~(\ref{eq:manyGaussID})
for massless scalars in AdS$_4$, which was first
considered in \cite{Okawa2015} with parameters 
$\{a_1,\sigma_1,R_1\}=\{1,1/4,0\}$,
$\{a_2,\sigma_2,R_2\}=\{1/4,1/4,\tan(\pi/8)\}$, and
$\{a_3,\sigma_3,R_3\}=\{1/8,1/4,1\}$. We have simulated the collapse of
this initial data and find, as in \cite{Okawa2015}, that it is apparently
stable for small amplitudes.  To understand why this data is stable, we
performed a spectral decomposition as a function of time. 
The top panel of figure \ref{fig:CardosoEnergy} shows the total 
energy up to mode $j_{max}$, while the bottom panel shows
the energy in each of the lowest four modes, both for
$\epsilon=1$. We find that approximately 82 per cent of the energy
is in the lowest mode, essentially independent of time. 
This suggests that this initial data is a
perturbation about a single mode, which
is known to be stable \cite{Bizon2011,Buchel2013,Maliborski2013}.
For the same simulation we plot the upper envelope of
$\Pi^2(t,x=0)$ in the top left
panel of figure \ref{fig:CardosoPi}. At least
for the duration of our simulation, there is no increase in the
Ricci scalar at the origin, as is naively expected for stable
solutions.

Our initial data for $\epsilon=1$ is plotted in the top right panel
of figure \ref{fig:CardosoPi} and appears to match the bottom right
panel of figure 9 in \cite{Okawa2015}.  
We observe qualitatively similar behavior to \cite{Okawa2015}, but 
our quantitative results differ significantly.  
For large amplitudes we observe rapid horizon formation, but, with
decreasing $\epsilon$, we find a small region where the
collapse time quickly increases and then decreases again.
Finally, near $\epsilon=5.756$ we observe another rapid
increase in formation time. Yet another small decrease in
collapse time is observed followed by another increase. It is
unclear if this behavior will recur indefinitely. In contrast, 
\cite{Okawa2015} observes a sudden increase in collapse time at
$\epsilon\approx1.75$ and no earlier increase (see their figure 8). 

\section{Collapse of massive fields in AdS$_5$}\label{s:massive}

We now turn to an overview of the behavior of massive scalar fields
in AdS.  For specificity, we work in AdS$_5$, which corresponds to a
dual 4D gauge theory.

Our primary motivation is to consider different values of the dimensionless
parameters $\lambda\mu$, $\ell\mu$, and $\lambda/\ell$ and to determine
when the behavior of massive fields diverges from that of massless fields.
For $\ell\mu<1$ (light fields), we consider initial conditions with 
width satisfying $\lambda\mu<\lambda/\ell<1$,  
$\lambda\mu<1<\lambda/\ell$, and $1<\lambda\mu<\lambda/\ell$.  Similarly,
for $\ell\mu>1$ (heavy fields), we consider the $\lambda/\ell<\lambda\mu<1$,
$\lambda/\ell<1<\lambda\mu$, and $1<\lambda/\ell<\lambda\mu$ cases.
In most cases, we find that decreasing initial amplitude leads to 
monotonically increasing horizon formation times, though the behavior may
differ in detail from massless scalars.  
However, as for massless scalars, we also find evidence
of stability against horizon formation at small amplitudes for 
$\lambda\sim\ell$, when a single eigenmode dominates the spectral distribution.
We have also uncovered an apparently distinct class of (quasi-)stable
initial conditions, which are not dominated by any single scalar eigenmode.
We discuss those in more detail in section \ref{s:quasistable}.

We consider multiple classes of initial conditions,
demonstrating that the behavior displayed is robust.

\subsection{Overview}\label{s:times}

For narrow pulses, $\lambda$ less than both $1/\mu$ and $\ell$, we 
expect the scalar fields to act effectively massless, whereas wider pulses
may exhibit different behavior.  More precisely, since
\cite{Brady1997,Garfinkle2003}
found a phase transition at $\lambda\mu\sim 1$, we also may expect a 
transition in behavior as the initial pulse width increases past either
the AdS radius or Compton wavelength.  However, because the scalar field
can disperse and then reflect back to the origin (recall that massive fields
are confined by a gravitational potential in AdS, even though they cannot
travel to the AdS boundary), the initial distinction in pulse width may be 
erased after sufficient reflections, leading to common behavior at low
amplitudes (indeed, this is suggested by the perturbative analysis of
\cite{Buchel:2014dba}).  The AdS radius also sets the scale of the 
gravitational potential, so the cross-over between light and heavy fields
and narrow initial pulses compared to $\ell$ may also change the behavior
of the gravitational collapse.  Heuristically, we expect the behavior of 
the collapse to be controlled by which type of energy dominates:
gradient energy, scalar potential, or gravitational potential.

\begin{figure}[!t]
 \begin{subfigure}{0.48\textwidth}
  \includegraphics[width=\textwidth]{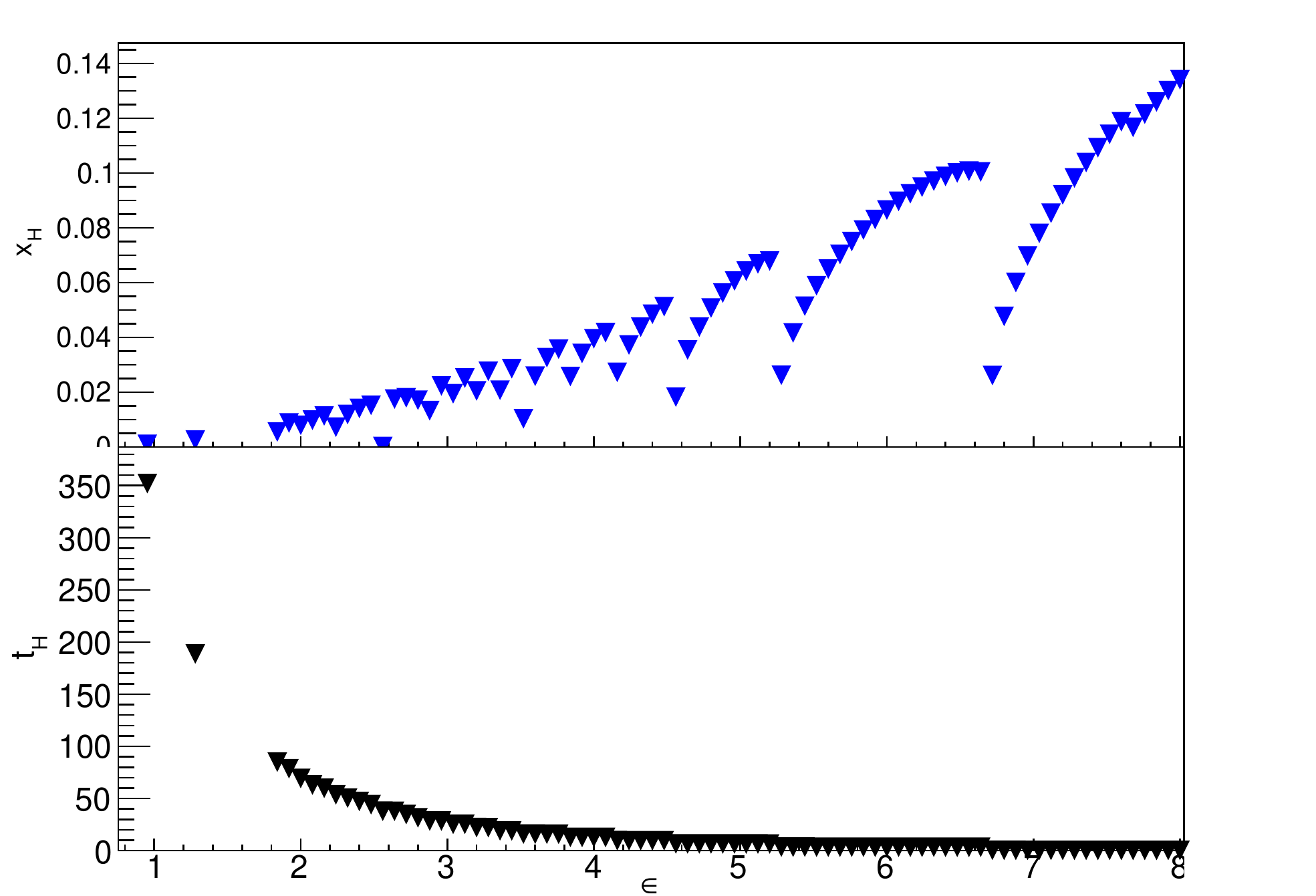}
  \caption{\label{f:m05w03PG} $x_H$ and $t_H$ vs $\epsilon$ for $\sigma=0.3\ell$}
  \end{subfigure}\hfill
  \begin{subfigure}{0.48\textwidth}
    \includegraphics[width=\textwidth]{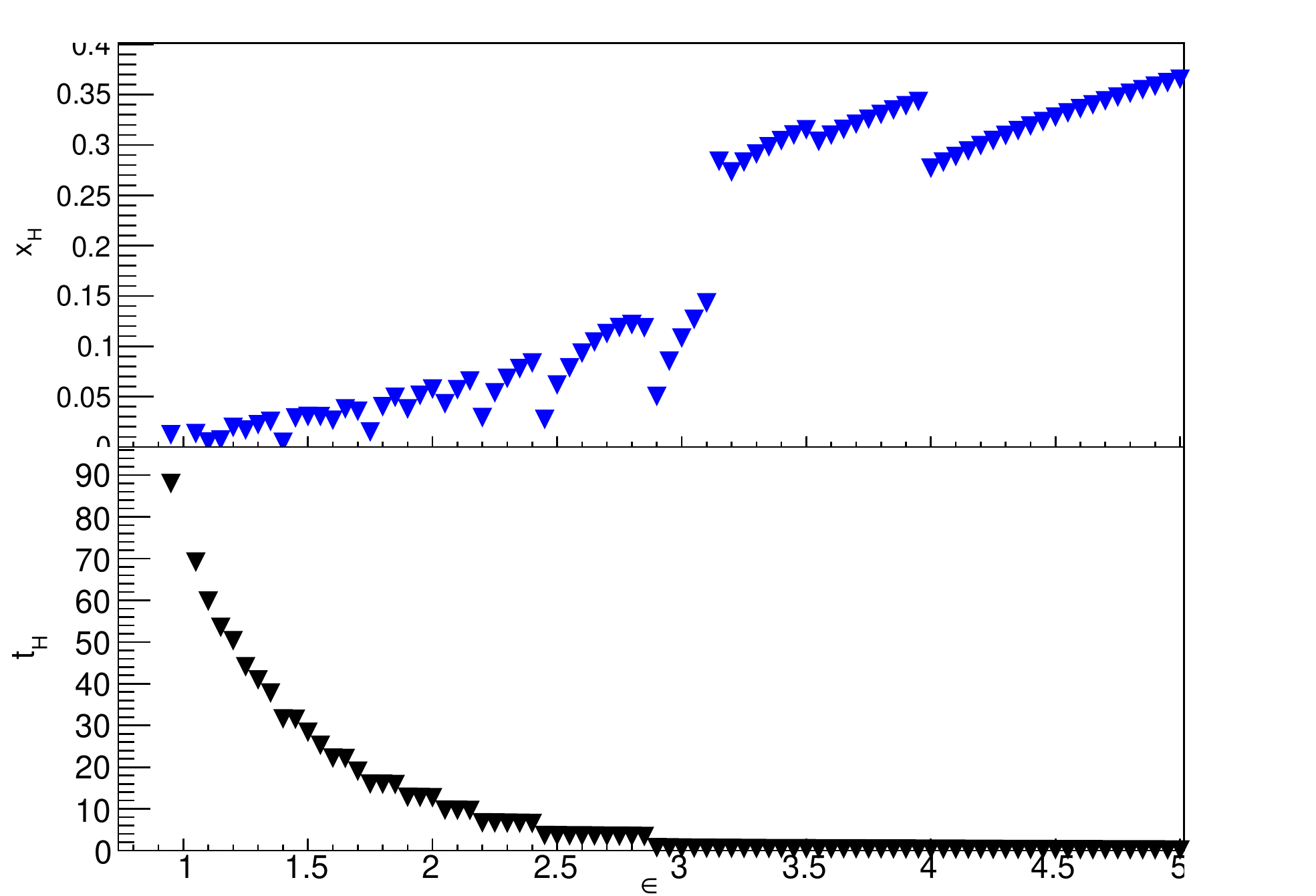}
    \caption{\label{f:m05w06PG} $x_H$ and $t_H$ vs $\epsilon$ for $\sigma=0.6\ell$}
  \end{subfigure}
  \begin{subfigure}{0.48\textwidth}
    \includegraphics[width=\textwidth]{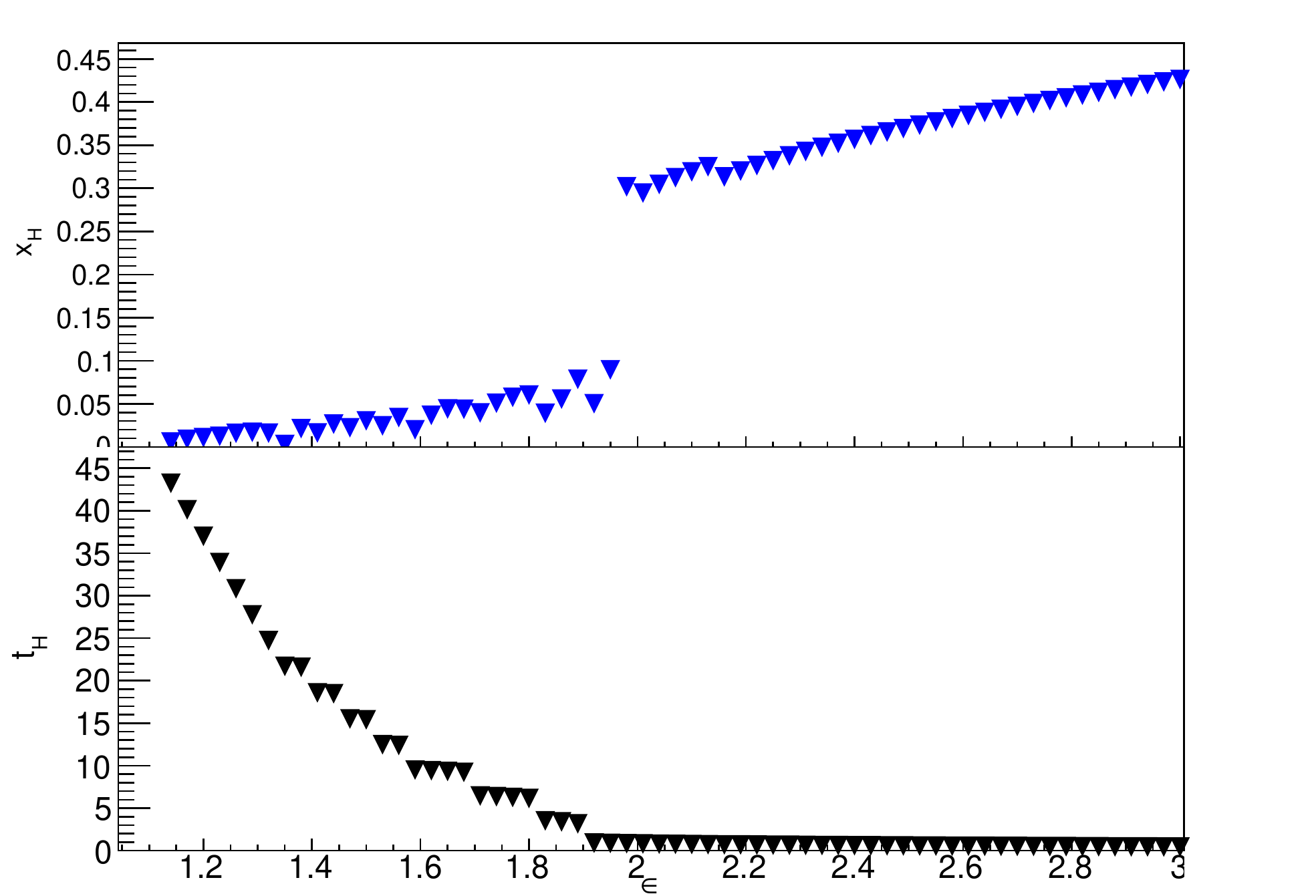}
    \caption{\label{f:m05w08PG} $x_H$ and $t_H$ vs $\epsilon$ for $\sigma=0.8\ell$}
  \end{subfigure}\hfill
  \begin{subfigure}{0.48\textwidth}
    \includegraphics[width=\textwidth]{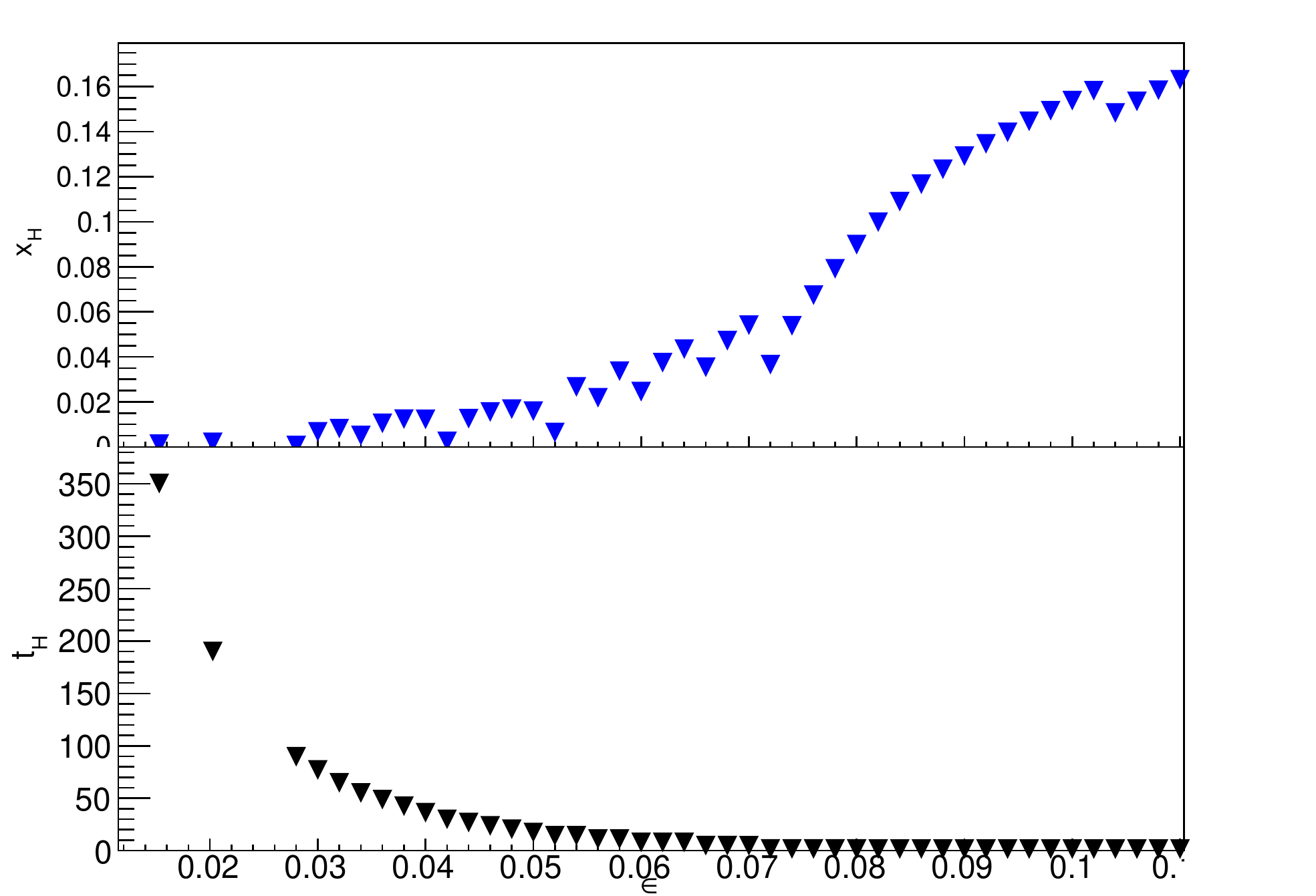}
    \caption{\label{f:m05w4PG} $x_H$ and $t_H$ vs $\epsilon$ for $\sigma=4\ell$}
  \end{subfigure}
  \caption{\label{f:mass05plots} Scaling of initial horizon radius $x_H$ and
    formation time $t_H$ for mass $\mu=0.5/\ell$}
\end{figure}

We consider first light scalars, picking $\mu=0.5/\ell$ for specificity.
To start, we choose Gaussian initial data in $\Pi$ as 
given in equation (\ref{eq:PiGaussianID}) for a variety of widths $\sigma$.
As expected, narrow initial pulses ($\lambda=2\sigma=0.6\ell$) 
lead to gravitational collapse behavior
similar to that of massless fields, as shown in figure \ref{f:m05w03PG}.
In particular, the horizon formation time forms slightly
sloped ``steps'' as a function of pulse amplitude, with each step separated
by a jump of $\Delta t_H$ increasing from approximately 2.8 at large amplitudes
to $\pi$ at small amplitudes.  The initial horizon radius $x_H$ follows the
characteristic arc pattern familiar from \cite{Bizon2011} with one arc per
step in $t_H$ (though the steps occur rapidly enough at small amplitude
that the arcs are difficult to resolve).  The behavior for pulses wide compared
to the AdS scale, shown in figure \ref{f:m05w4PG}
for $\lambda=8\ell$, is similar.  The arcs in $x_H$ are less distinct at
low amplitudes, and more investigation would be required to determine the 
behavior of the horizon radius in this case.  Additionally, the steps in
$t_H$ are more steeply sloped, leading to a noticeably larger collapse time
even for amplitudes that collapse promptly (without reflections from the
boundary).  At intermediate widths, $\ell<\lambda<1/\mu$, however, a
different structure emerges.  Specifically, in the amplitude regime where
the scalar field undergoes prompt collapse, the initial horizon radius
undergoes a sudden decrease (not associated with a step in $t_H$).
This behavior is apparent in figure \ref{f:m05w06PG} for 
$\sigma=0.6\ell$ and in figure \ref{f:m05w08PG} for 
$\sigma=0.8\ell$.  We present a further discussion of this phenomenon in
section \ref{s:shrink} below.

\begin{figure}[!t]
  \begin{subfigure}{0.48\textwidth}
    \includegraphics[width=\textwidth]{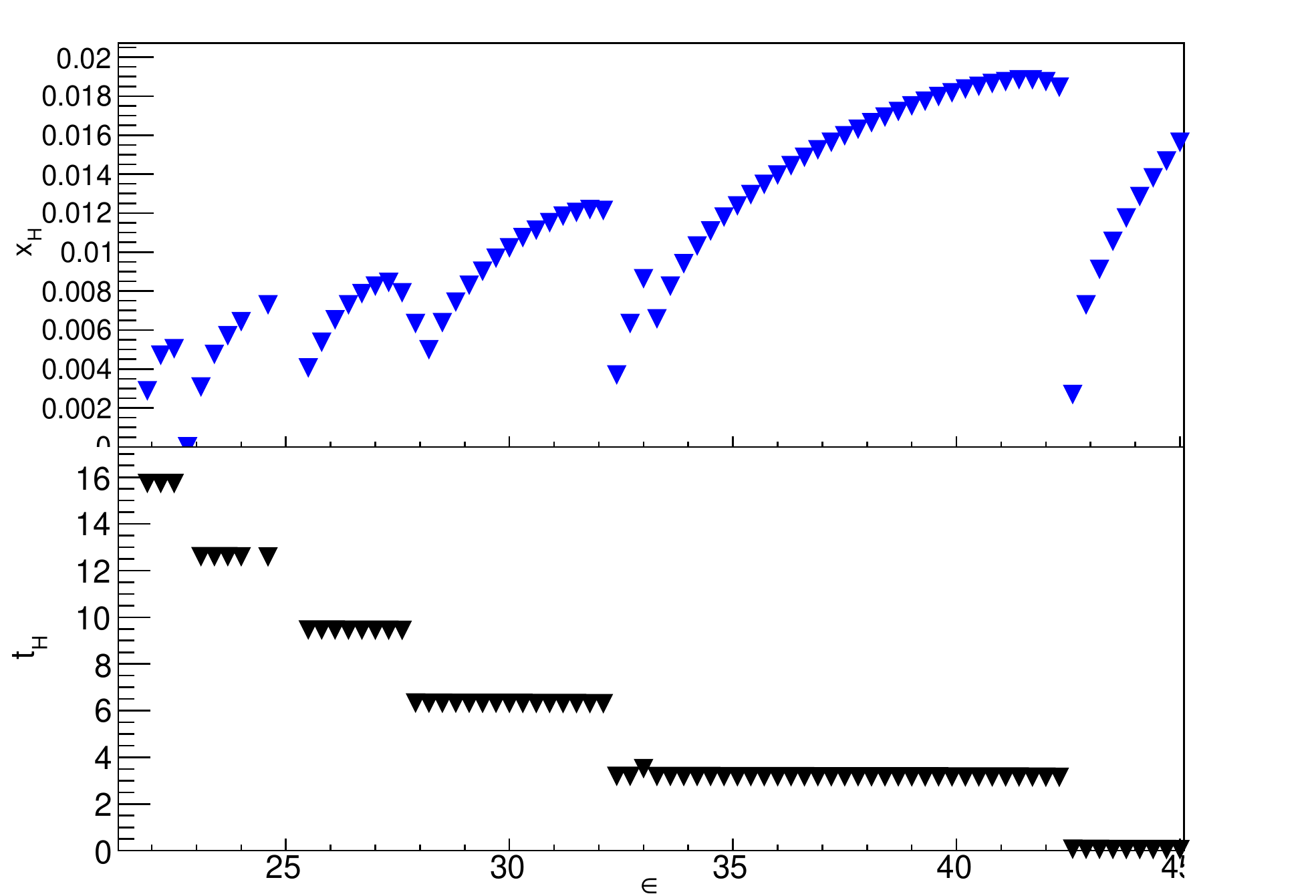}
    \caption{\label{f:m5w005PG} $x_H$ and $t_H$ vs $\epsilon$ for $\sigma=0.05\ell$}
  \end{subfigure}\hfill
  \begin{subfigure}{0.48\textwidth}
    \includegraphics[width=\textwidth]{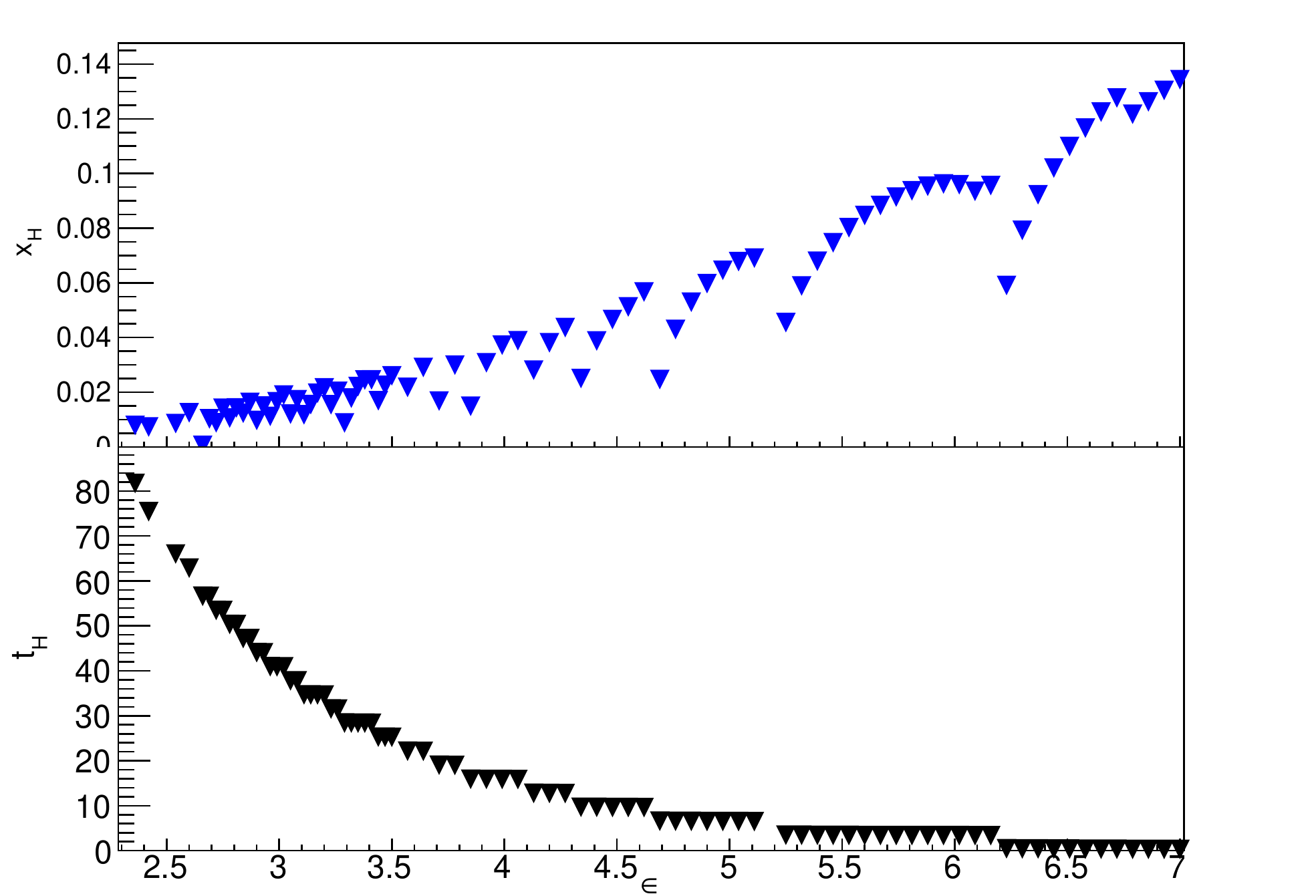}
    \caption{\label{f:m5w03PG} $x_H$ and $t_H$ vs $\epsilon$ for $\sigma=0.3\ell$}
  \end{subfigure}
  \begin{subfigure}{0.48\textwidth}
    \includegraphics[width=\textwidth]{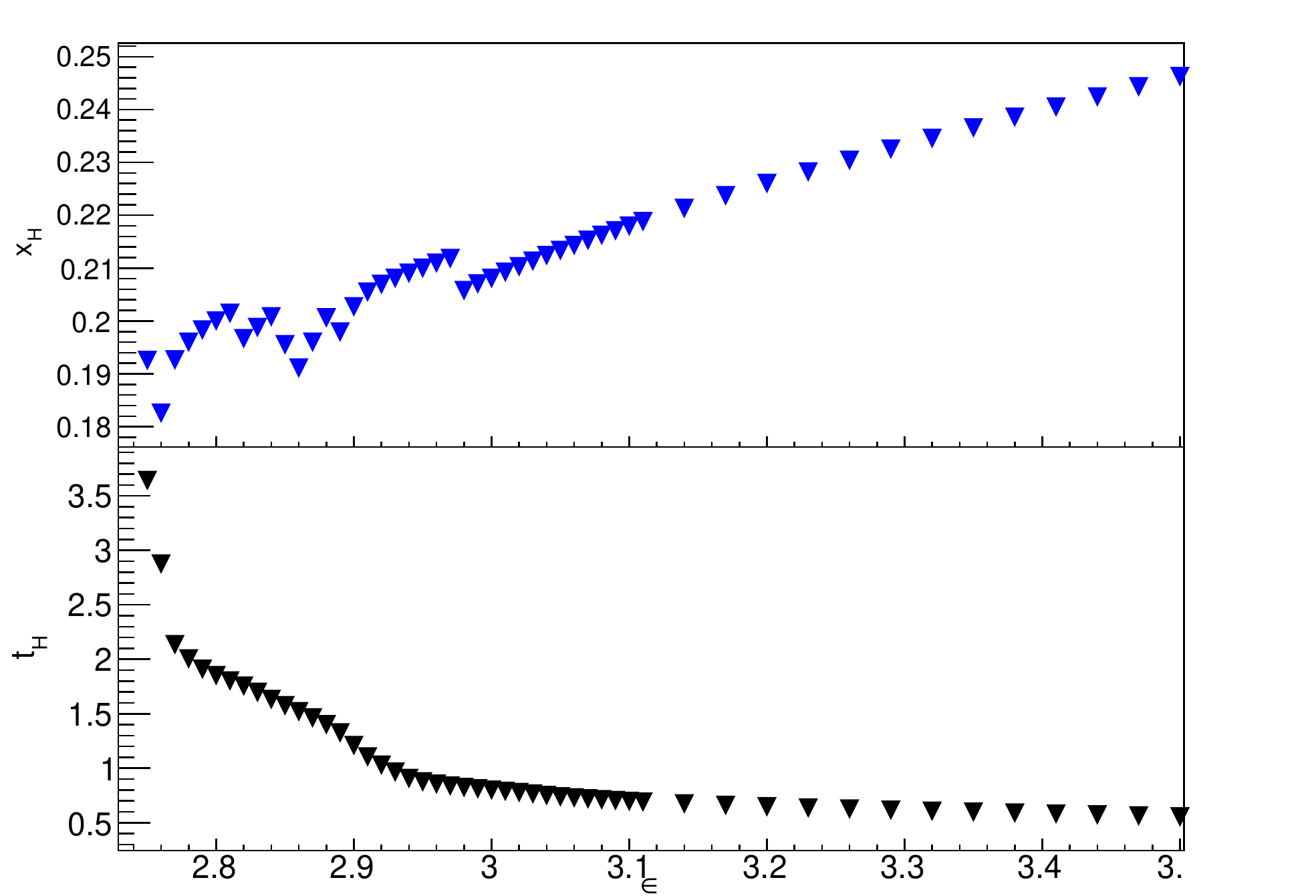}
    \caption{\label{f:m5w05PG} $x_H$ and $t_H$ vs $\epsilon$ for $\sigma=0.5\ell$}
  \end{subfigure}\hfill
  \begin{subfigure}{0.48\textwidth}
    \includegraphics[width=\textwidth]{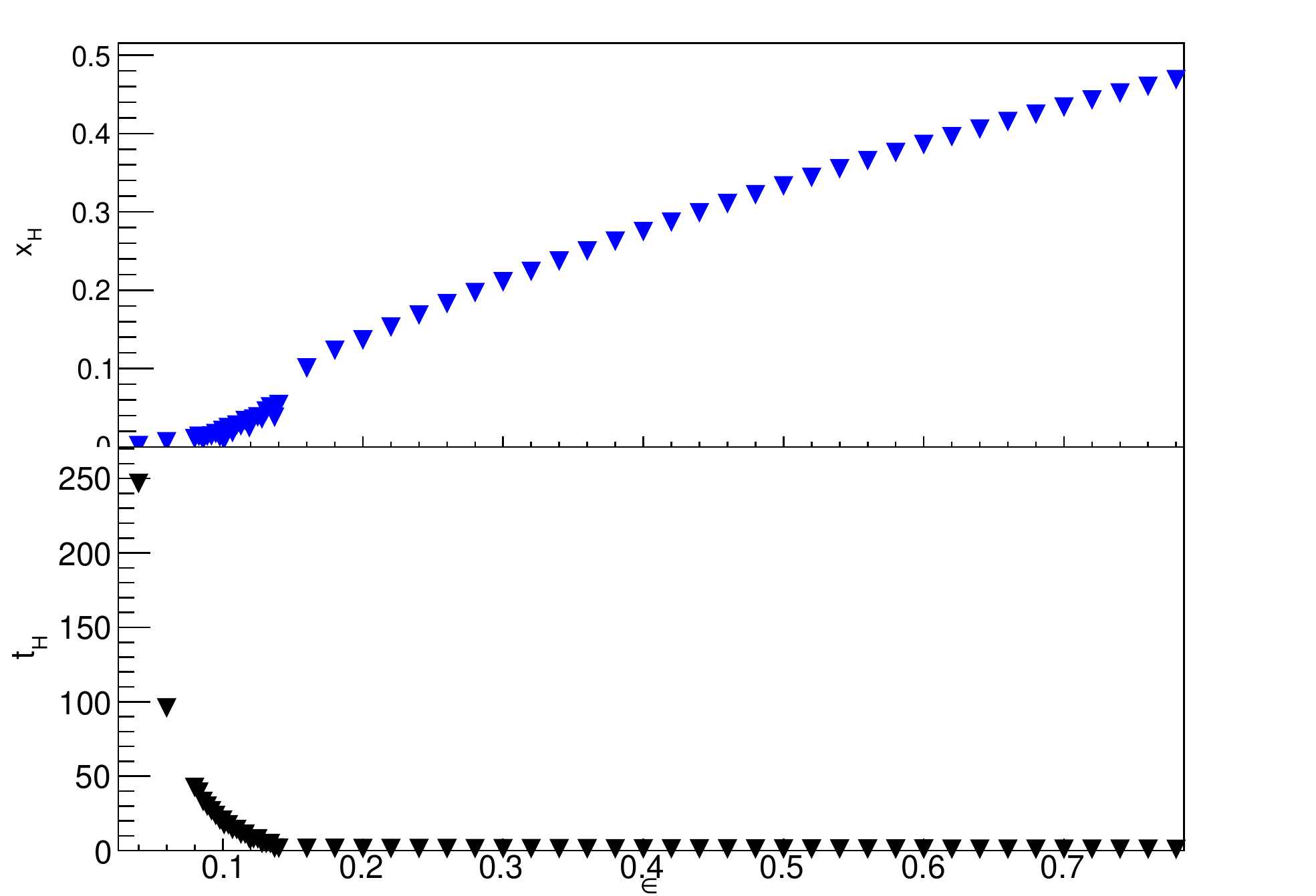}
    \caption{\label{f:m5w2PG} $x_H$ and $t_H$ vs $\epsilon$ for $\sigma=2\ell$}
  \end{subfigure}
  \caption{\label{f:mass5plots} Scaling of initial horizon radius $x_H$ and
    formation time $t_H$ for mass $\mu=5/\ell$}
\end{figure}

\begin{figure}[!t]
\begin{subfigure}{0.48\textwidth}
\includegraphics[width=\textwidth]{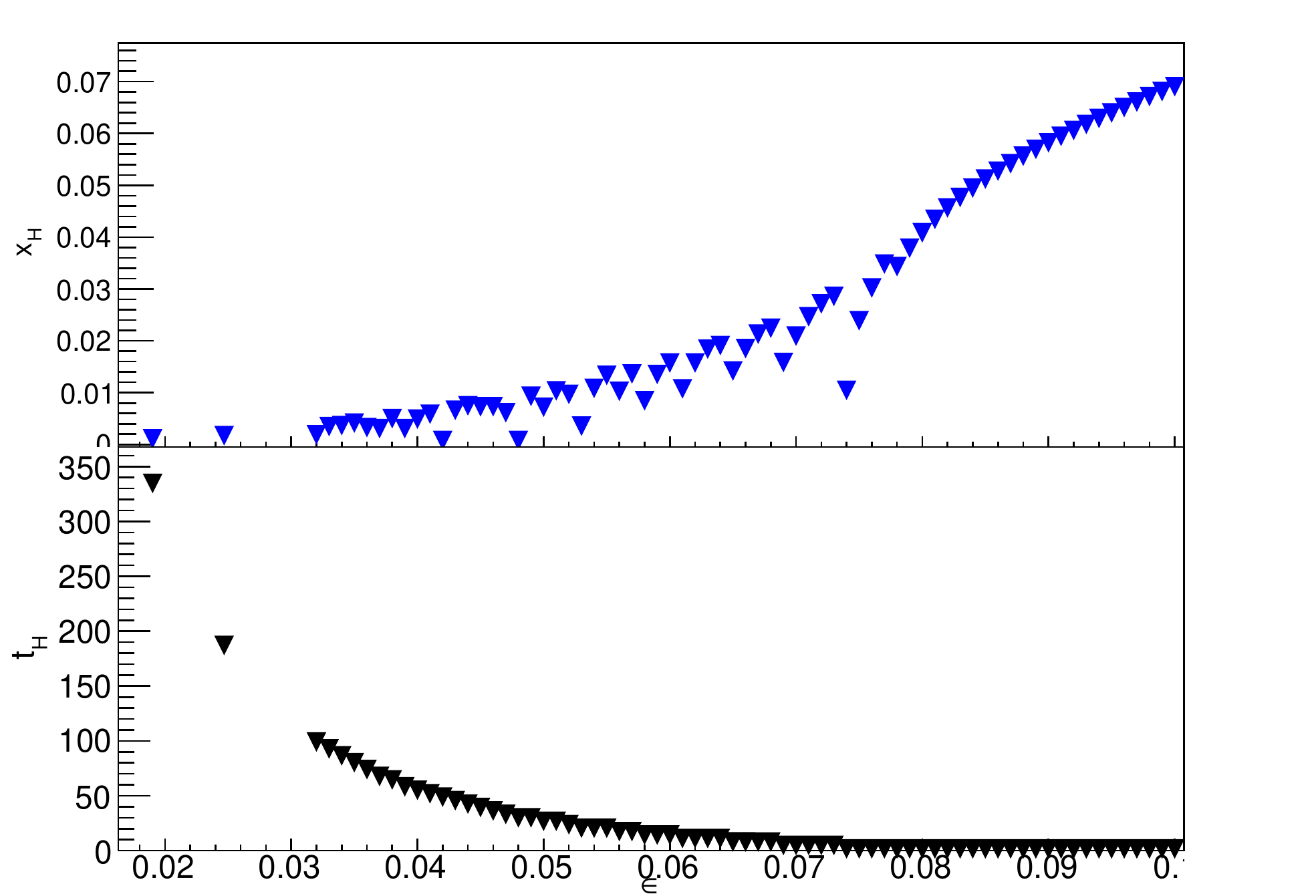}
\caption{\label{f:m10w2t} $\mu=10/\ell$, $\sigma=2\ell$}
\end{subfigure}\hfill
\begin{subfigure}{0.48\textwidth}
\includegraphics[width=\textwidth]{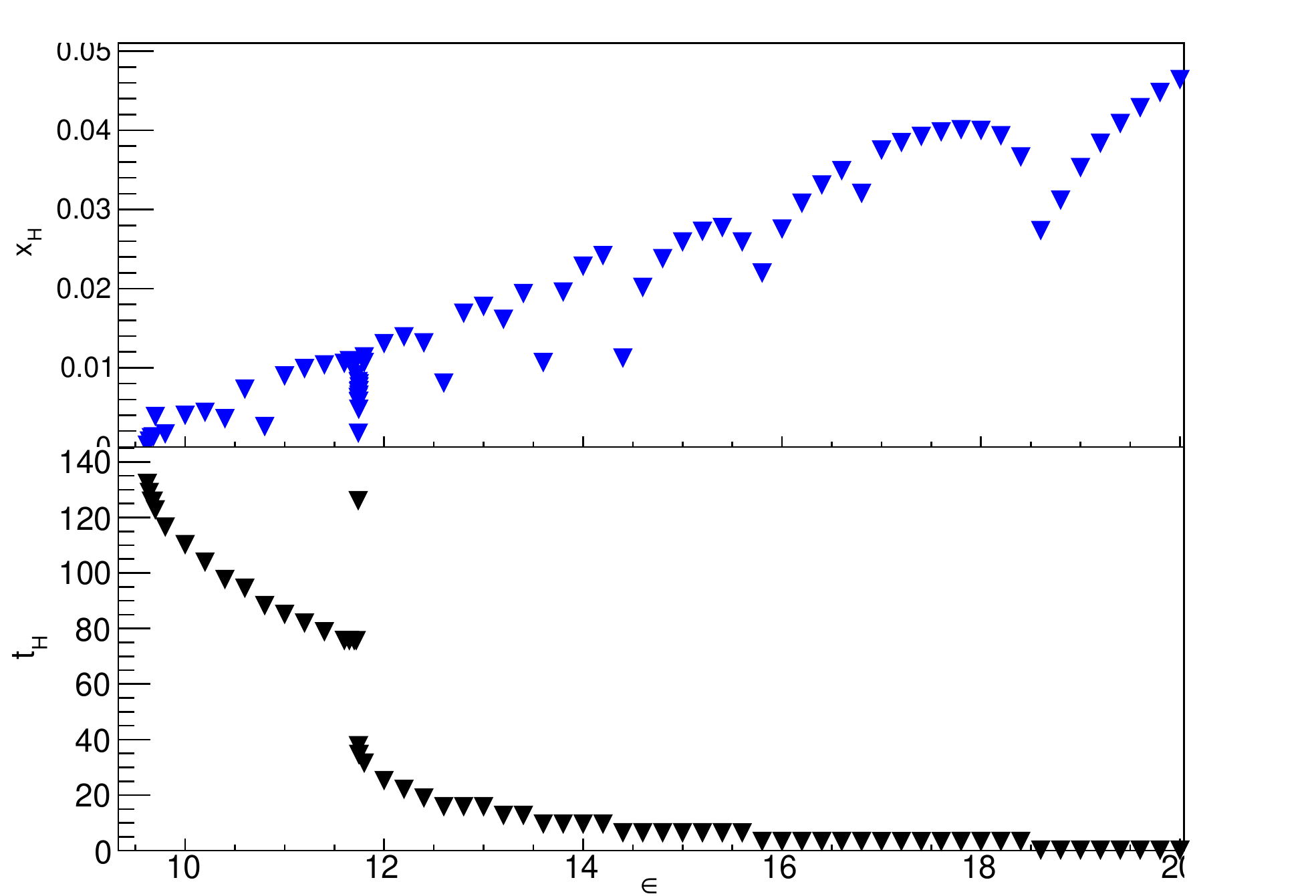}
\caption{\label{f:m20w01t} $\mu=20/\ell$, $\sigma=0.1\ell$}
\end{subfigure}
\caption{\label{f:heavy}  $x_H$ and $t_H$ vs $\epsilon$ for higher masses}
\end{figure}

We now turn to the case of heavy scalars; figure \ref{f:mass5plots} 
shows our results for a field of mass $\mu=5/\ell$.  For narrow pulses
$\lambda<1/\mu$, we find again, as expected, behavior similar to massless
fields for the amplitudes we study; results are shown in figure
\ref{f:m5w005PG} for $\sigma=0.05\ell$.
However, the case of heavy scalars and narrow initial data is numerically 
challenging, likely due to the narrow field pulse propagating close the 
boundary, where the $\phi/\cos^2(x)$ term in the scalar equation of
motion (\ref{eq:scalarField}) becomes subject to large numerical error
(since both numerator and denominator are small).
Wider initial pulses are less numerically challenging.  Very wide initial
pulses ($\lambda>\ell$) as in figure \ref{f:m5w2PG}
show a lengthened collapse time for prompt collapse surpassing $t_H=\pi/2$,
which corresponds to the length of time needed for the broadly distributed
field to collapse toward the center of AdS.  The subsequent steps in 
$t_H$ appear more compressed as a function of $\epsilon$, leading to a 
rapid increase in horizon formation time with decreasing amplitude.
This is apparent to some extent also in figure \ref{f:m10w2t} for mass
$\mu=10/\ell$.  We also consider intermediate widths $1/\mu<\lambda\leq\ell$.
At $\sigma=0.5\ell$ as in figure \ref{f:m5w05PG}, the
step pattern in $t_H$ has mostly disappeared, replaced by a continuously
varying behavior down to a critical amplitude $\epsilon\approx 2.75$.  
(There appear to be steps in $t_H$ just above this amplitude, but, since the
step is $\Delta t_H<\pi$, it seems that the continuous behavior has simply
become steep.)  We have not been able to form a horizon below this 
critical amplitude, even while allowing the simulations to run to $t=500$
(see section \ref{s:analysis} below for further information).  As it
turns out, we should expect this initial data to be stable; over 91\%
of the energy of the initial data is carried by the $j=0$ eigenmode, so this
is a single-mode-dominated solution.  

At a somewhat lower widths still in the intermediate range, we find more
interesting behavior that becomes more apparent at larger masses.  
For $\mu=5/\ell$, $\sigma=0.3\ell$, we find behavior very similar to the
massless scalar, at least for solutions with $t_H\lesssim 85\ell$.
However, at the larger mass $\mu=20/\ell$ and intermediate width 
$\sigma=0.1\ell$, the horizon formation time $t_H$ has a sudden narrow
jump before decreasing again; see figure \ref{f:m20w01t}.  At lower
amplitudes, $t_H$ increases rapidly again (faster than $\epsilon^{-2}$), 
leading to apparently stable behavior.
Note that this behavior is distinct from single-mode-dominated solutions
as exemplified in figure \ref{f:m5w05PG} --- in fact, as discussed in 
section \ref{s:quasistable} below, the energy is distributed democratically
throughout several eigenmodes.  In distinction to the single-mode-dominated 
oscillon and their quasi-periodic generalizations, our results
suggest the presence of stable \textit{oscillaton} solutions for heavy
scalar fields.  
For appropriate values of width, then, our initial data 
can approach the oscillaton solutions, leading to an extended collapse time.
This effect may also be a manifestation of the dynamical mass gap found
by \cite{Brady1997} in a regime where the mass term dominates over gradient
energy but the pulse is too narrow to be deformed by the AdS curvature.
We revisit these issues in more detail in section \ref{s:quasistable}.

\subsection{Low amplitudes}\label{s:analysis}

As mentioned in the introduction, one of the most important questions 
is the dependence of $t_H$ on $\epsilon$.  At small amplitudes, 
self-gravitation is suppressed by two powers of $\epsilon$, so gravitational
collapse should not occur until a time of order $\epsilon^{-2}$ has past.
In fact, the scaling symmetry 
of the perturbation theory \cite{Balasubramanian2014,Craps2014}
makes this argument precise; if a solution remains perturbative until a time
$t_0$, then a solution with lower amplitude will also remain perturbative
until at least time $t_0/\epsilon^2$.  Since gravitational 
collapse should require non-perturbative gravitational physics, 
then $t_H\gtrsim 1/\epsilon^2$.  While first presented for massless scalars,
this argument generalizes to massive scalars and Einstein-Gauss-Bonnet
gravity \cite{Buchel:2014dba}.  Since it would be noteworthy if gravitational
collapse happened on a faster time scale (in the perturbative regime), we 
investigate the functional dependence of $t_H$ on $\epsilon$ in 
some of our simulations.  Similarly, if $t_H$ grows faster than $\epsilon^{-2}$
with decreasing amplitude, that is an indication of quasi-stability (if not
absolute stability).

Specifically, we fit $t_H=a\epsilon^p+b$ to low amplitude data points, both
with and without the constraint $b=0$, and take rough agreement between
the values of $p$ in the two fits as an indicator that we are in
the perturbative regime.
We have also checked that the results of the fit are robust against removal
of some of the low amplitude data points.
Results appear in table
\ref{t:fits} and are generally consistent with $t_H\propto \epsilon^{-2}$
scaling for the initial data shown in the table.  Due to time constraints,
we have not carried out calculations for all our initial data into the
perturbative regime; fits to $t_H$ from these initial data are characterized
by a disagreement between the values of $p$ for fits with $b=0$ and $b$
unconstrained.  From inspection, it appears that initial data that
collapses with $t_H\lesssim 100$ is typically 
not yet in the perturbative regime.

\begin{table}[!t]
\begin{center}\begin{small}
\begin{tabular}{|c|c|c|c|c|}
\hline
&$\mu=0.5,\sigma=0.3$ &$\mu=0.5,\sigma=4$ & $\mu=5,\sigma=2$& 
$\mu=10,\sigma=2$\\ 
\hline
$a$& 318& 0.026 & 0.077& 0.027 \\
$p$& -2.18& -2.27& -2.51& -2.38\\
\hline
$a$& 325& 0.061 & 0.446& 0.095\\ 
$p$& -2.06& -2.08& -1.99& -2.08\\
$b$& -7.67& -13.0 & -25.4& -20.4\\ 
\hline
\end{tabular}
\end{small}\end{center}
\caption{\label{t:fits} Fits to $t_H=a\epsilon^p+b$ for indicated
parameters.  First line constrains $b=0$.}
\end{table}

We also noted that scalars with mass $\mu=5/\ell$ and initial data width
$\sigma=0.5\ell$ lead to apparently stable evolution below a critical
amplitude $\epsilon\approx 2.75$, and we have been unable to induce 
gravitational collapse below this amplitude.  To support this interpretation,
we have run simulations at amplitudes $\epsilon=2.74, 2.67, 2.49$ to
time $t=500$ at a base resolution of $n=14$.  The ADM mass is conserved to 
2 parts in $10^{12}$ over this time for all these simulations, and we
have also run convergence tests (to $t=235,500,275$ respectively) to verify
agreement at higher resolution.
As we noted above, we expect this initial data to be stable at low 
amplitudes, since it is single-mode dominated, and it remains dominated by
the $j=0$ mode to late times.  Specifically, we find that over 80\% of the
energy remains in the $j=0$ mode to $t=500\ell$ for all three amplitudes, and
the high $j$ spectrum decays exponentially.  Interestingly, the
initial data appears to become stable outside of the perturbative 
regime --- the Ricci scalar at the origin does not respect the
perturbative scaling symmetry at these three amplitudes, and the fraction 
of the energy held in the $j=0$ mode averaged over long times increases with
decreasing amplitude.

\subsection{Robustness against change of profile shape}\label{s:profiles}

We have also examined the stability behavior of other initial data to 
determine the robustness of results.  Specifically, we have considered
an ingoing pulse as in eq.~(\ref{eq:phiGaussianID}) and 
two-mode initial data (\ref{eq:twoModeID}), both in AdS$_5$.  

\begin{figure}[!t]
\begin{subfigure}{0.48\textwidth}
\includegraphics[width=\textwidth]{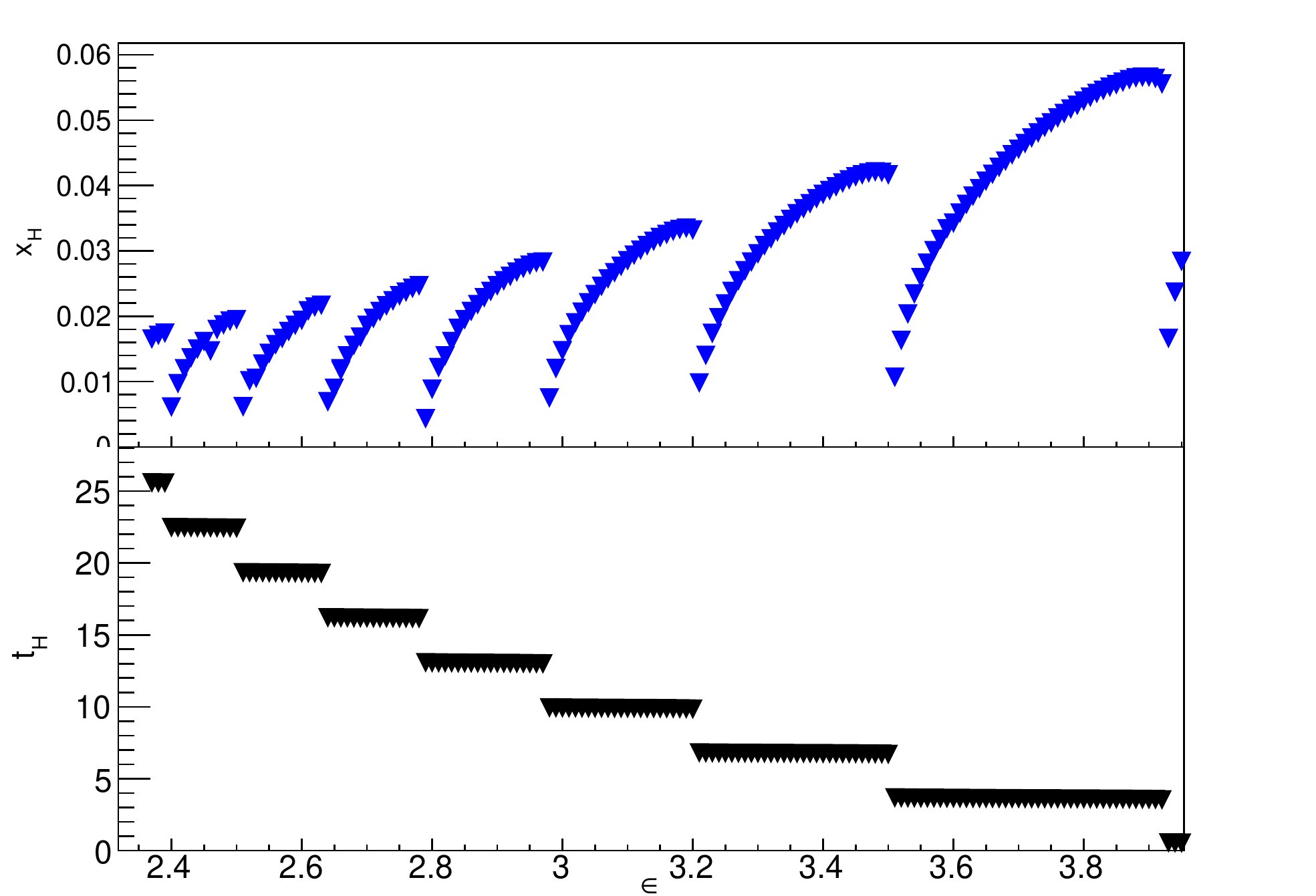}
\caption{\label{f:phim05w03}$\mu=0.5/\ell$, $\sigma=0.3\ell$}
\end{subfigure}\hfill
\begin{subfigure}{0.48\textwidth}
\includegraphics[width=\textwidth]{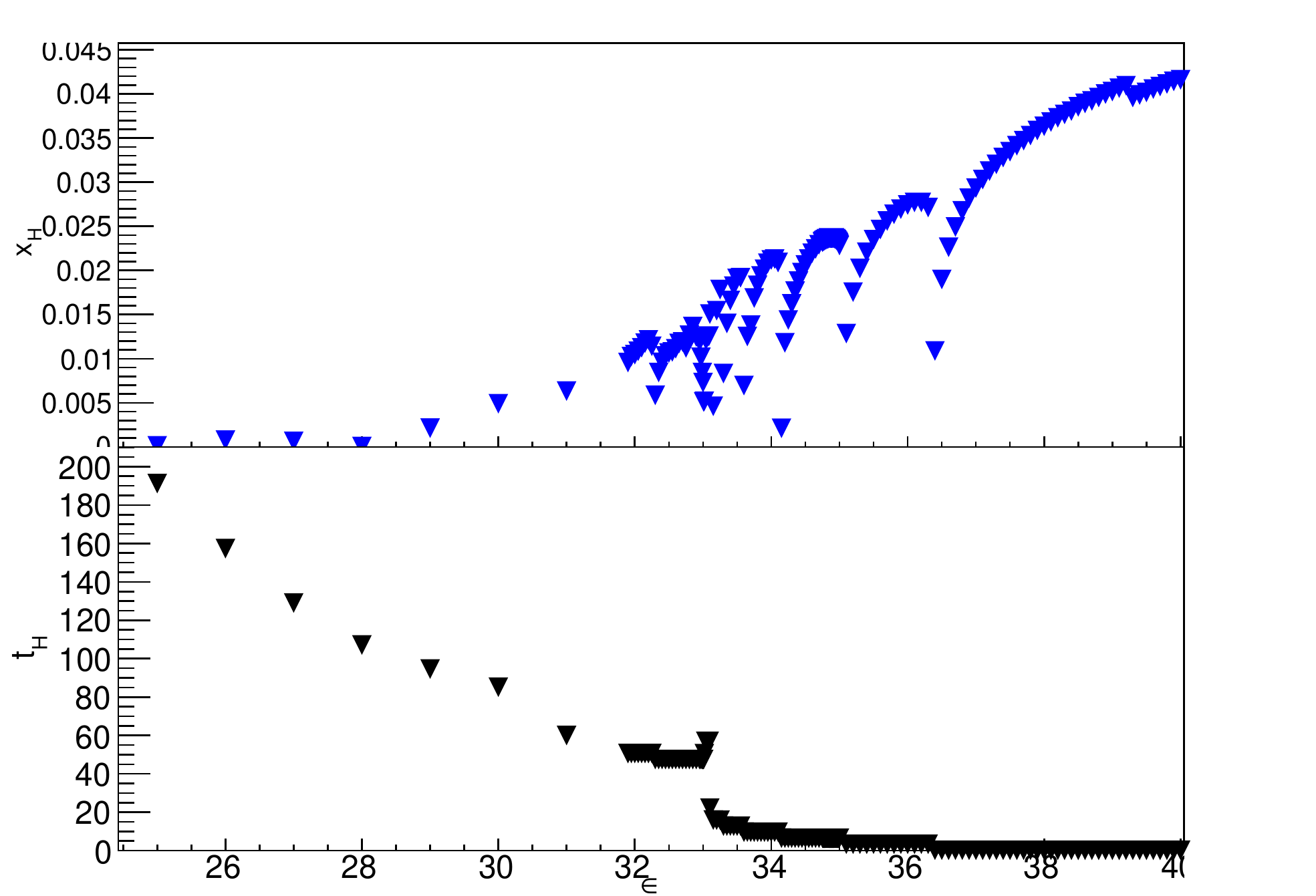}
\caption{\label{f:phim20w01}$\mu=20/\ell$, $\sigma=0.1\ell$}
\end{subfigure}
\caption{\label{f:phiGaussian}$x_H$ and $t_H$ vs $\epsilon$ for 
ingoing pulse initial data}
\end{figure}

We considered two values of $(\mu,\sigma)$ for comparison of the ingoing
pulse to the Gaussian initial data in $\Pi$, $\mu=0.5$ with $\sigma=0.3$
and $\mu=20$ with $\sigma=0.1$.  The initial horizon radii and formation
times are shown in figure \ref{f:phiGaussian} and share the basic 
characteristics of the corresponding mass and width in figures 
\ref{f:m05w03PG} and \ref{f:m20w01t}.  In particular, the $\mu=20,\sigma=0.1$
case has the same striking sudden increase and reduction in $t_H$
as the amplitude decreases.  This provides clear numerical evidence that
our results are robust against change in the shape of initial data.

Since single mode oscillons and perturbations around them are
stable \cite{Bizon2011,Balasubramanian2014}, 
the simplest possible data that could produce a horizon is
two-mode initial data.  We choose the modes to have
equal energy so they are as far away from a single-mode-dominated solution as
possible and for comparison with
previous work \cite{Bizon2011,Balasubramanian2014} in AdS$_4$ and
\cite{Bizon2015} in AdS$_5$.
With this choice, the characteristic width $\lambda$ of the initial data
is fixed for a given mass; changing the ratio $\lambda\mu$ of the width to 
the Compton wavelength requires considering several masses. 
We consider mass values $\mu=0,0.5,1.0$ and $\sqrt{10}$ and estimate the 
width by fitting a Gaussian to the profile (the massless case was 
previously studied in \cite{Bizon2015}). For $\mu=0.5$,
$\lambda\mu<1$; for $\mu=1$, $\lambda\mu\approx 1$;  and for
$\mu=\sqrt{10}$, $\lambda\mu>1$.  The initial horizon radius $x_H$ and 
horizon formation time $t_H$ in these cases are qualitatively similar to
the results shown in section \ref{s:times} above.

\begin{figure}[!t]
\begin{center} \includegraphics[scale=0.4]{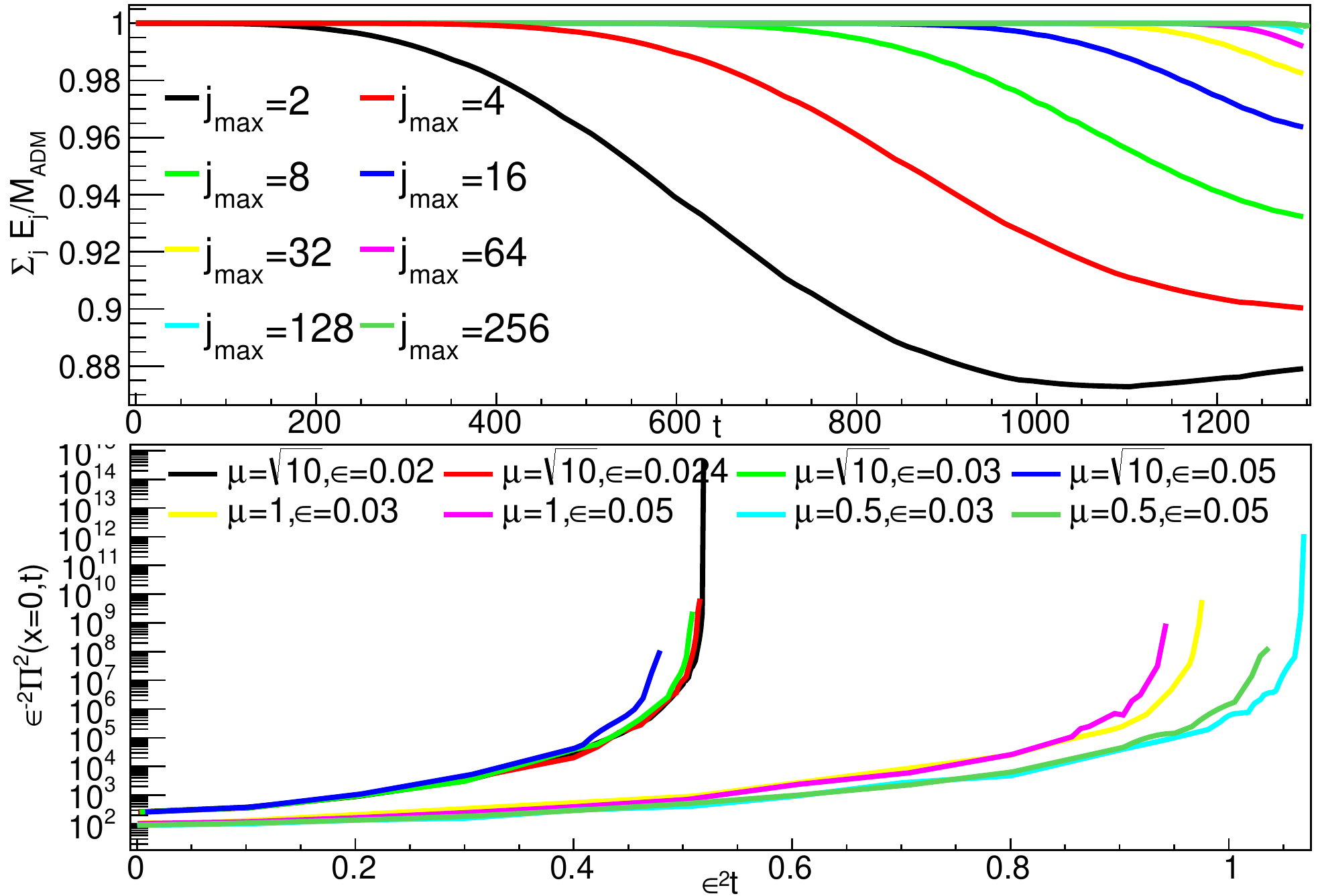}
\end{center}
  \caption{\label{fig:twoModeDecomps}
    Two-mode initial data.
    Top panel: Energy decomposition for $\epsilon=0.02$
    and $\mu=\sqrt{10}$. Bottom panel: $\epsilon^{-2}\Pi^2$
    for several evolutions at different masses and amplitudes. }
\end{figure}

In particular, we find no evidence that these solutions are close to
a stable region.  For $\mu=0,1/2,1$ and $\sqrt{10}$, we find only a 
direct cascade of energy to higher modes for the duration of our
simulations, in agreement with the results
of \cite{Bizon2015}.  The top panel of figure \ref{fig:twoModeDecomps} 
shows the turbulent energy cascade for the $\mu=\sqrt{10}$ and
$\epsilon=0.02$.  
While we cannot rule out that inverse cascades
appear at lower amplitudes, we note that this initial data already appears
to be in the perturbative regime, as $\Pi^2(t,x=0)$ obeys the 
perturbative scaling law,\footnote{Note that $\Pi^2$ is not the Ricci
  scalar at the origin when considering massive scalars. However, we
  have verified that the plots are qualitatively identical if one
  plots the Ricci scalar instead.} as we see in the bottom panel of figure
\ref{fig:twoModeDecomps}.  
This suggests that AdS$_5$ may have smaller stable regions
around single mode data than AdS$_4$, at least for the values of $\mu$ we
considered.  
While most of the behavior is similar for massive and massless scalars 
with two-mode initial data, $\Pi^2(t,x=0)$ displays one interesting 
difference: it increases smoothly with small oscillations for
massive scalars, while for massless scalars it increases in a
piecewise linear fashion.

\begin{figure}[!t]
\begin{center} \includegraphics[scale=0.4]{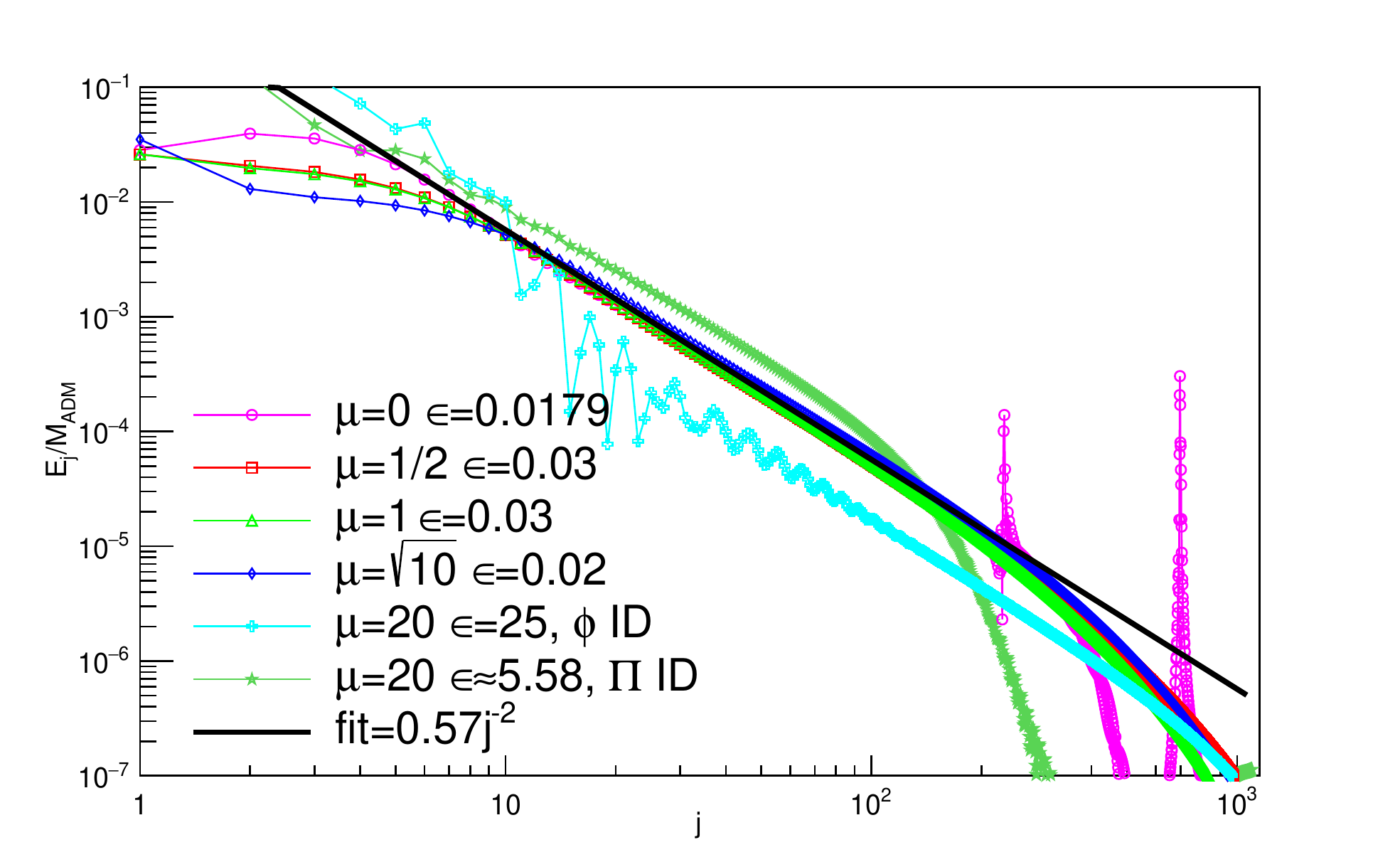}
\end{center}
  \caption{\label{fig:energyDecay}
    Energy spectra late in evolution. Black line shows
    $E_j\propto j^{-2}$. Initial data are two-mode except $\Pi$ ID as 
   eq.~(\ref{eq:PiGaussianID}) and $\phi$ ID as 
   eq.~(\ref{eq:phiGaussianID}).}
\end{figure}

We also consider the late-time energy spectrum for all the types of 
initial data we study.
In collapsing solutions, \cite{Maliborski2013b} 
provides numerical evidence that the spectrum takes the form
is $E_j\propto j^{-\alpha}$ where $\alpha=6/5+4(d-3)/5$.  We show the
late-time spectrum in figure \ref{fig:energyDecay} for several masses 
of two-mode
initial data as well as ingoing wave and Gaussian in $\Pi$ initial data. 
These data have been evolved to long times ($t\approx 1100$ for 
two-mode initial data and $t\approx 350$ for the others).
Remarkably the exponent $\alpha$ appears to be independent of
both the type of initial data and the scalar mass. This suggests
that it is the gravitational physics that sets the decay rate, rather than
the matter content.

\section{Initial horizon radius decrease}\label{s:shrink}

\begin{figure}[!t]
\begin{subfigure}{0.48\textwidth}
\includegraphics[width=\textwidth]{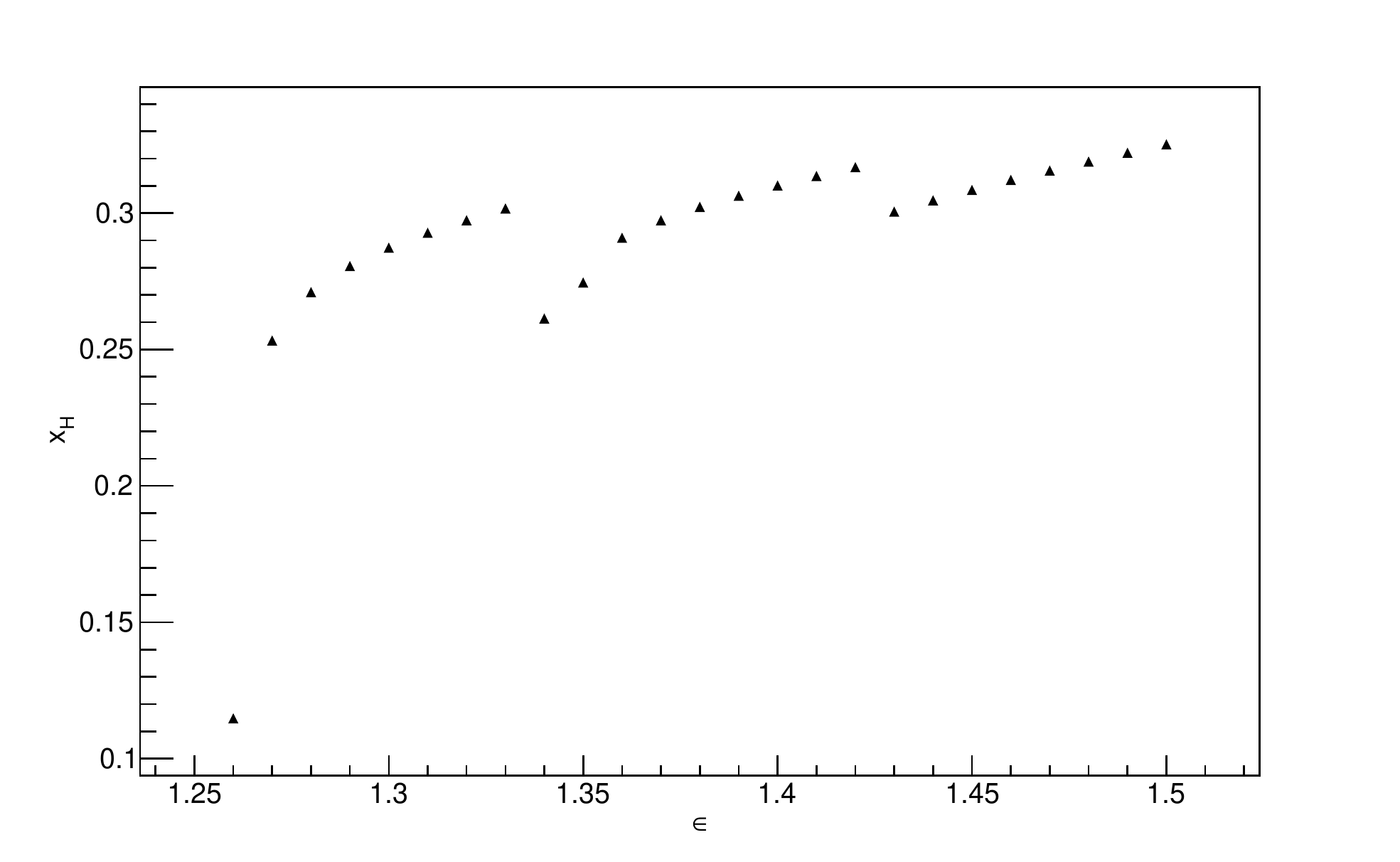}
\caption{\label{f:m01w1r} Discontinuities of $x_H$}
\end{subfigure}
\begin{subfigure}{0.48\textwidth}
\includegraphics[width=\textwidth]{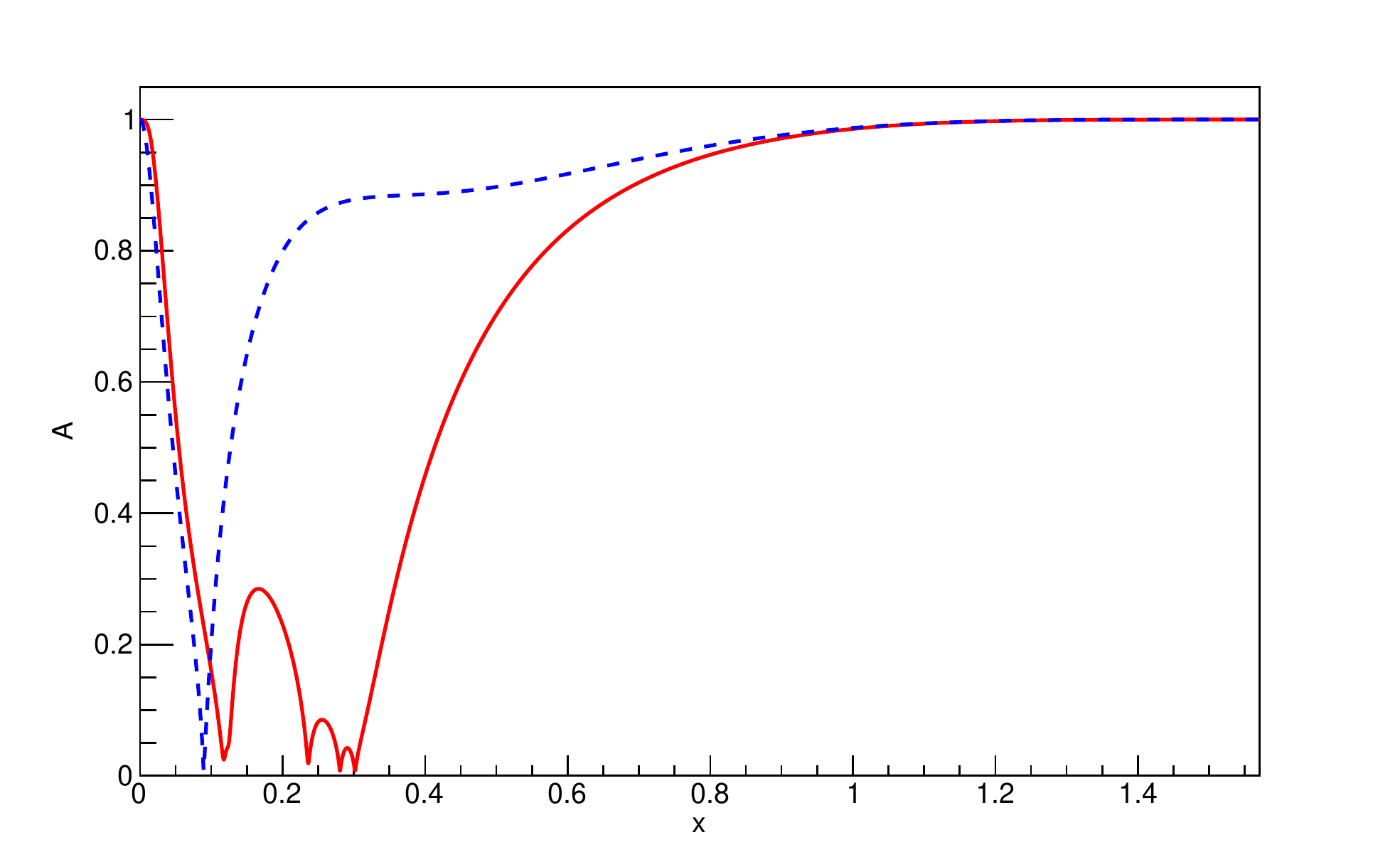}
\caption{\label{f:horizonmins} Horizon function $A$ for $\mu=0.5/\ell$,
$\sigma=0.8\ell$}
\end{subfigure}
\caption{\label{f:horizonjumps} (a) More evidence of 
discontinuities in horizon radius for $\mu=0.1/\ell$, $\sigma=\ell$
(b) Horizon function $A$ just prior to horizon formation.  Red solid
curve is $\epsilon=1.98$, blue dashed is $\epsilon=1.95$. Resolution is
$n=14$.}
\end{figure}

In this section we present a more detailed discussion of the rapid 
decrease in the initial horizon radius for length hierarchy 
$\ell\lesssim\lambda<1/\mu$, as seen in figures \ref{f:m05w06PG} and
\ref{f:m05w08PG}.  As shown in figure \ref{f:m01w1r},
we have also found similar behavior for 
a mass $\mu=0.1/\ell$ and width $\sigma=\ell$.  While the decrease in
$x_H$ is not as dramatic in this case, the left-most data point does show a 
clear drop, indicating that this behavior may be common.  The astute
reader will also notice several sudden increases in the initial horizon
radius as $\epsilon$ decreases in figures \ref{f:m05w06PG}, \ref{f:m05w08PG},
and \ref{f:m01w1r}.  These increases, though again less striking,
are familiar from the original work of \cite{Bizon2011} on massless scalar
collapse in AdS$_4$.  

Both of these phenomena are related to the existence of multiple local
minima in the metric function $A$, which we use as a determinant of horizon
formation; a typical-looking example of this function just prior to
$t_H$ for $\mu=0.5/\ell$, $\sigma=0.8\ell$, $\epsilon=1.98$ is shown in
figure \ref{f:horizonmins} (solid red curve).\footnote{Similar functional
forms of $A$ with multiple minima were also found for the case of massless
scalar collapse in Einstein-Gauss-Bonnet gravity \cite{Deppe2015}.}  
Recall that we terminate our calculations and declare horizon
formation when $A$ decreases below a resolution-dependent threshold;
the minimum in $A$ first reaches the threshold at $(t_H,x_H)$.
When $A$ has multiple local minima (which are associated with shells of infalling
matter), the initial horizon radius $x_H$ is the
position of the first local minimum to reach the threshold.
It is easy to see how decreasing the initial amplitude might lead to a 
jump in $x_H$, then: at some value of $\epsilon$, the innermost shell of
matter is dense enough to push the inner minimum of $A$ below the 
threshold, but, at a slightly lower value of $\epsilon$, the inner minimum
does not contain quite enough mass to decrease past the threshold. 
Horizon formation must wait until another shell of mass approaches the 
origin, but the jump in mass corresponds to a jump in the Schwarzschild radius.
It is worth noting that $t_H$ increases negligibly in this process; the
formation of the inner minimum is associated with growth in the metric
function $-\delta$, leading to time dilation that allows the outer mass
shells to fall inward while very little proper time $t$ at the origin passes.
Previous literature \cite{Deppe2015}
has confirmed that this behavior remains under
increasing resolution, which corresponds to decreasing the threshold,
but the location of the jump may not be robust as decreasing the threshold
can cause the jump to shift to higher amplitudes.

The reverse behavior --- a sudden decrease in the horizon radius --- is
more curious.  For a fixed resolution ($n=14$ in this case), 
decreasing the amplitude from
$\epsilon=1.98$ to $1.95$ changes the metric function $A$ from the 
multi-minimum red solid curve to the single-minimum blue dashed curve in
figure \ref{f:horizonmins}.  Counter-intuitively, the minimum for the lower
amplitude is deeper than the inner minimum of the higher-amplitude curve.
This behavior at fixed amplitude
is \textit{not} robust against decreasing the threshold for
horizon formation (which, in our analysis, occurs with increased resolution);
we have verified for $\epsilon=1.95$ that the initial minimum in $A$ 
(pictured in figure \ref{f:horizonmins} for resolution $n=14$) does not
pass the threshold for resolution $n=16$ and that a second local minimum of
$A$ appears; the sudden decrease in horizon radius should shift to lower
amplitudes.  However, to be clear, this is a real feature of the solutions
and not a convergence issue --- the $n=14$ and higher resolution $n=15,16$
solutions agree to roundoff, at least until the horizon threshold is reached
for $n=14$. We have also verified convergence of 
the solution at lower resolution (base resolutions $n=10,12$).
This is yet another way in which gravitational collapse in
AdS is sensitive to initial conditions and also numerical algorithms.
These results also stress that it is crucial to use the same parameters 
when comparing simulations to each other --- while most comparisons will not
evince this type of sensitivity, we have not been able to predict when 
such sensitivity will appear.  In vernacular, it is important to compare
apples to apples.

\begin{figure}[!t]
\begin{subfigure}{0.48\textwidth}
\includegraphics[width=\textwidth]{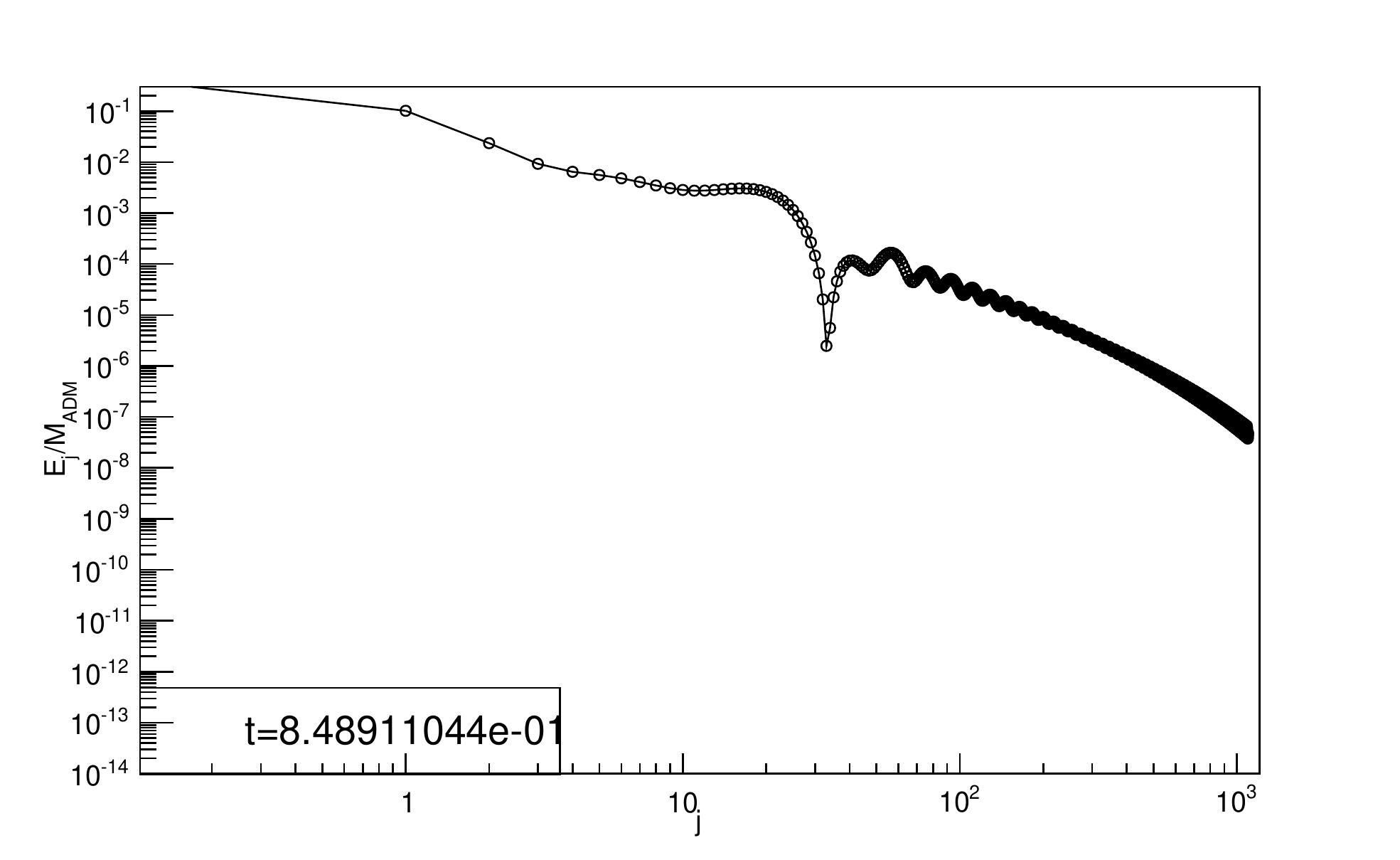}
\caption{\label{f:spectrum195} $\epsilon=1.95$}
\end{subfigure}
\begin{subfigure}{0.48\textwidth}
\includegraphics[width=\textwidth]{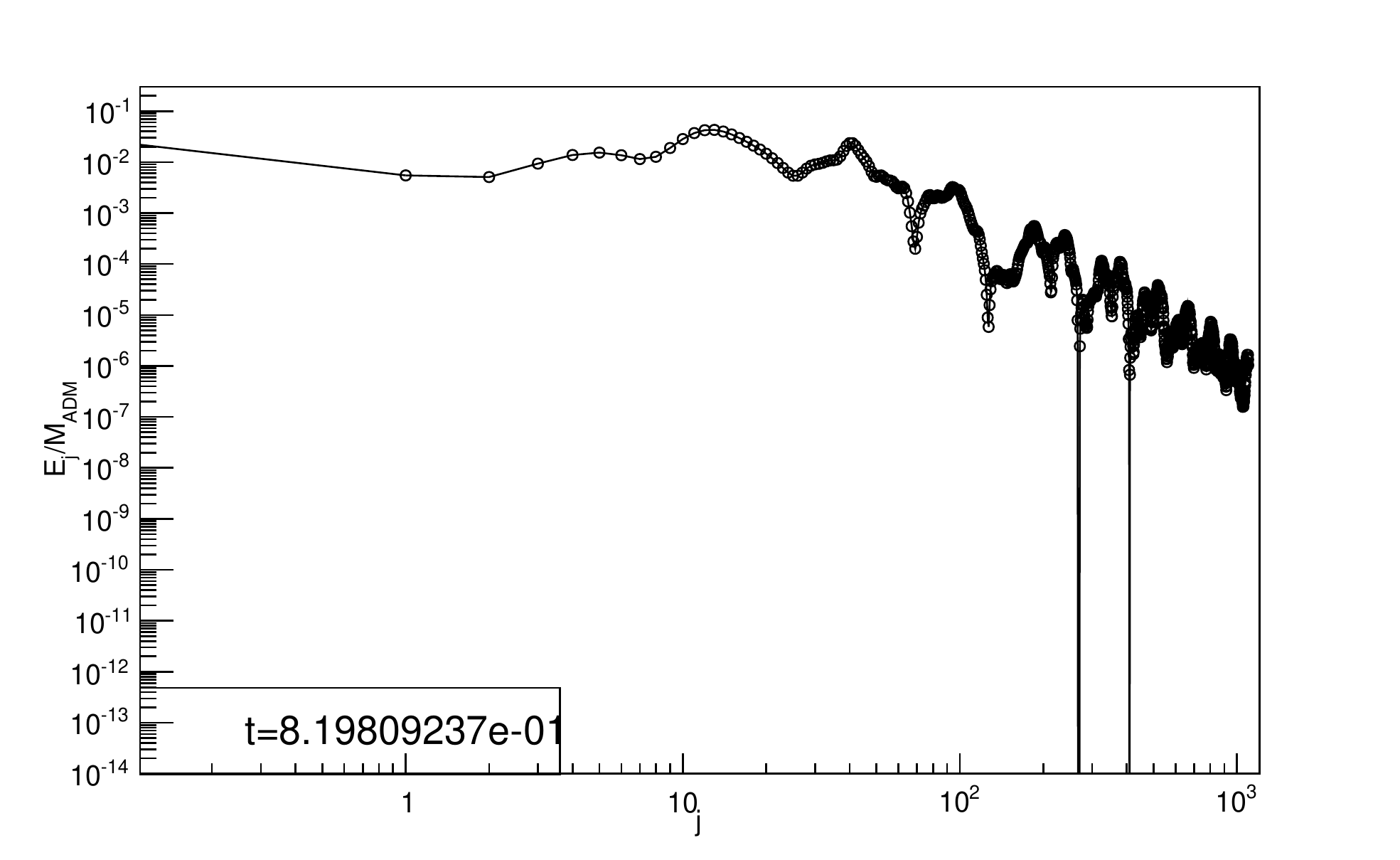}
\caption{\label{f:spectrum198} $\epsilon=1.98$}
\end{subfigure}
\caption{\label{f:shrinkspectra} Energy spectra at horizon formation
for $\mu=0.5/\ell$, $\sigma=0.8\ell$ at resolution $n=14$.}
\end{figure}

It is at first glance unclear whether we should interpret this phenomenon in 
the dual gauge theory; the initial horizon radius depends on the spatial 
slicing used to describe the gravitational collapse, 
and nothing physical in the boundary CFT should depend on a gauge choice in
that way.  In addition, the sharp decrease in $x_H$ apparently depends 
sensitively on the threshold for horizon formation.
However, let us momentarily give in to temptation and give our results an
interpretation in the boundary theory.  
Since we take approximate horizon formation as
corresponding to partial thermalization, we can think of the mass inside
the initial horizon as corresponding to the fraction of the boundary energy
that is approximately thermalized.  If we take these results at face value,
they tell us that decreasing the initial amplitude past a critical value
leads to approximate thermalization with less energy transferred into
higher modes.  In figure \ref{f:shrinkspectra}, we display energy spectra
for our samples with $\mu=0.5/\ell$, $\sigma=0.8\ell$, and amplitudes
$\epsilon=1.95$ and $1.98$ which support this view to some degree.  
Specifically, for $\epsilon=1.95$, a significantly greater fraction of 
the energy remains 
in the two lowest modes, indicating that a smaller amount of energy has
approximately thermalized.  Conversely, the energy fraction in higher
modes ($j\geq 100$, for example) is significantly larger for the 
$\epsilon=1.98$ evolution.

The boundary time of horizon formation provides a possible explanation for
this phenomenon.  Although both amplitudes lead to 
similar values of $t_H$ as measured in terms of proper time at the origin, 
they have quite different conformal time $\tilde t_H$ at the boundary 
due to time dilation between the inner minimum of
$A$ and the boundary.  At resolution $n=14$, we have 
$t_H\approx 0.85$ and $\tilde t_H\approx 2.28$ for 
$\epsilon=1.95$ as compared to
$t_H\approx 0.82$ and $\tilde t_H\approx 4.04$ for $\epsilon=1.98$.  In 
essence, our measure of thermalization ($A$ passing a fixed threshold) has
allowed the higher amplitude solution a longer time to thermalize in the
boundary theory.  Indeed, if we increase resolution to $n=16$, the
$\epsilon=1.95$ solution develops more minima in $A$, and the energy spectrum
\ref{f:spectrum195} evolves to resemble figure \ref{f:spectrum198} more
closely.  For future work, it would be interesting to determine what other
simple indicators beside a threshold for $A$ make accurate measures of 
partial thermalization in the boundary theory.

\section{Quasi-stable solutions}\label{s:quasistable}

In section \ref{s:times}, we noted that massive scalars with intermediate
pulse widths $1/\mu<\lambda<\ell$ exhibit a curious behavior of $t_H$
as the amplitude decreases, which we specifically observed in figure
\ref{f:m20w01t} for $\mu=20,\sigma=0.1$ with initial data given by
eq.~(\ref{eq:PiGaussianID}).  At large amplitudes, $t_H$ increases in the 
typical piecewise (roughly) constant fashion.  The increase then becomes
quite rapid, followed by a sudden jump in $t_H$ by more than a factor of
3.  Very quickly, $t_H$ decreases again, followed by a steady but rapid
increase.  This type of behavior has appeared in the previous literature
for massless \cite{Buchel2013} and massive \cite{Okawa2015} scalars.  In
the massless scalar case, the corresponding initial data leads to stable,
single-mode-dominated oscillon solutions at low amplitudes; in this 
section, we provide preliminary evidence that the corresponding ID 
for massive scalars may represent a novel class of stable solutions ---
oscillatons --- with energy spread democratically through multiple modes.
While this behavior also appears for ingoing wave initial data as in
figure \ref{f:phim20w01}, in this section we will restrict to ID in
the form (\ref{eq:PiGaussianID}) for simplicity.

At the lowest amplitudes we have studied for $\mu=20,\sigma=0.1$, we
find that $t_H$ rapidly increases as $\epsilon$ decreases, which provides
some support for possible stability at arbitrarily small amplitude.  
These calculations require very high resolution ($n\geq 17$) to remain
convergent until collapse, and we have carried out convergence testing
to be confident in our calculations.  
In table \ref{t:fits20} below, we show fits $t_H=a\epsilon^p+b$
for the indicated amplitude ranges (which are entirely below the
large jump in $t_H$ noted in section \ref{s:massive}); 
due to computational complexity, 
these fits all have $\epsilon\geq 9.62$. Except for the smallest amplitude
range $9.62\leq\epsilon\leq 9.8$, we find a disagreement in $p$ values
for fits constrained to have $b=0$ or not, and we also find that the 
smaller ranges have increasingly negative values for the power $p$ when $b=0$.  
This indicates that $t_H=a\epsilon^p+b$ does not provide a good fit to $t_H$;
we have tried also fits of the form $t_H=a(\epsilon-\epsilon_*)^p+b$, 
but those also fail to provide a good fit (in fact, common fitting 
algorithms do not find sensible fits at all).  
The lesson of table \ref{t:fits20} is not that the horizon formation time
follows a specific power law in amplitude.  What is evident, though,
is that $t_H$ increases much more rapidly than $\epsilon^{-2}$ 
with decreasing amplitude, so this initial data appears to be quasi-stable,
that is, stable over longer time scales than expected from generic 
arguments about perturbation theory.  Given constraints on computing time,
we have not yet ascertained whether $t_H$ undergoes repeated jumps
such as the one seen near $\epsilon=11.74$ in figure \ref{f:m20w01t} or
reaches a critical amplitude below which horizons do not (apparently) 
form.  However, it seems reasonable to conclude that 
the low amplitude behavior indicates (quasi-)stability over longer than normal 
time scales at the least.

\begin{table}[!t]
\begin{center}\begin{small}
\begin{tabular}{|c|c|c|c|c|}
\hline
& $\epsilon\leq 11.72$ & $\epsilon\leq 11.4$ & $\epsilon\leq 10.6$& 
$\epsilon\leq 9.8$\\ \hline
$a$ &76000 & $1.3\times 10^5$& $3.2\times 10^5$& $8.2\times 10^6$\\
$p$ & -2.82& -3.05 & -3.45 & -4.88\\
\hline
$a$ & $6.2\times 10^6$& $6.1\times 10^6$& $5.9\times 10^6$ &$7.1\times 10^6$\\
$p$ & -4.94 & -4.92 & -4.87 & -4.67\\
$b$ & 41.4& 38.5& 32.3& -49.2\\
\hline
\end{tabular}
\end{small}\end{center}
\caption{\label{t:fits20} Fits to $t_H=a\epsilon^p+b$ for $\mu=20,\sigma=0.1$
for the indicated amplitude range.  First line constrains $b=0$.}
\end{table}

\begin{figure}[!t]
  \centering
  \begin{subfigure}[t]{0.48\textwidth}
    \includegraphics[width=\textwidth]{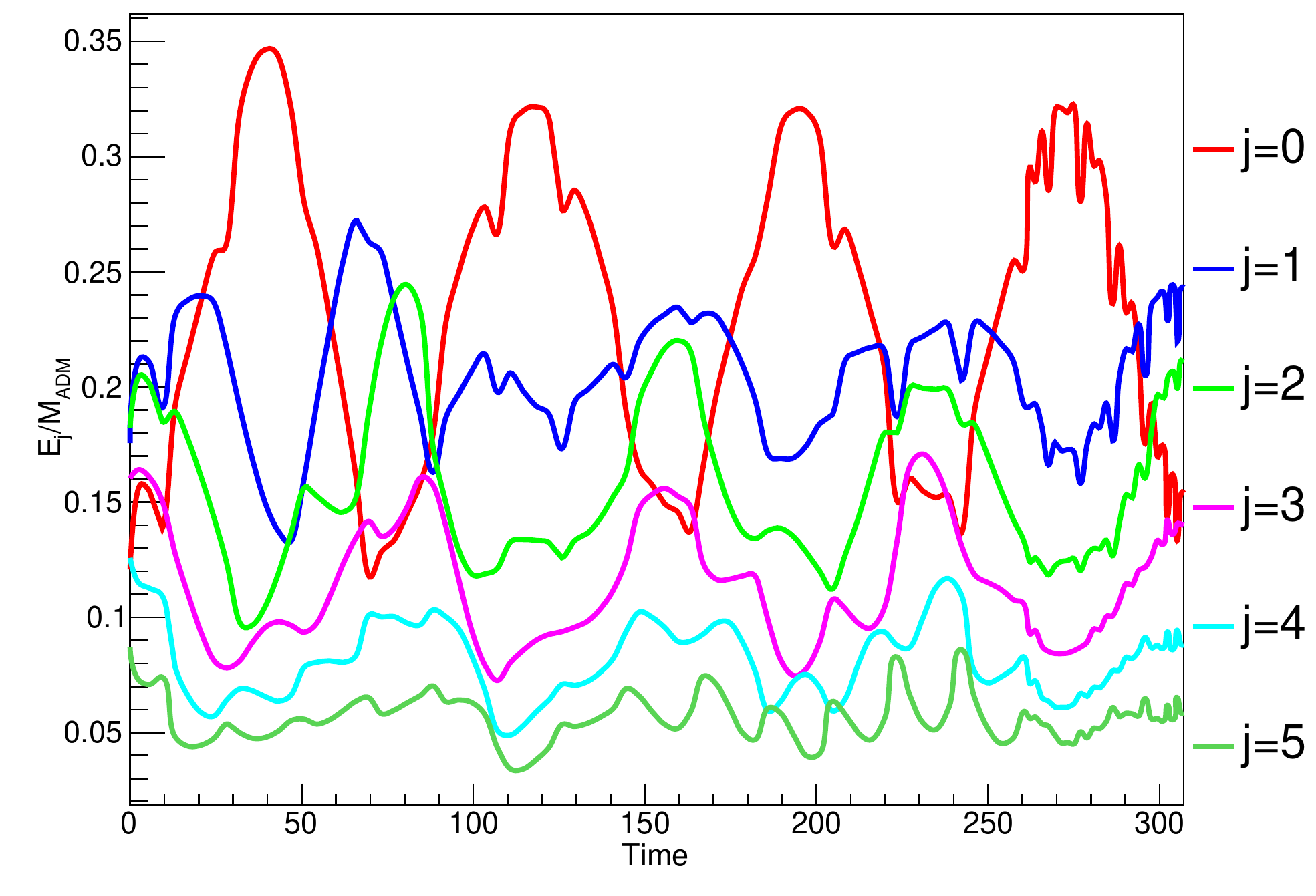}
    \caption{$\hat E_j$ for the lowest six modes.}
    \label{fig:mu20PiEnergy}
  \end{subfigure}\hfill
  \begin{subfigure}[t]{0.48\textwidth}
    \includegraphics[width=\textwidth]{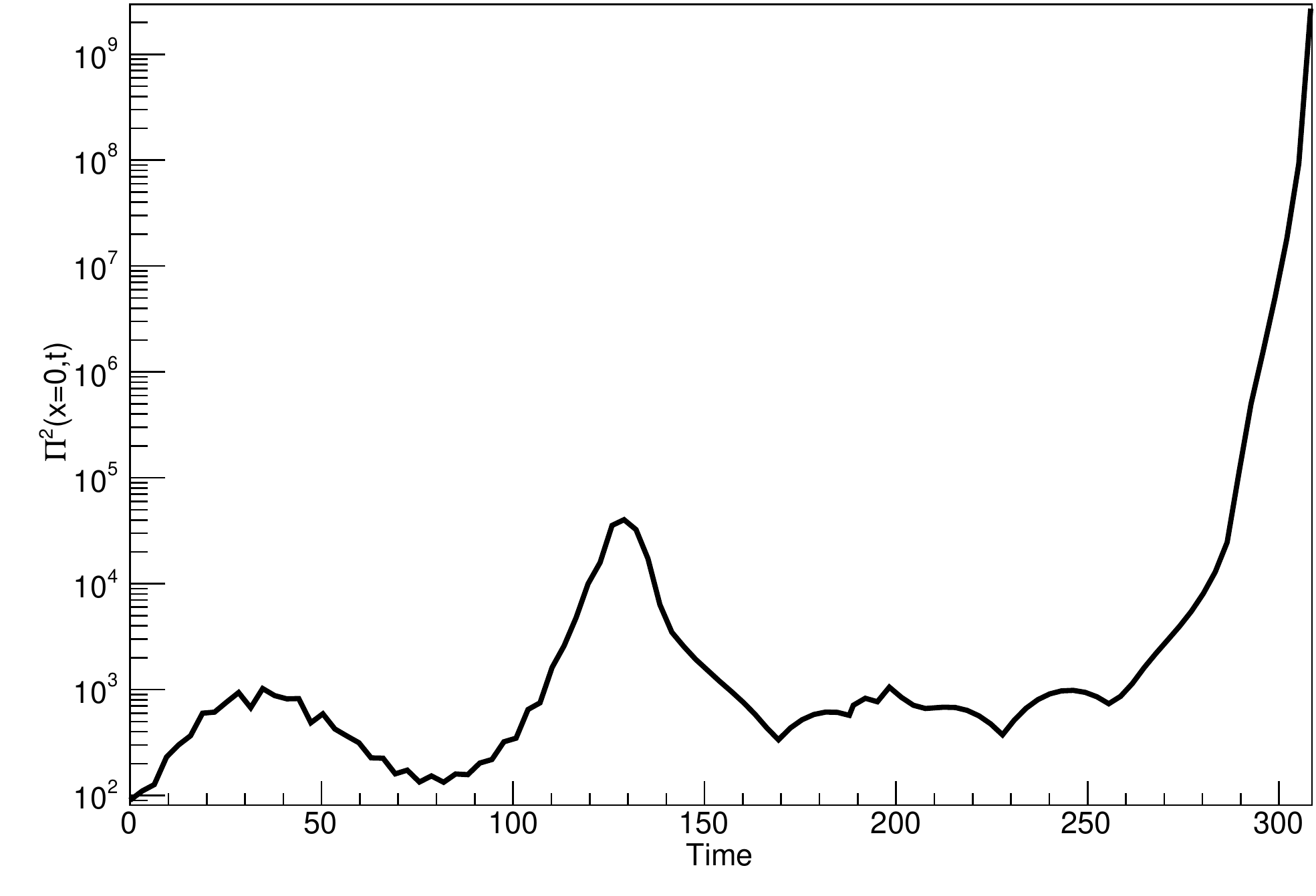}
    \caption{Upper envelope of $\Pi^2(t,x=0)$}
    \label{fig:mu20PiPi2}
  \end{subfigure}
  \caption{Evolution with $\mu=20/\ell$, $\sigma=\ell/10$, and
    $\epsilon=9.5$. }
  \label{fig:mu20PiA9_5}
\end{figure}

We plot the energy in the lowest six
modes alongside the upper envelope of
$\Pi^2(t,x=0)$ in figure \ref{fig:mu20PiA9_5} for $\epsilon=9.5$. An
inverse energy cascade occurs at
$t\approx 35$ and then again at $t\approx 130$ with a rapid increase
in $\Pi^2$
occurring at $t\approx 307$. There is a pattern of increases and decreases
in $\Pi^2$ and an approximate recurrence in the spectrum before collapse, 
both consistent with direct and indirect energy cascades.  This is 
similar to the orbits around quasi-periodic
solutions studied by \cite{Green:2015dsa} for massless scalars; 
however, the spectrum in 
figure \ref{fig:mu20PiEnergy} is distinct in that the the $j=0,1,2$ modes 
all carry the greatest share of the energy at some point in time
(compare to the spectrum of equal-energy two-mode ID evolution in
figure \ref{fig:twoModeAds4Energy}). Interestingly, during the final
growth in $\Pi^2$ ($t\approx311$), $\sim99$\% of the energy is in the
lowest 16 modest.
 
The presence of inverse cascades tend to coincide with rapid
increase in formation time and has been used to argue indirectly for
stability in certain solutions (for example, \cite{Balasubramanian2014}).
As figure \ref{fig:mu20PiPi2} shows, it is possible for collapse
to occur even long after an inverse cascade with no obvious way of
predicting whether or not a rapid cascade of energy to higher modes
will occur later in the evolution. As we found for two-mode
data in section \ref{s:twomode}, the evolution may be apparently
quasi-periodic for long times and then abruptly change, possibly
leading to horizon formation.

\begin{figure}[!t]
\begin{center}
\includegraphics[scale=0.4]{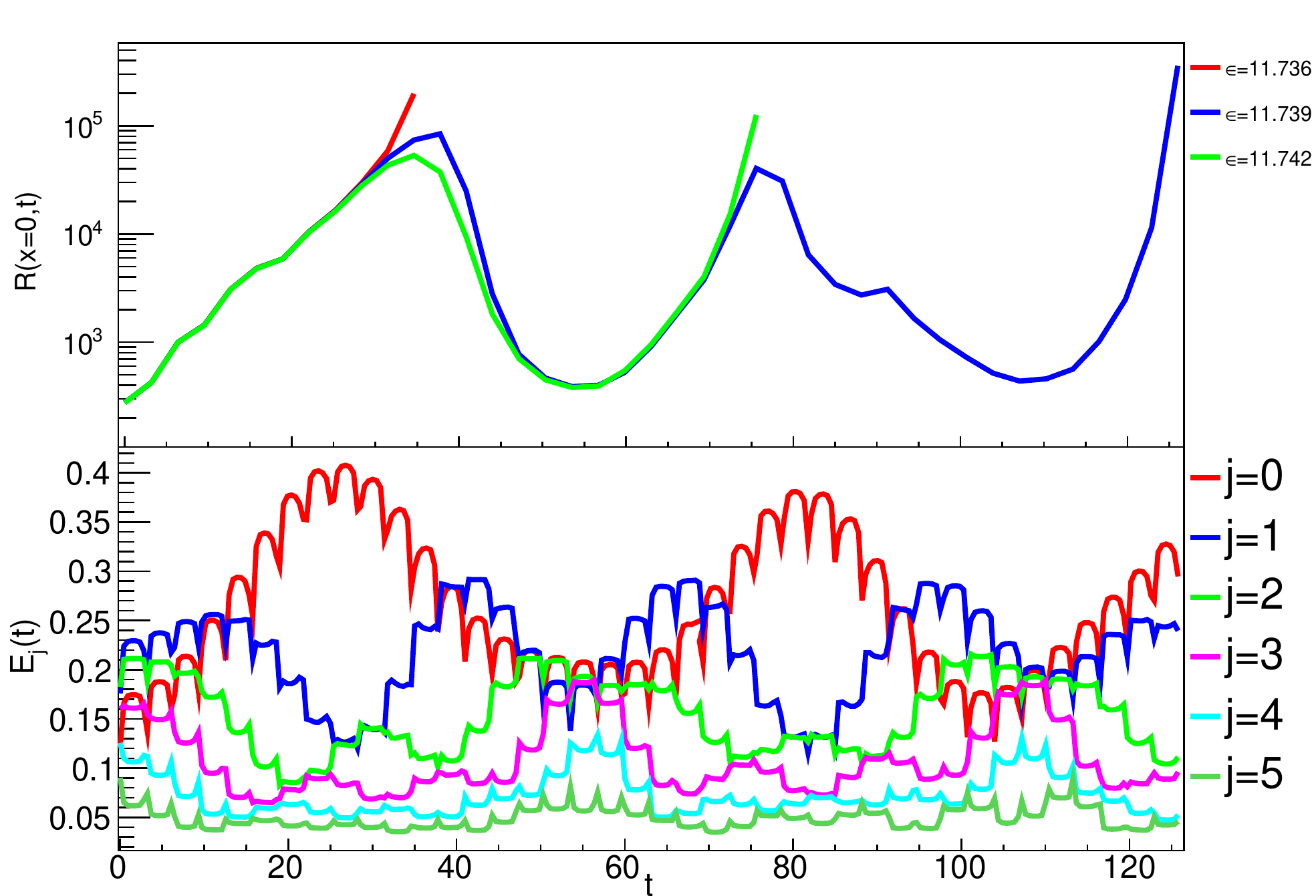}
\end{center}
\caption{\label{f:mu20jump}Evolution of $\mu=20/\ell,\sigma=\ell/10$ ID 
for $\epsilon=11.742,11.739,11.736$. Top panel: upper envelope of 
$\mathcal{R}(t,x=0)$ Bottom panel: $\hat E_j$ for lowest 6 modes for
$\epsilon=11.739$.}
\end{figure}

Another example of this type of behavior is found in the region near
$\epsilon\approx 11.74$, where $t_H$ jumps from approximately 37.7 to 
126 and back to 75.6.  We have carried out convergence testing on 
numerical evolution of three amplitudes $\epsilon=11.742$ ($t_H\approx 37.73$),
$\epsilon=11.739$ ($t_H\approx 126.02$), and $\epsilon=11.736$ 
($t_H\approx 78.71$) to verify the expected 4th-order convergence. The
data for the $\epsilon=11.739$ simulation is presented in appendix
\ref{s:convergence} as an example of our convergence testing.  Furthermore,
it is worth noting that we find the long-lived ($t_H\approx 126.02$)
behavior for more than one amplitude value.
As we show in figure \ref{f:mu20jump}, evolution at these
three amplitudes is in remarkable agreement until immediately before
horizon formation.  While that is to some extent to be expected due to
the very small changes in amplitude (about 0.06\% over the whole region),
it is striking how unpredictable horizon formation seems in a comparison
of the three amplitudes.  In fact, the curvature for
the intermediate amplitude, which has the largest $t_H$, starts growing 
but then turns over, allowing a lower amplitude evolution to form a horizon
first.  Figure \ref{f:mu20jump} also shows the low $j$ part of the spectrum,
which evinces a similar pattern of recurrences as the lower amplitude 
($\epsilon=9.5$) evolution shown in figure \ref{fig:mu20PiEnergy}, 
though the higher-amplitude evolutions have a number of higher-frequency
oscillations on top of the long-term recurrence pattern.
Once again, the spectra for the three amplitudes near $\epsilon=11.74$
are difficult to distinguish
except just before one of the evolutions forms a horizon.

%%%%%%%%%%%%%%
To investigate how common this behavior is for massive scalars,
we also study extremely massive scalars with $\mu=100/\ell$. 
This data is numerically much more difficult to
evolve than the less massive fields. Therefore, all simulations 
discussed here were typically run with of $2^{14}$ or more
grid points. We then ran simulations at higher resolutions to test if the
results agreed, as well as did global convergence tests of the
longer simulations. We find that the solutions that collapse at
``late'' times (after $t\approx 15$) require $2^{17}$ or more grid
points to be in the convergent regime.
Running simulations to several hundred time units at this resolution
is a daunting task, so we present only preliminary results here. 
See Appendix \ref{s:convergence} for a
detailed discussion of the convergence of our methods.

\begin{figure}[!t]
  \centering
  \begin{subfigure}[t]{0.48\textwidth}
    \includegraphics[width=\textwidth]{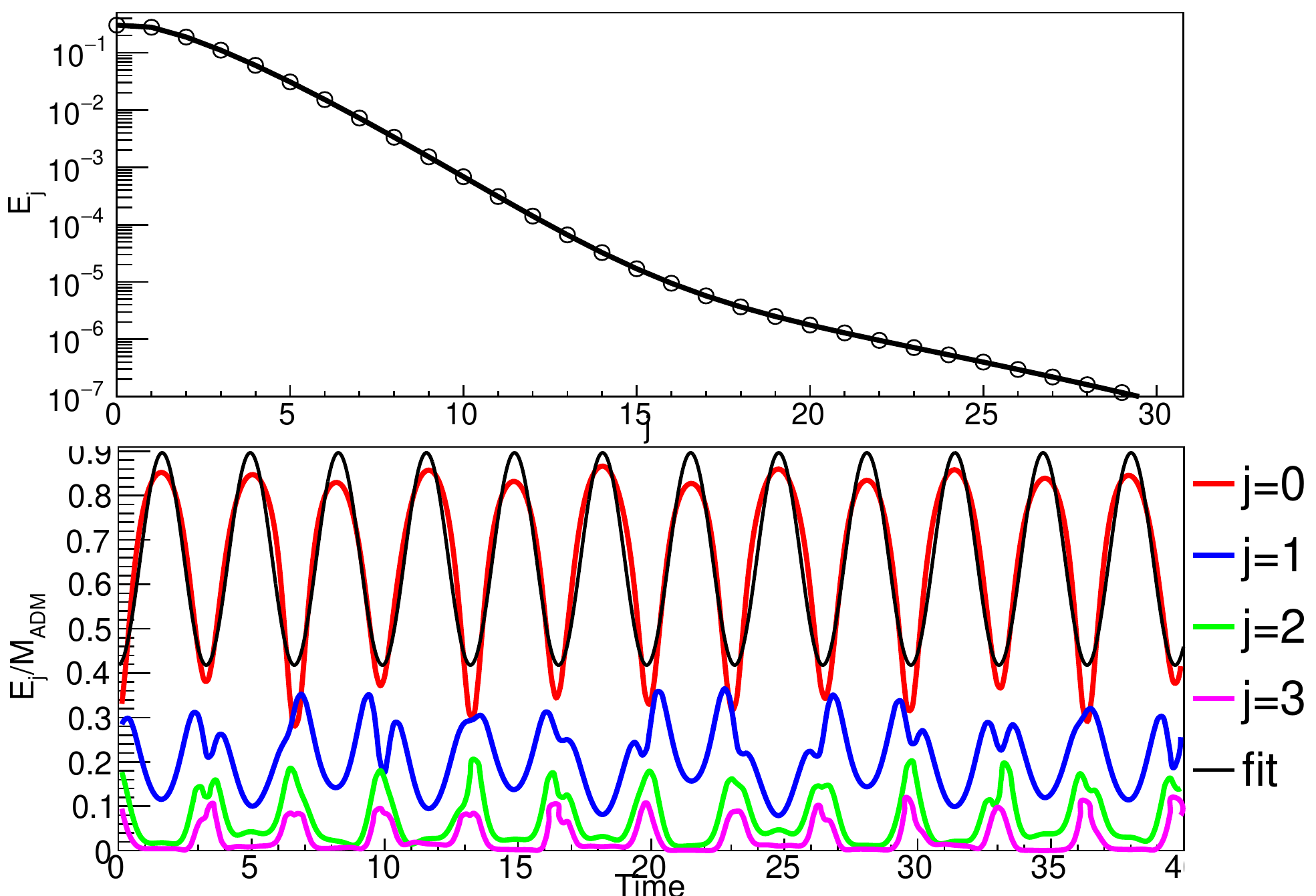}
    \caption{Top panel: $\hat E_j(t=0)$.
      Bottom panel: evolution of $\hat E_j$ in the
      lowest four modes for $\epsilon=15$. Fit to $\hat E_0$ is
      $p_0\cos(\Omega t+p_1)+p_2$. }
    \label{fig:PiGauss100Energy}
  \end{subfigure}\hfill
  \begin{subfigure}[t]{0.48\textwidth}
    \includegraphics[width=\textwidth]{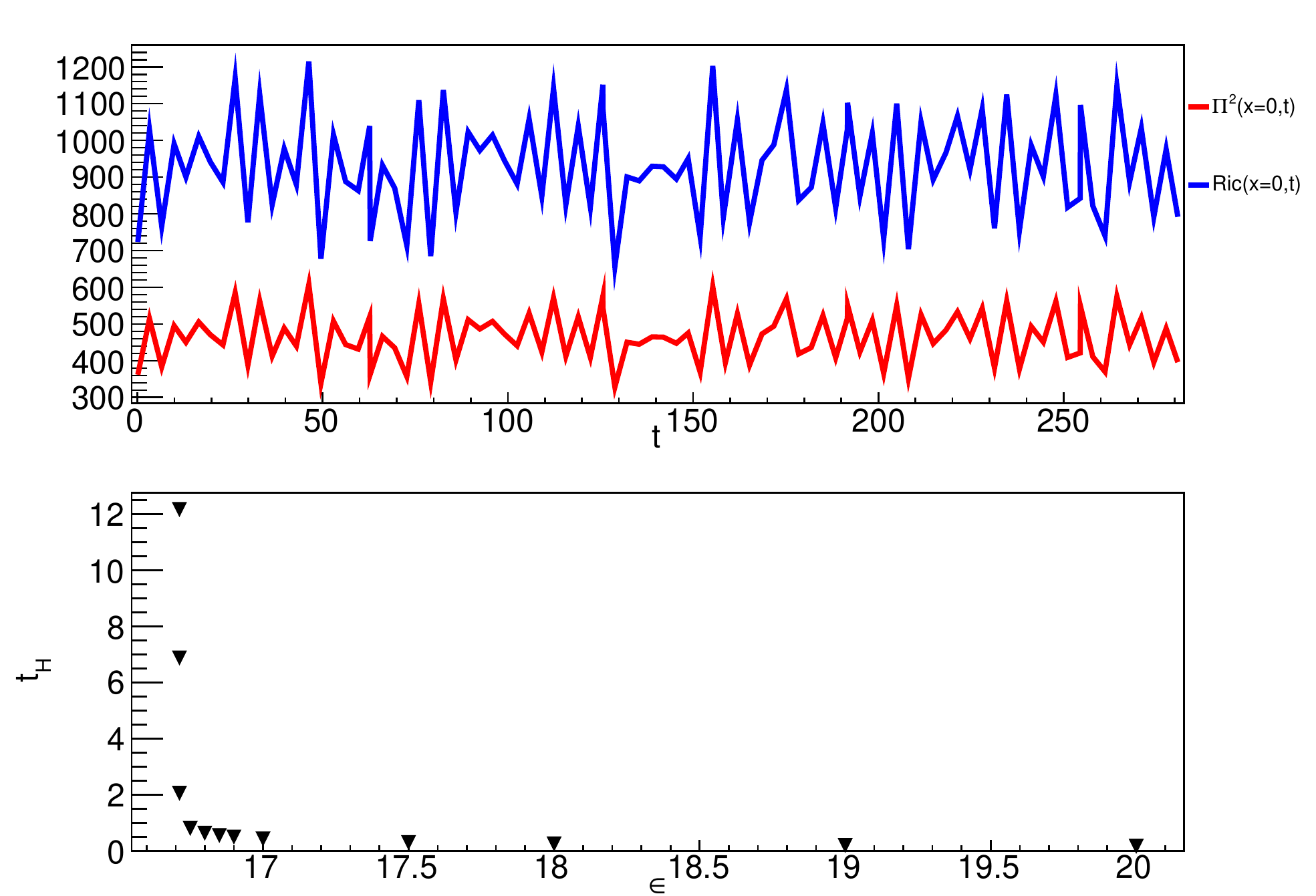}
    \caption{Top panel:  upper envelope across every
      seven reflections from the boundary for
      $\Pi^2(t,x=0)$ and $\mathcal{R}(t,x=0)$.
      Bottom panel: $t_H$ vs $\epsilon$.}
    \label{fig:PiGauss100Time}
  \end{subfigure}
  \caption{Evolution for $\mu=100$, $\sigma=1/16$. 
    The reduced upper envelope is plotted because the upper envelopes 
    vary rapidly.}
  \label{fig:PiGauss100}
\end{figure}

We are interested in data that is not a perturbation about a single
mode, so we choose a narrow pulse of $\sigma=1/16$; the width of the
Gaussian that best fits $e_0(x)$ for $\mu=100$ 
is $\sigma\approx 7/50$ (for $\mu=20$, it is $\sigma\approx3/10$).
We plot the energy spectrum at $t=0$ in the top panel
of figure \ref{fig:PiGauss100Energy}. The two lowest modes have nearly
equal energy, and an exponential decay in the higher modes, so this is
not single-mode dominated, at least initially.
We also show the evolution of the energy in the 
lowest four modes in the bottom panel of figure \ref{fig:PiGauss100Energy}.  
Compared to other solutions (such as for $\mu=20$ discussed above),
we find a strikingly periodic and rapid transfer of energy into and out of 
the lowest mode. Based on a fit $p_0\cos(\Omega t+p_1)+p_2$ to 
$E_0$,\footnote{We use $\Omega$ to distinguish from the eigenfrequencies 
$\omega_j$.} we find a frequency approaching $\Omega\approx 2$ as the
amplitude is decreased. Interestingly,
we also find that the amplitude of the oscillations decrease for
smaller perturbations.

As we have noted, the stable solutions found so far are perturbations
about single mode solutions, typically with exponentially decaying spectra
at higher modes. In figure \ref{fig:PiGauss100Time}, we provide 
some suggestive evidence that this initial data may be
stable despite initially having nearly equal energies in the $j=0,1$ modes
(and significant energy in the next two modes). We plot the
horizon formation time in the bottom panel, and $\Pi^2(t,x=0)$ and
$\mathcal{R}(t,x=0)$ for $\epsilon=15$ in the top panel. Tracking
the rapid increase in formation time has proven surprisingly
difficult, which means that whether or not this data truly is stable
is an open question. Nevertheless it is clear that increasing the mass
of the scalar results in surprisingly different dynamics from what is
seen in the massless case.

\section{Discussion}\label{s:discussion}

The question of stability of perturbations of AdS against gravitational 
collapse (either absolutely or on time scales of order $\epsilon^{-2}$ for
amplitude $\epsilon$) is intriguing both as a problem in gravity and 
through its connection to thermalization in strongly-coupled gauge theories
via the AdS/CFT correspondence.  We have investigated what classes of
initial data might lead to stable evolution for both massless and massive
scalar fields.

As reviewed in the introduction, it is well-established that perturbations
around single eigenmode oscillons (or boson stars) are stable, at least on
long time scales.  We propose, therefore, that one criterion for stability
is that the energy spectrum is initially dominated by a single scalar 
eigenmode; this is compatible with the conjecture of \cite{Green:2015dsa}
that stable solutions orbit quasi-periodic solutions in configuration
space.  It is simple to check that most of the stable solutions for massless 
scalars present in extant literature are, in fact, single-mode-dominated.
We have provided a check that the stable three-Gaussian initial data
of \cite{Okawa2015} are single-mode-dominated, as well.  In addition, 
we studied the somewhat controversial equal-energy two-mode initial data
discussed also in \cite{Balasubramanian2014,Buchel2015,Green:2015dsa,Bizon2014a}
and and find that claims of stability based on amplitudes studied in the 
literature so far are premature, though stability at smaller amplitudes 
remains an open question.
Specifically, for the amplitudes of initial data studied in the 
extant literature, according to a fixed
approximate measure of horizon formation, gravitational collapse occurs
at a time proportional to $\epsilon^{-2}$.  We conclude that the disagreement
in the literature over stability of this initial data at small amplitudes
is due to the use of insufficient resolution in the numerical solutions of
\cite{Balasubramanian2014}. It is also important to note that the
perturbative TTF solutions require high resolution (in the sense of requiring
a large number of eigenmodes); even a small fraction of energy in high
$j$ modes can trigger gravitational collapse since horizon formation is an
essentially local process.  Furthermore, other than comparison to a 
numerical solution of the fully nonlinear problem, there is no simple 
diagnostic in the literature for the failure of a perturbative solution, 
so it is not clear that perturbative methods are a reliable gauge of 
stability even over a fixed time scale.

Starting in section \ref{s:massive}, we present a thorough overview of
gravitational collapse for massive fields in AdS$_5$.  We consider 
evolution for the six regions of parameter space determined by whether
each dimensionless ratio  $\lambda\mu,\mu\ell,\lambda/\ell$ is larger or 
smaller than unity and confirm in some examples that our qualitative 
results for each parameter range are robust against changes in the initial
data.  We also confirm that initial data in most of these parameter ranges
collapse with $t_H\propto\epsilon^{-2}$, as expected from perturbation theory.
In the process, we further discovered two types of novel behavior related to
thermalization and stability for initial data with intermediate widths.

First, for scalars with mass less than the AdS scale, we noted, in addition
to the more familiar sudden increases in the initial horizon radius $x_H$,
also the appearance of sudden decreases in $x_H$ with decreasing amplitude.
Both phenomena are related to the threshold for the metric function $A(t,x)$ 
used to determine approximate horizon formation and the appearance of 
multiple local minima in $A$ at a fixed time.  In some cases, as the
amplitude decreases, the first-formed minimum of $A$ never reaches the 
threshold, but subsequently formed local minima created by additional 
infalling matter at larger $x$ can reach the threshold.  This leads to
sudden increases in $x_H$.  However, we have shown that the opposite case
can occur; sometimes decreasing the amplitude of the initial data actually
allows the first-formed local minimum of $A$ to decrease beyond a threshold
that it cannot reach at a slightly higher amplitude.  
This puts a spotlight on how approximate horizon formation is
defined due to the sensitivity of results to the threshold. It also 
motivates the development of other criteria either for 
approximate horizon formation or thermalization in the dual field theory.

Our most important result is to provide evidence of a new type of
(quasi-)stable field profile for massive fields and intermediate
width $1/\mu<\lambda<\ell$.  For $\mu=100, \sigma=1/16$, there is an 
apparent critical amplitude, below which we have not been able to find
horizon formation.  For $\mu=20, \sigma=1/10$, we find with decreasing
amplitude first the usual increase of $t_H$, then a sudden jump 
from $t_H\approx 37.7$ to $t_H\approx 126$ followed by a decrease to 
$t_H\approx 75.6$, then a rapid increase in $t_H$.  While we have not yet
ascertained whether there might be a critical amplitude, our results do
argue for stability over time scales of order $\epsilon^{-2}$.
In both cases, the energy spectrum is \textit{not} dominated by a single
mode, unlike for other known stable solutions, but rather is initially 
spread democratically through several modes, possibly indicating the 
existence of a quasi-stable multi-mode oscillaton solution.  Following the 
hypothesis \cite{Green:2015dsa} that stable evolutions are orbits of 
quasi-periodic solutions, it would be interesting to determine if 
quasi-periodic solutions of the perturbation theory for a massive scalar
allow energy to be distributed more evenly between modes.  An alternate
possibility is that the ``island of stability'' for perturbations around
single-mode oscillons or quasi-periodic solutions is much wider for
heavy scalars in an AdS manifestation of the mass gap found in asymptotically
flat spacetime \cite{Brady1997}.
Another important point is that our $\mu=20, \sigma=1/10$ solutions in
the region when $t_H$ jumps from $37.7\to 126\to 75.6$ show a significant
agreement up until the time of collapse, even though the final horizon
formation time differs greatly.  There is no clear indication
in advance whether a given amplitude will collapse
sooner or later.\footnote{To re-emphasize, these results have been validated 
by convergence testing and show the expected 4th order convergence up to
collapse.}
We leave a more careful study of these potentially novel stable solutions
for future work. 

In summary, there are two major lessons from our work.  First is that,
while there is by now a well-developed leading-order perturbation theory
for scalar gravitational collapse in AdS, comparison to numerical solutions
of the full nonlinear theory shows that it is very difficult to predict
when the perturbative theory will break down.  We have given several 
explicit examples of this difficulty; as a result, it is difficult to know
how long a perturbative solution remains reliable (for example in the sense
of \cite{Dimitrakopoulos:2015pwa}).  Second, all the known stable or 
quasi-stable
evolutions of massless scalars in AdS are dominated by a single eigenmode
of the scalar on a fixed AdS background, whereas quasi-stable evolutions
of the massive scalar can have energy democratically spread through several
eigenmodes.  These massive scalar solutions call for future investigation.

\acknowledgments

We gratefully acknowledge many helpful conversations with P.~Bizo\'n,
A.~Buchel, S.~Green, L.~Kidder, G.~Kunstatter, L.~Lehner,
A.~Rostworowski, and S.~Teukolsky.
The work of ND is supported in part by NSF Grants PHY-1306125 and
AST-1333129 at Cornell University, and by a grant from the Sherman
Fairchild Foundation.
The work of AF is supported by the Natural Sciences
and Engineering Research Council of Canada Discovery Grant program.  
This research was enabled in part by support provided by WestGrid 
(www.westgrid.ca) and Compute Canada Calcul Canada (www.computecanada.ca).

\appendix

\section{Convergence tests and code validation}\label{s:convergence}

\begin{figure}[!t]
\begin{center}\includegraphics[scale=0.5]{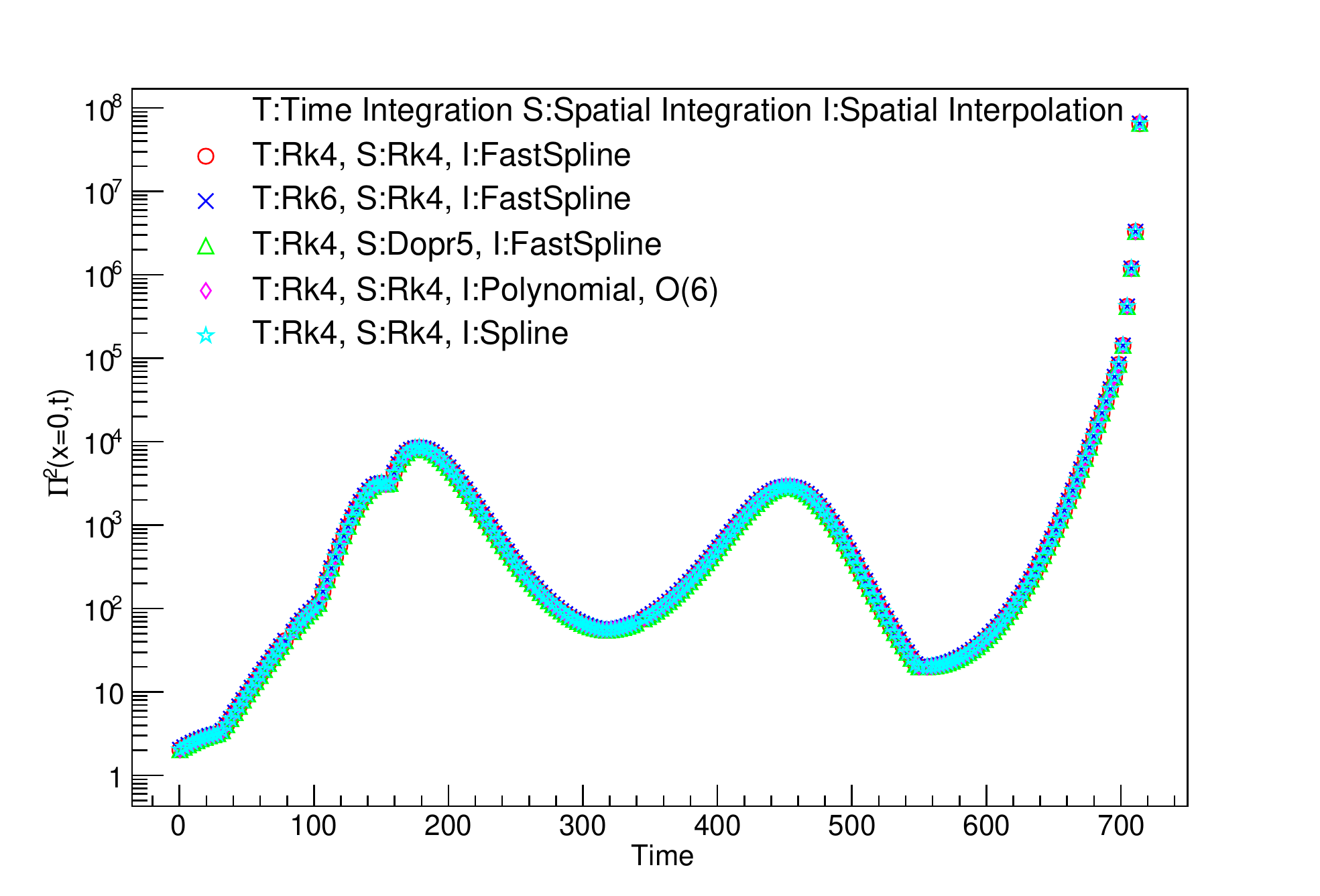}
\end{center}
\caption{\label{f:consistency}$\Pi^2(t,x=0)$ as in figure 
\ref{fig:twoModeAds4Pi} with $\epsilon=0.109$ for several numerical methods}
\end{figure}

As mentioned at the end of section \ref{s:methods}, in addition to the
standard RK4 spatial integration, we also use a fifth order DoPr method. 
The DoPr reduces
the resolution needed to bring massive solutions to collapse. When using
RK4, we often need a spatial resolution of $2^{18}$ grid points 
near the end of long 
simulations (or those near critical points) to prevent crashes.
On the other hand, with DoPr, it often suffices to use a resolution of 
$2^{15}$ grid points, but DoPr is significantly slower than RK4 at the same
resolution.  We have also tried different spatial integrators.  
To validate our code, we have checked that several different methods give
consistent results over a long simulation of equal-energy two-mode 
ID with $\epsilon=0.109$, as discussed in section \ref{s:twomode}.
As shown in figure \ref{f:consistency}, these methods show a remarkable
agreement for $\Pi^2(t,x=0)$.  Furthermore, they agree on $t_H$ with a 
relative error of $4\times 10^{-5}$ and on $x_H$ to within one grid point.

There are two quantities we monitor for consistency of the solutions. 
The system (\ref{eq:scalarField})
has a constraint, $\mathcal{C}:=\phi_{,x}-\Phi=0$, and the ADM mass should
be conserved by the time evolution.  Since explicit
convergence tests for each simulation are too demanding, we typically 
monitor the ADM mass and $\mathcal{C}$ for consistency. 
For small and zero $\mu$, this is usually sufficient because
we find that the mass rapidly decreases
as convergence is lost  (see figure \ref{fig:massloss}).
This happens because the pulses slip between the mesh
if insufficient resolution is used.
However, as $\mu$ is increased the
ADM mass is conserved well and $\mathcal{C}$ will remain small even
when convergence is lost.  A careful
study of the evolution when this occurs shows that it is the formation
of several small minima in $A$ that are under resolved. These minima
each reach a different value and the narrower, farther out (radially)
ones are ultimately those that trigger horizon formation. We find that
in many cases at least $2^{17}$ grid points are required to resolve
these minima, making the computations incredibly expensive. Nevertheless
for particularly long or interesting simulations, we have run explicit
convergence tests to provide credibility for our results.

We perform an array of convergence tests to determine
the reliability of our results, which are presented here for $\mu=20$,
$\sigma=0.1$, and amplitude $\epsilon=11.74$.  This solution has a large
$t_H$ compared to slightly larger or smaller amplitudes, so it is an 
interesting test case for convergence and consistency of the code.

\begin{figure}[!t]
  \centering
  \begin{subfigure}[t]{0.31\textwidth}
    \includegraphics[width=\textwidth]{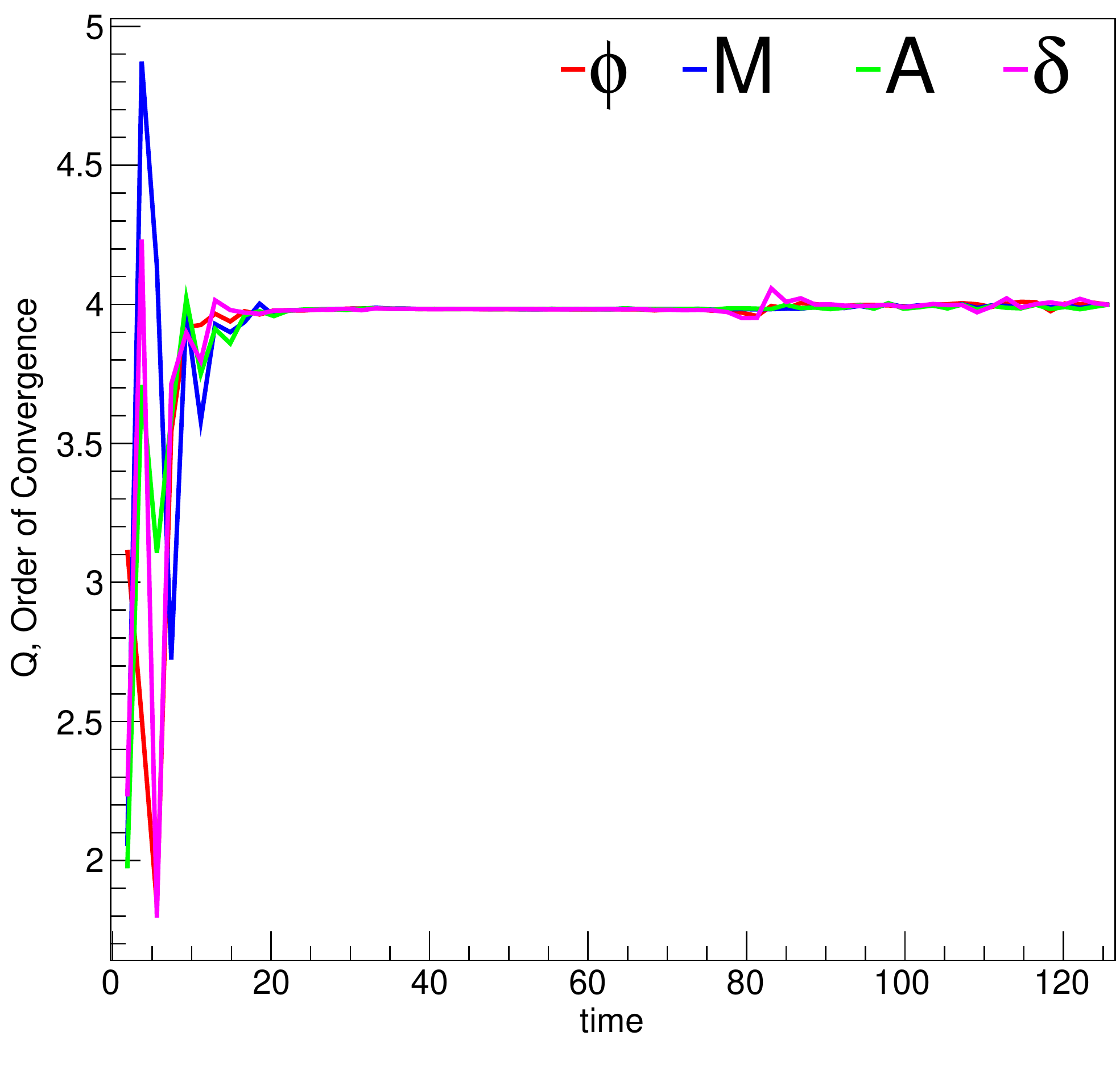}
    \caption{Global order of convergence of $\phi, M, A,\delta$}
    \label{fig:convergence}
  \end{subfigure}\hfill
  \begin{subfigure}[t]{0.31\textwidth}
    \includegraphics[width=\textwidth]{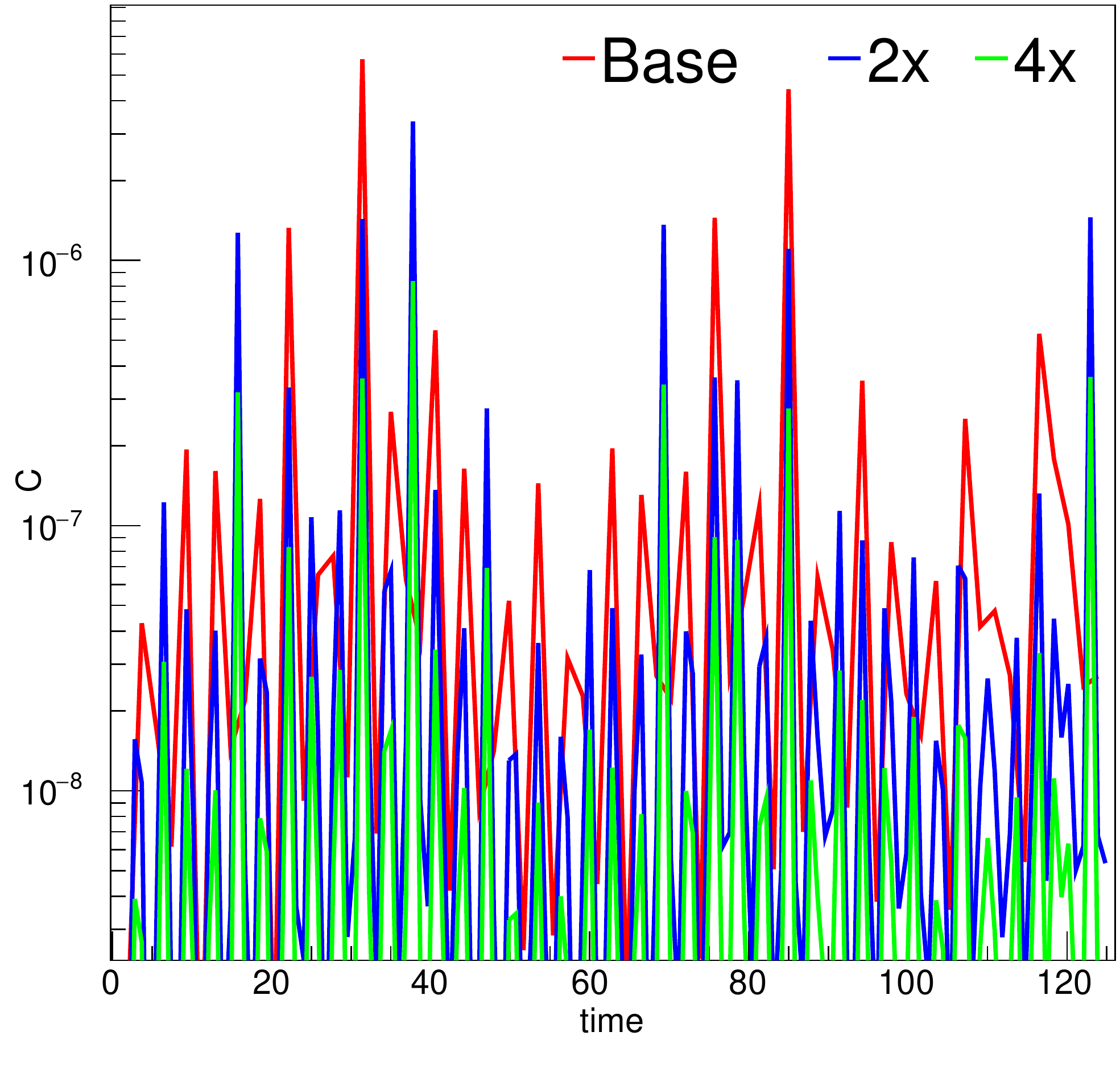}
    \caption{Constraint violations $\mathcal{C}$}
    \label{fig:constraintConv}
  \end{subfigure}\hfill
  \begin{subfigure}[t]{0.31\textwidth}
    \includegraphics[width=\textwidth]{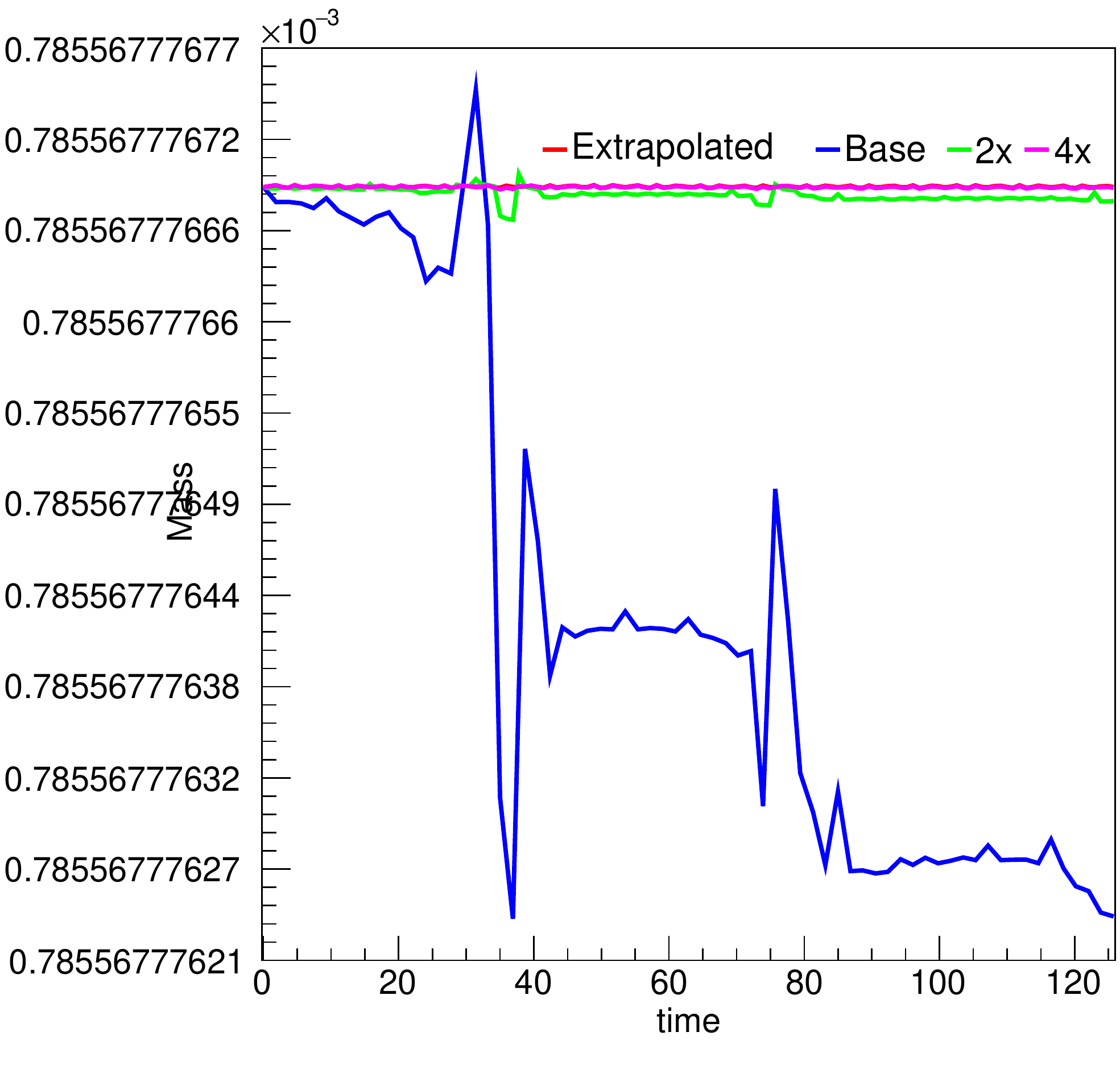}
    \caption{ADM mass and Richardson extrapolated mass for 
      coarsest resolution}
    \label{fig:massConv}
  \end{subfigure}
  \begin{subfigure}[t]{0.31\textwidth}
    \includegraphics[width=\textwidth]{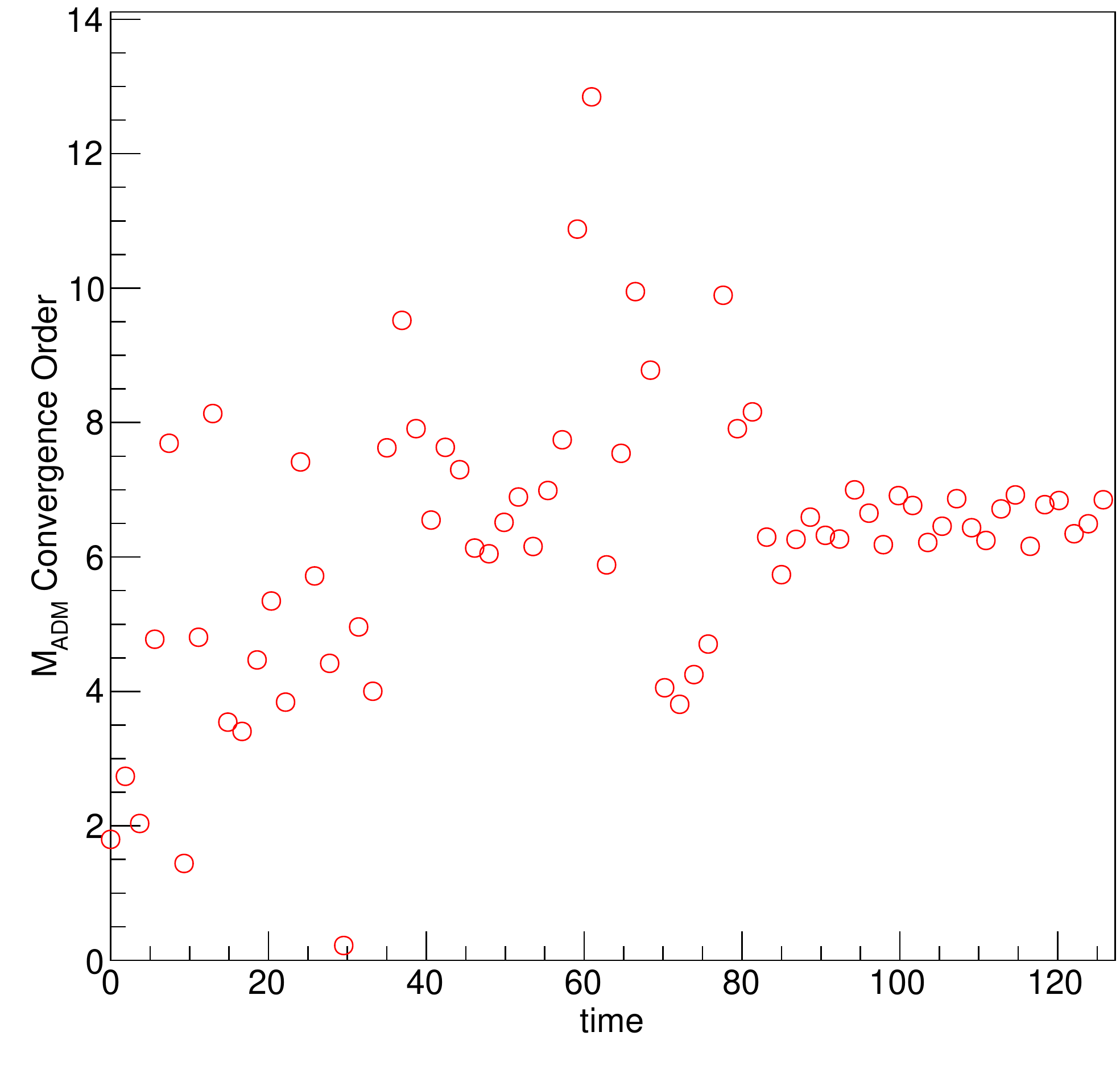}
    \caption{Convergence of ADM mass}
    \label{fig:AdmMassConv}
  \end{subfigure}\hfill
  \begin{subfigure}[t]{0.31\textwidth}
    \includegraphics[width=\textwidth]{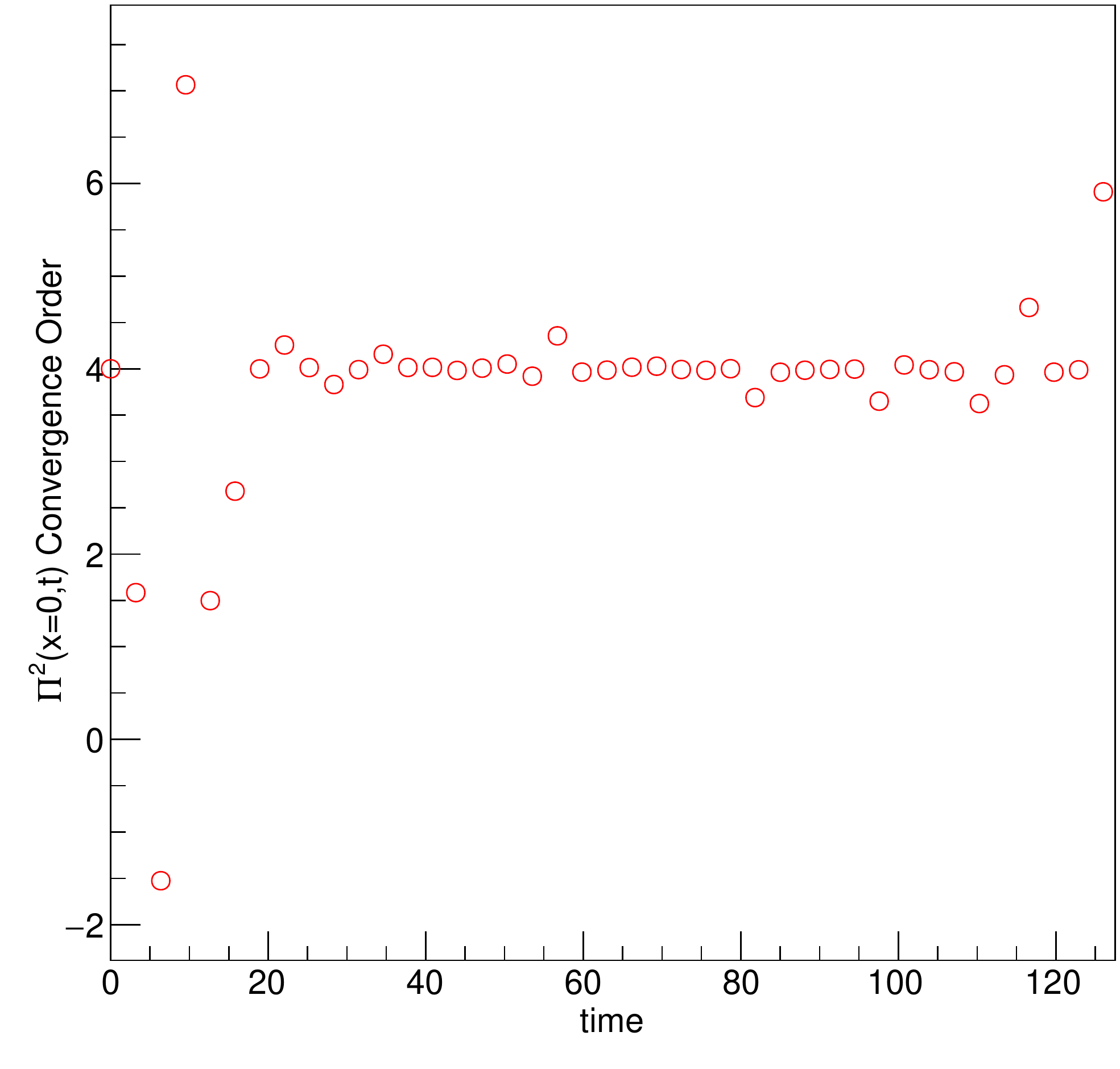}
    \caption{Order of convergence of $\Pi^2(x=0,t)$}
    \label{fig:PiOriginConv}
  \end{subfigure}\hfill
  \begin{subfigure}[t]{0.31\textwidth}
    \includegraphics[width=\textwidth]{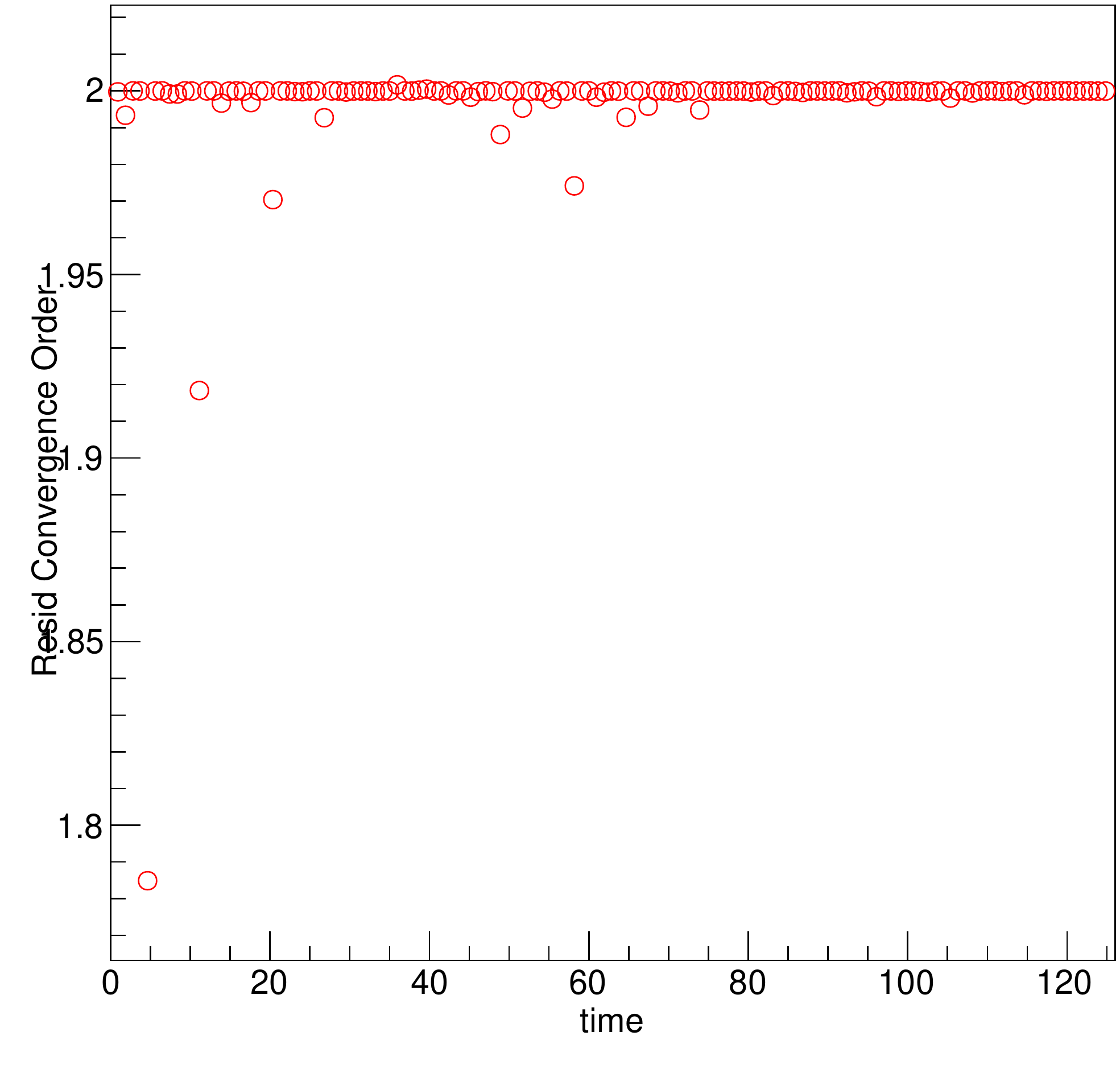}
    \caption{Order of convergence of constraint violation $\mathcal{C}$.}
    \label{fig:ResidConv}
  \end{subfigure}
  \caption{Several measures of the accuracy and convergence
  of our numerical methods. The base resolution is $2^{14}+1$
  with refinement by factors of two.
  The test was performed using the DoPr spatial integrator. 
  The initial data used eq.~(\ref{eq:PiGaussianID}) with 
  $\mu=20,\sigma=0.1,\epsilon=11.74$.}
  \label{fig:convPlots}
\end{figure}

Figure \ref{fig:convergence} demonstrates that the scalar field,
mass function, and metric functions $A$ and $\delta$ are
fourth-order convergent (in $L^2$ norm) through multiple reflections off
the outer boundary, and figure \ref{fig:constraintConv} shows that the
constraint residual $\mathcal{C}$ decreases with increasing resolution.
Figure \ref{fig:massConv} shows that the ADM mass is well-conserved
for the duration of our simulation. We plot the ADM mass at all three
resolutions as well as the Richardson extrapolated value.\footnote{See
\cite{evans1989frontiers} for a discussion of this technique as a
measure of consistency with adaptive mesh refinement.}
The order of convergence of the ADM mass is
plotted in figure \ref{fig:AdmMassConv} and exhibits
approximately sixth order convergence. Because of the variable step
size method used with the DoPr integration, the ADM mass has a higher
order of convergence than expected solely from the time evolution.
Since $\Pi^2(t,x=0)$ has
been used to detect the onset of the turbulent instability. We
show convergence of this quantity in figure \ref{fig:PiOriginConv}.
Finally, we plot the convergence of the
constraint residual in figure \ref{fig:ResidConv}, which demonstrates
that the during the evolution we stay on the constraint sub-manifold.
In post processing we used a second order method, unlike in the actual
time evolution where we use a fourth order method. The results of our
testing demonstrate that our code is convergent and consistent even
for long simulations with several reflections off the outer boundary,
as well as for large scalar field mass values.

\providecommand{\href}[2]{#2}\begingroup\raggedright\endgroup

\end{document}